\documentclass[mathleft
]{an}
\usepackage{times}
\usepackage{graphics,graphicx}
\usepackage{natbib,geometry}
\usepackage[section] {placeins}
\usepackage{amsmath,amssymb,amstext,amscd,bm}

\begin{document}

\Pagespan{789}{}
\Yearpublication{2015}%
\Yearsubmission{2015}%
\Month{11}%
\Volume{999}%
\Issue{88}%

\title{A Dynamical Evolution Study of  40 2MASS Open Clusters}

\author{Orhan  G\"une\c{s}\inst{1,}\inst{2}\fnmsep\thanks{Corresponding author:
  \email{or.gunes@gmail.com}\newline}
\and  Y\"uksel Karata\c{s}\inst{2}
\and  Charles Bonatto\inst{3}
}
\titlerunning{Instructions for authors}
\authorrunning{T.H.E. Editor \& G.H. Ostwriter}
\institute{
Department of Astronomy and Space Sciences, Faculty of Arts and Sciences,
Erciyes University, Talas Yolu, 38039, Kayseri, Turkey 
\and 
Istanbul University, Science Faculty, Department of Astronomy and 
Space Sciences, 34119, \"Universite-Istanbul, Turkey, \email{karatas@istanbul.edu.tr}
\and 
Universidade Federal do Rio Grande do Sul, 
Departamento de Astronomia, CP\,15051, RS, Porto Alegre 91501-970, 
Brazil, \email{charles@if.ufrgs.br}
}



\abstract{
We investigate the dynamical evolution of 40 open 
clusters (OCs) by means of their astrophysical parameters derived 
from field-decontaminated 2MASS photometry. We find a bifurcation 
in the planes core radius vs. age and cluster radius vs. age, in 
which part of the clusters appear to expand with time probably due 
to the presence of stellar black holes while others seem to shrink 
due to dynamical relaxation. \\
Mass functions (MFs) are built for 3$/$4 of the sample (31 OCs), 
which are used to search for indications of mass segregation and 
external dynamical processes by means of relations among 
astrophysical, structural and evolutionary parameters. 
We detect a flattening of MF slopes ocurring at the evolutionary 
parameters $\tau_{core}\leq 32$ and $\tau_{overall}\leq 30$, 
respectively. \\
Within the uncertainties involved, the overall MF slopes of 14 
out of 31 OCs with $m_{overall} > 500~M_{\odot}$ are consistent with 
Kroupa's initial mass function, implying little or no dynamical 
evolution for these clusters. The remaining 17 OCs with MF slopes 
departing from that of Kroupa show mild/large scale mass 
segregation due to dynamical evolution.}

\maketitle

\section{Introduction}

The internal dynamical processes of open clusters (OCs) are  
mass loss during stellar evolution, mass segregation and evaporation 
of its stellar content with time. Tidal interactions with the Galaxy's disc and bulge, 
as well as collisions with Giant Molecular clouds (hereafter GMCs) are 
the main external dynamical effects upon OCs. Because of these dynamical interactions, as 
clusters age, their structures are subject to considerable 
changes, and may even be dissolved in the Galactic field. 
A massive cluster can be dissolved by central 
tidal effects in $\approx$ 50 Myr \cite{por02,ber01}. This time 
is much shorter than $\sim 1\; Gyr$  found for most OCs 
within the Solar circle \cite{bon06a}.  
Interactions with the galactic disc, the tidal pull of the 
Galactic bulge and collisions with GMCs destroy more easily 
the poorly-populated OCs, on a time-scale of $10^{8}$ yr , 
particularly inside the Solar circle \cite{ber01}.  

A cluster loses low-mass stars from its outer regions into the field by stellar 
evaporation.  As a result of this mass segregation, low-mass stars are transferred from its 
core to the cluster's outskirts while massive stars accumulate in 
the core \cite{bon05,sch06}. This results in a flat mass function (MF hereafter) in 
the core and steep one in the halo.
These external and internal dynamical processes play different 
roles, depending on the location of an OC with respect to the 
Solar circle: old OCs with Age $>1$ Gyr tend to be concentrated in 
the anti-centre, a region with a low density of GMCs \cite{van80, 
cam09}. Tidal shocks from the Galaxy and from GMCs and observational  
incompleteness or biases are responsible for the scarcity of OCs in direction 
to the Galactic center \cite{bon07a}. Due to absorption and crowding 
in regions dominated by disc and bulge stars, the  OCs' observational completeness 
is decreased.  With the effect of tidal interaction, an OC heats and its stars 
gain kinetic energy, which leads to an increase in the evaporation rate.

In this paper we have considered 40 OCs with 2MASS JH${K_{s}}$ photometric data, which are
selected in respective to the cluster location and the age (Age$\ge100$ Myr) from WEBDA OC and Dias et al. catalogues \cite{mer92,dia12}.
These OCs have been considered to study their dynamical evolution, particularly in dependence of their 
location in the Galaxy.  We state that our sample is relatively small but our work have the advantage of being
based on a uniform database, in the sense that we determine the parameters following
the same methods, based on the same kind of photometry.
The robust structural parameters have been  
derived from high-contrast stellar radial density profiles 
following the method of  \cite{bon07a}, and the ages were derived 
from a fit of isochrones to decontaminated colour-magnitude 
diagrams of the 2MASS JH${K_{s}}$ photometric data. 
As can be seen from the WEBDA database, CCD-based CMDs of these 40 OCs 
are also available. We stress that the CMDs presented here go fainter than is available there.

From our sample of young, intermediate and old OCs ($100\; Myr \leq Age \leq 5\; Gyr$), 
the relations between the dynamical evolution indicators  and cluster radius 
(R$_{RDP}$, hereafter), core radius (R$_{core}$, hereafter), mass, mass function slope $\chi$, 
mass density $\rho$, evolutionary parameter $\tau$-, and the parameters $(Age, d, R_{GC}, z)$ 
have been derived  and compared with the values given in the 
literature. Here, d, R$_{GC}$, and z denote  heliocentric, 
galactocentric (hereafter R$_{GC}$), and Galactic plane distances, 
respectively. Such relations have been studied by  \cite{lyn82}, 
\cite{jan94},\cite{nil02}, \cite{tad02}, \cite{bon05}, 
\cite{sch06}, \cite{sha06},  \cite{bon07a}, \cite{mn07}, 
\cite{buk11}, and \cite{cam09}. In this paper, 
R$_{\odot}$$=7.2\pm0.3$ kpc  which is based on the 
updated distances of Galactic globular clusters \cite{bic06b}, is taken through this paper.

This paper is organised as follows: the  selection of the OCs is 
presented in Section 2. In Section~3 the 2MASS JH${K_{s}}$ 
photometry and the field star decontamination algorithm (employed 
in the CMD analyses) are given. The derivations of astrophysical 
and structural parameters, mass and mass functions, relaxation 
time and evolutionary parameter are presented  in Sections~4 to 6. 
Section 7 is devoted to Results, which 
contain the following subsections: 7.1 the relation 
between $R_{RDP}$ and $R_{core}$, 7.2 relations of cluster dimensions 
with distance and age, 7.3 the relations between $R_{RDP}$ and $Age$ 
and $R_{core}$ with $Age$, 7.4 the relations $R_{RDP}$ with 
$R_{GC}$ and $R_{core}$ with $R_{GC}$, 7.5 the spatial 
distribution of the 40 OCs in the Galaxy, 7.6 relations between 
the overall mass with ($R_{RDP}$, $R_{core}$) and with $(Age, 
R_{GC})$, 7.7 the relations between the mass density with $MF$ 
slopes, $Age$, $R_{RDP}$ and with $R_{GC}$, 7.8 the relation 
between the $MF$ slopes and the evolutionary parameter, and a 
comparison with Kroupa's IMF. Conclusions are presented in Section~8.

\section{Open cluster sample and Spatial distribution}

We applied two criteria to select the OCs for our work from WEBDA 
OC and Dias et al. catalogues \cite{mer92,dia12}. Namely, the 
cluster location in the Galaxy and their ages, see  Fig.~1. 
In order to study dynamical evolution of middle- and older-age 
OCs,  40 OCs with $100\; Myr \leq Age \leq 5\; Gyr$ as a  
function of the Galactic location (see Fig. 1, slices I-IV) from 
the 2MASS data base are considered. 
The location criteria is important because 
the longevity/survival rate of the OCs are related to the Galactic 
slices  inside/outside the Solar circle. Over 40 OCs in WEBDA OC  
\cite{mer92} and \cite{dia12} catalogue have been considered. The 
OCs which were not appropriate to the decontamination technique 
of field stars were eliminated by  examining their decontamination 
surface density distributions (see sect.~3). Thus the sample size  
resulted to be 40 OCs. We are aware that the sample is not large but we 
intended that the sample with robust parameters 
would be significant to address the dynamical problems mentioned 
earlier. From the 40 OCs, we have also studied the relations 
between the parameters (Age, d, R$_{GC}$, $z$) and  dynamical 
indicators (R$_{core}$,~R$_{RDP}$, $m$, $\chi$, $\tau$).

\begin{figure}      
\centering
\includegraphics*[width = 7cm, height = 7cm]{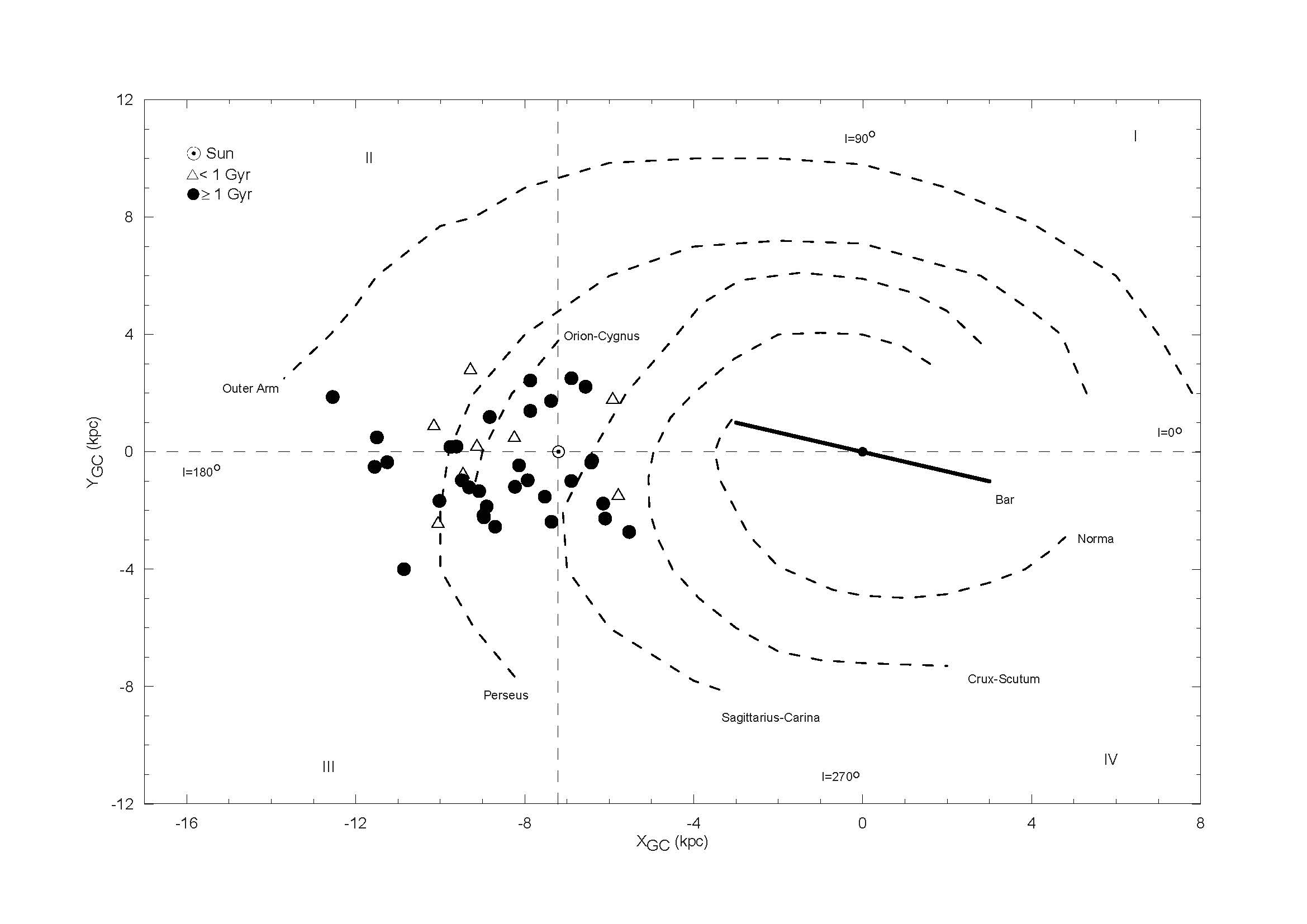}
\caption {Spatial distribution (X,Y) of 40 OCs.  
Open triangles and filled circles represent OCs with ages younger than 1 Gyr 
and older than 1 Gyr, respectively. The schematic projection of the Galaxy is seen 
from the North pole. The Sun's distance to the 
Galactic center is taken to be 7.2 kpc of \cite{bic06b}.} 
\end{figure}

1148 OCs out of 2000 OCs in \cite{dia12} catalogue, which can be 
considered the most representative for our purpose have 
age determinations. As can be seen from Table 1, there are 13 OCs 
with $200\; Myr \leq Age < 1\; Gyr$  (3.5 \%), 26 OCs  with $1\; 
Gyr \leq Age <5 Gyr$ (14.5 \%) and one with $Age \le 200\; Myr$ in 
our sample. 

\begin{table} 
\tiny
\centering
\tiny
\caption{Comparison of our cluster sample to the catalogue of 
\cite{dia12} for the age data.}
\begin{tabular}{cccc}
\hline
Age (Myr) & N~(This~work) & N(Dias) & Percentage (\%) \\
\hline
Age $<$               200   & 1     & 579   & 0.17 \\
200 $\leq$ Age $<$    1000  & 13    & 373   & 3 \\
1000 $\leq$ Age $<$   5000  & 26    & 179   & 15 \\
5000 $\leq$ Age $\leq$10000 & -     & 17    & - \\
\hline
Total & 40    & 1148  & 3 \\
\hline
\end{tabular}
\end{table}

The spatial distribution in (X, Y) plane together spiral arms \footnote{(X, Y) is a right handed 
Cartesian coordinate system with the Sun on its center, 
with the X axis pointing towards the Galactic centre and the Y axis pointing in 
the disc rotation direction.} of the 40 OCs is displayed in Fig.~1.
As seen from the Fig.~1, our sample comprises the OCs of four Galactic slices (I-IV). 
Note that the number of OCs towards the anti-center in Fig.~1 is larger than the ones toward the Galactic center directions. 
Six out of eight OCs with Age $ < 1\, $ Gyr fall in the Galactic anticenter directions, 
whereas the remaining two occupy the Galactic center direction. 
This is because the OCs in Galactic center directions cannot be observed due to strong absorption, 
crowding or were systematically dissolved by different tidal effects such as high frequency of collisions with GMCs \citep{gie06}.

The majority of OCs with Age $\ge$ 1 Gyr lies outside the Solar circle.
From Fig.~1, one readily sees that the number of OCs inside 
the Solar radius is biased in direction of the Galactic center. The 
reason is that the inner Galaxy clusters cannot be observed because of 
strong absorption and crowding, or because they have been dissolved by 
a  combination of tidal effects. In a good measure the latter is caused 
by the expected higher frequency of collisions with GMCs in that direction 
\cite{gie06,cam09}. From an inspection of Fig.~1, there 
are more OCs in the anticentre direction than in the opposite direction, 
in agreement with \cite{van80}, who find that the OCs with Age $\ge$ 
1.0~Gyr tend to be concentrated in the anticentre, which is a region with 
lower density of GMCs. Our sample has small statistics to draw 
significant conclusions in that respect. However, statistically, working with a representative
sub-population of the Galactic OCs minimizes the occurrence of biases in the analyses.

Finally, to put the present OC sample in context, in Fig.~2 we compare some observational data
together with fundamental parameters (derived in subsequent sections) with the corresponding
ones found in OC databases. This analysis is also important for checking for the presence of
systematic biases in our sample. For this analysis we use the parameters derived by
\cite{kha13} for 3006 OCs. The advantage of their work is that the parameters
follow from a systematic and uniform analysis. Since \cite{kha13} do not
provide cluster mass, we take such values from \cite{pis08}, although for a
smaller number of OCs, 236.

\begin{figure}     
\centering
\includegraphics*[width = 7cm, height = 11cm]{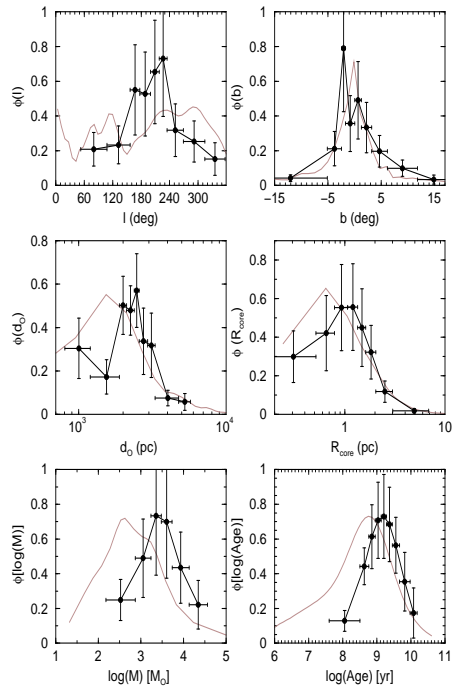}
\caption {Normalized distribution functions of our OC sample (circles)
compared to those of \cite{kha13} and \cite{pis08} (solid
line).}
\end{figure}

Our analysis compares distribution functions of the several parameters between both
sets,  as is seen from Fig.~2. Uncertainties in the parameters have been incorporated 
into the respective distribution function. And, since the samples differ significantly in the number of OCs, the distribution 
functions have been scaled to provide the best visual comparison between both.
The top panels of Fig.~2 show how the OCs distribute with respect to the Galactic longitude
(left) and latitude (right). Clearly, most of our sample corresponds to clusters directed
towards the 2nd and 3rd Galactic quadrants. Regarding Galactic latitude, our sample tend
to avoid the plane. In terms of distance from the Sun (middle-left), our sample is
somewhat consistent with that of \cite{kha13}, particularly for distances
in excess of 2~kpc. The same applies to the core radius (middle-right) for R$_{core}$ $>$ 1pc;
below this threshold, our sample appears to contain a lower fraction of OCs than that
in \cite{kha13}. Regarding mass, both distributions have a similar shape,
but with a shift of $\approx$ 0.7 dex between the peaks, which suggests that our sample
occupies the high-mass wing observed in \cite{pis08} distribution. The age
distributions also have similar shapes, with our sample consisting essentially of clusters
older than 100 Myrs. Thus, we can conclude that the 40 OCs dealt with here are a
representative sub-sample of the Galactic OC population, with no systematic biases.

\section{The 2MASS photometry and the field-star decontamination}

We have used JH${K_{s}}$ photometry of 2MASS\footnote{The Two Micron All Sky Survey Catalogue, 
available at \textit{http://www.ipac.caltech.edu/2mass/releases/allsky/}} 
to find the apparently cluster members of 40 OCs \citep{skr06}. We used  
VizieR\footnote{http://vizier.u-strasbg.fr/viz-bin/VizieR?-source=II/246.}
to extract  the near infrared (NIR) (J, H, and ${K_{s}}$ 2MASS) photometry for a large-area centered
on each cluster, which is essential to build the RDPs with a high contrast relative to 
the background, and for a better field star decontamination. 2MASS provides an all-sky 
coverage with the spatial and photometric uniformity required for high star count statistics. 
For the photometric constraint, the 2MASS magnitude extractions have been restricted 
to stars with errors smaller than 0.2 mag in JH${K_{s}}$  magnitudes.
The extraction radii of 40 OCs have been chosen by visual inspection on the DSS-I 
image\footnote{Extracted from the Canadian Astronomy Data Centre (CADC),  
at  \textit{http://ledas-www.star.le.ac.uk/DSSimage/}},  and taking into account the RDP, 
in the sense that the profile must become relatively stable in the outer region. 
As an example we show only the DSS-I image of Pismis~19 in Fig.~3.

The technique used here for determining the cluster members of the 40 
OCs is known as the field star decontamination procedure coupled to 
the 2MASS JH${K_{s}}$ photometry, and it was succesfully used by \cite
{bon07a,bon07b,bon08} and more recently by \cite{gun12}, This decontamination procedure was 
applied to the 40 OCs discussed here. This technique samples photometric 
properties of the stars in a neighbour comparison field considered free of 
cluster stars to (statistically) remove the contamining field stars from 
the cluster stars with help of the colour magnitude diagrams (CMD).

\begin{figure}   
\centering
\includegraphics*[width = 7cm, height = 7cm]{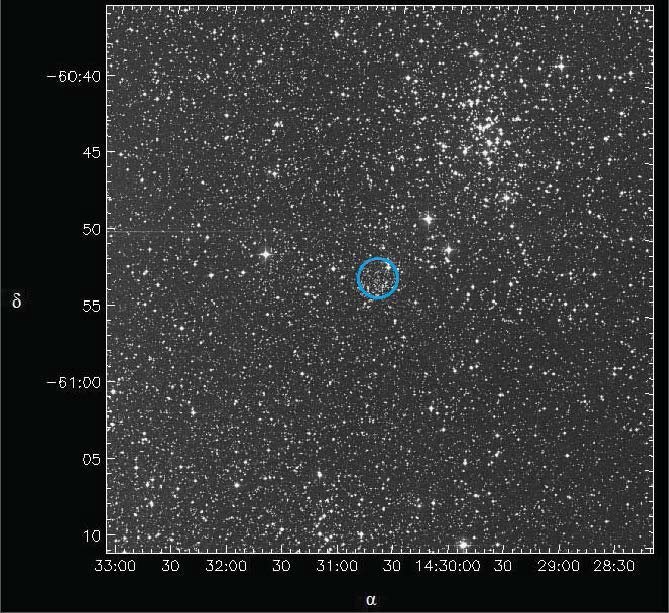}
\caption {The image of Pismis~19 in DSS-I $25\,^\prime$x $25\,^\prime$.}
\end{figure}

Firstly, the stellar surface densities $\sigma(stars\,\rm arcmin^{-2})$ 
and the  surface isopleths of both the raw and decontamination data of 
40 OCs, computed for a mesh size of $3^{\prime}\times3^{\prime}$ and 
centred on the galactic coordinates of Table~2 (see Supplementary 
material section) If necessary, we have re-determined them in this work 
(see below). Here, isopleth is star density map. These maps have been used to maximise the contrast of the 
cluster against the background. In Figs.~4 and 5 we show the result for 
Pismis~19 as an example. The central stellar density excesses are 
significant in the decontamination surface-density distributions, as is 
seen in Fig.~5 for Pismis~19.

\clearpage
\begin{table*} 
\renewcommand\thetable{2}
\centering
\caption{Literature (left columns) and presently optimised (right columns) Equatorial and Galactic coordinates of 40 OCs.}
\begin{tabular}{lcccccccc}
\hline
{Cluster} & ${\alpha}$(2000) & ${\delta}$(2000) & \textit{l} & \textit{b} & ${\alpha}$(2000) & ${\delta}$(2000) & \textit{l} & \textit{b} \\
          & (h  m  s) & ($^{o}$    $'$   $''$) & ( $^{o}$ ) & ( $^{o}$ ) & (h  m  s) & ($^{o}$    $'$   $''$) & ( $^{o}$ ) & ( $^{o}$ ) \\
\hline
NGC 436     & 01 15 58   & 58 48 42  & 126.11 & -3.91  & 01 15 58    & 58 48 42     & 126.11 & -3.91 \\
King 5      & 03 14 45   & 52 41 12  & 143.78 & -4.29  & 03 14 45    & 52 41 12     & 143.78 & -4.29 \\
NGC 1513    & 04 09 57   & 49 30 54  & 152.59 & -1.57  & 04 09 50    & 49 31 17     & 152.57 & -1.58 \\
Be 15       & 05 02 06   & 44 30 43  & 162.26 & 1.62   & 05 02 06    & 44 30 43     & 162.26 & 1.62 \\
NGC 1798    & 05 11 39   & 47 41 30  & 160.70 & 4.85   & 05 11 39    & 47 41 30     & 160.70 & 4.85 \\
Be 17       & 05 20 36   & 30 36 00  & 175.65 & -3.65  & 05 20 38    & 30 34 28     & 175.67 & -3.66 \\
NGC 1907    & 05 28 05   & 35 19 30  & 172.62 & 0.31   & 05 28 09    & 35 18 20     & 172.64 & 0.31 \\
NGC 2112    & 05 53 45   & 00 24 36  & 205.87 & -12.62 & 05 53 51    & 00 25 44     & 205.87 & -12.58 \\
Koposov 12  & 06 00 56   & 35 16 36  & 176.16 & 6.00   & 06 00 56    & 35 16 36     & 176.16 & 6.00 \\
NGC 2158    & 06 07 25   & 24 05 48  & 186.63 & 1.78   & 06 07 30    & 24 05 50     & 186.64 & 1.80 \\
Koposov 53  & 06 08 56   & 26 15 49  & 184.90 & 3.13   & 06 08 56    & 26 15 49     & 184.90 & 3.13 \\
NGC 2194    & 06 13 45   & 12 48 24  & 197.25 & -2.35  & 06 13 45    & 12 48 24     & 197.25 & -2.35 \\
NGC 2192    & 06 15 17   & 39 51 18  & 173.42 & 10.65  & 06 15 22    & 39 51 06     & 173.42 & 10.67 \\
NGC 2243    & 06 29 34   & -31 17 00 & 239.48 & -18.01 & 06 29 34    & -31 17 00    & 239.48 & -18.01 \\
Trumpler 5  & 06 36 42   & 09 26 00  & 202.86 & 1.05   & 06 36 36    & 09 25 21     & 202.86 & 1.02 \\
Col 110     & 06 38 24   & 02 01 00  & 209.65 & -1.98  & 06 38 35    & 02 01 30     & 209.66 & -1.93 \\
NGC 2262    & 06 39 38   & 01 08 36  & 210.57 & -2.10  & 06 39 38    & 01 08 36     & 210.57 & -2.10 \\
NGC 2286    & 06 47 40   &-03 08 54  & 215.31 & -2.27  & 06 47 43    & -03 10 20    & 215.33 & -2.27 \\
NGC 2309    & 06 56 03   & -07 10 30 & 219.84 & -2.24  & 06 56 02    & -07 11 05    & 219.85 & -2.25 \\
Tombaugh 2  & 07 03 05   & -20 49 00 & 232.83 & -6.88  & 07 03 05    & -20 49 00    & 232.83 & -6.88 \\
Be 36       & 07 16 06   & -13 06 00 & 227.38 & -0.59  & 07 16 24    & -13 11 23    & 227.49 & -0.56 \\
Haffner 8   & 07 23 24   & -12 20 00 & 227.53 & 1.34   & 07 23 09    & -12 16 12    & 227.45 & 1.32 \\
Mel 71      & 07 37 30   & -12 04 00 & 228.95 & 4.50   & 07 37 30    & -12 04 00    & 228.95 & 4.50 \\
NGC 2425    & 07 38 22   & -14 52 54 & 231.52 & 3.31   & 07 38 22    & -14 52 54    & 231.52 & 3.31 \\
NGC 2506    & 08 00 01   & -10 46 12 & 230.56 & 9.93   & 07 59 59    & -10 45 28    & 230.55 & 9.93 \\
Pismis 3    & 08 31 22   & -38 39 00 & 257.86 & 0.50   & 08 31 16    & -38 39 02    & 257.85 & 0.48 \\
NGC 2660    & 08 42 38   & -47 12 00 & 265.93 & -3.01  & 08 42 38    & -47 12 00    & 265.93 & -3.01 \\
NGC 3680    & 11 25 38   & -43 14 36 & 286.76 & 16.92  & 11 25 35    & -43 15 11    & 286.76 & 16.91 \\
Ru 96       & 11 50 38   & -62 08 23 & 295.89 & -0.10  & 11 50 37    & -62 09 04    & 295.89 & -0.11 \\
Ru 105      & 12 34 15   & -61 34 11 & 300.88 & 1.24   & 12 34 12    & -61 33 00    & 300.88 & 1.25 \\
Trumpler 20 & 12 39 34   & -60 37 00 & 301.48 & 2.22   & 12 39 34    & -60 37 00    & 301.48 & 2.22 \\
Pismis 19   & 14 30 40   & -60 53 00 & 314.71 & -0.30  & 14 30 40    & -60 53 00    & 314.71 & -0.30 \\
NGC 6134    & 16 27 46   & -49 09 06 & 334.92 & -0.20  & 16 27 46    & -49 09 06    & 334.92 & -0.20 \\
IC 4651     & 17 24 49   & -49 56 00 & 340.09 & -7.91  & 17 24 46    & -49 55 06    & 340.10 & -7.89 \\
NGC 6802    & 19 30 35   & 20 15 42  & 55.33  & 0.92   & 19 30 33    & 20 15 48     & 55.32  & 0.92 \\
NGC 6819    & 19 41 18   & 40 11 12  & 73.98  & 8.48   & 19 41 18    & 40 11 12     & 73.98  & 8.48 \\
Be 89       & 20 24 36   & 46 03 00  & 83.16  & 4.82   & 20 24 30    & 46 02 53     & 83.15  & 4.84 \\
NGC 6939    & 20 31 30   & 60 39 42  & 95.90  & 12.30  & 20 31 30    & 60 39 42     & 95.90  & 12.30 \\
NGC 7142    & 21 45 09   & 65 46 30  & 105.35 & 9.48   & 21 45 12    & 65 47 43     & 105.36 & 9.50 \\
NGC 7789    & 23 57 24   & 56 42 30  & 115.53 & -5.39  & 23 57 24    & 56 42 30     & 115.53 & -5.39 \\
\hline
\end{tabular}
\end{table*}

The stellar radial density profiles (RDP) were derived from the 
isopleth surfaces of each cluster, the coordinates were checked and 
the cluster radii were determined (e.g. Table 4). The residual 
background level of each RDP corresponds to the average number of 
CM-filtered stars measured in the comparison field. A wide external 
ring $(\Delta R=13'-70')$ centered in the cluster (Col.~11  of 
Tables~4 and S4) has been considered to eliminate field stars of the 
40 OCs. Stars within the cluster radii have been considered to be 
probable members.

The stellar radial density profile (RDP) of each cluster, 
built based on the JH${K_{s}}$ photometry extracted with the 
WEBDA\footnote{www.univie.ac.at/WEBDA-Mermilliod \& Paunzen 
(2003)} coordinates are displayed in Table~2 and have been 
computed to check cluster centering. In some cases the RDP 
built with the original cluster coordinates presented a dip 
at the center. Then, new central coordinates are searched 
after field star decontamination to maximise the star counts 
in the innermost RDP bin. From these RDPs, the cluster radii 
of 40 OCs are determined (Table 4). The stellar RDP is the 
projected number of stars per area around the cluster centre. 
To avoid oversampling near the centre and undersampling for 
large radii, the RDPs are built by counting stars in concentric 
rings of increasing width with distance to the centre. The number 
and width of rings are optimised so that the resulting RDPs have 
adequate spatial resolution with moderate $1\sigma$ Poission 
errors. The residual background level of each RDP corresponds 
to the average number of CM-filtered stars measured in the 
comparison field.

\begin{figure}    
\centering
\includegraphics*[width = 7cm, height = 7cm]{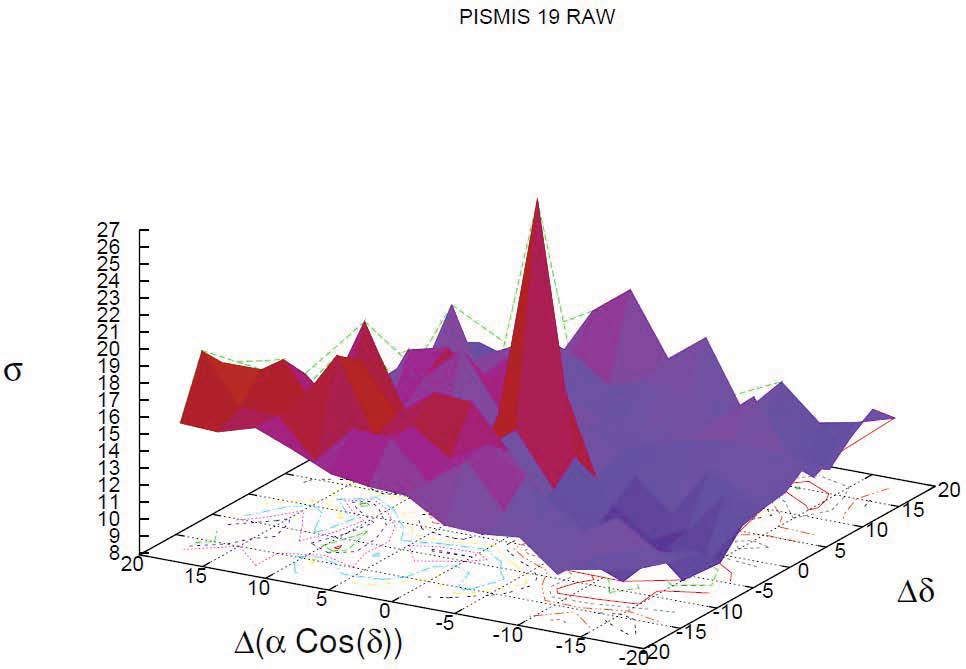}
\includegraphics*[width = 7cm, height = 7cm]{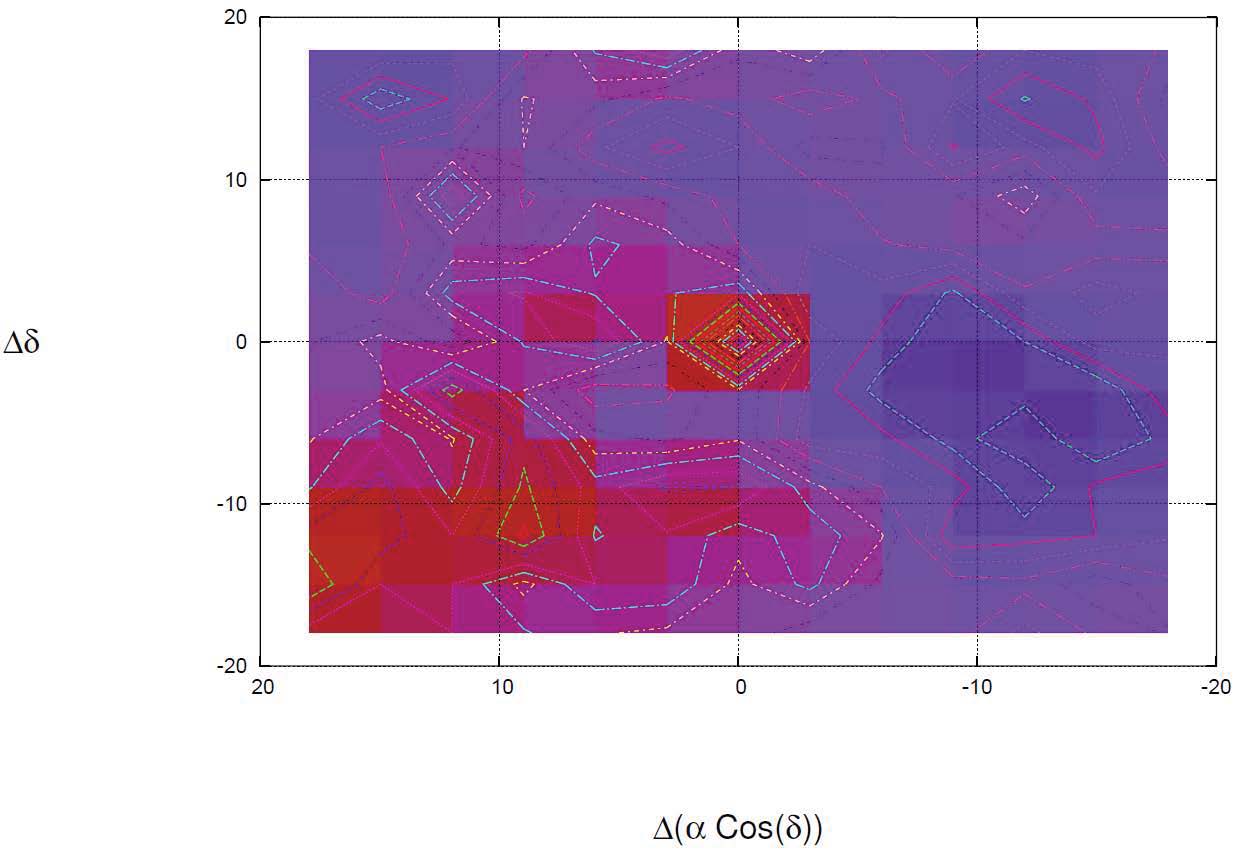}
\caption {For observed (raw) photometry, top panel: stellar surface$-$density 
$\sigma (stars\,\rm arcmin^{-2}$) of Pismis~19, computed for a mesh size of 
$3^\prime\times3^\prime$, centred on the coordinates in Table 2. 
Bottom panel : The corresponding isopleth surface.} 
\end{figure}

\begin{figure}      
\centering
\includegraphics*[width = 6.5cm, height = 7cm]{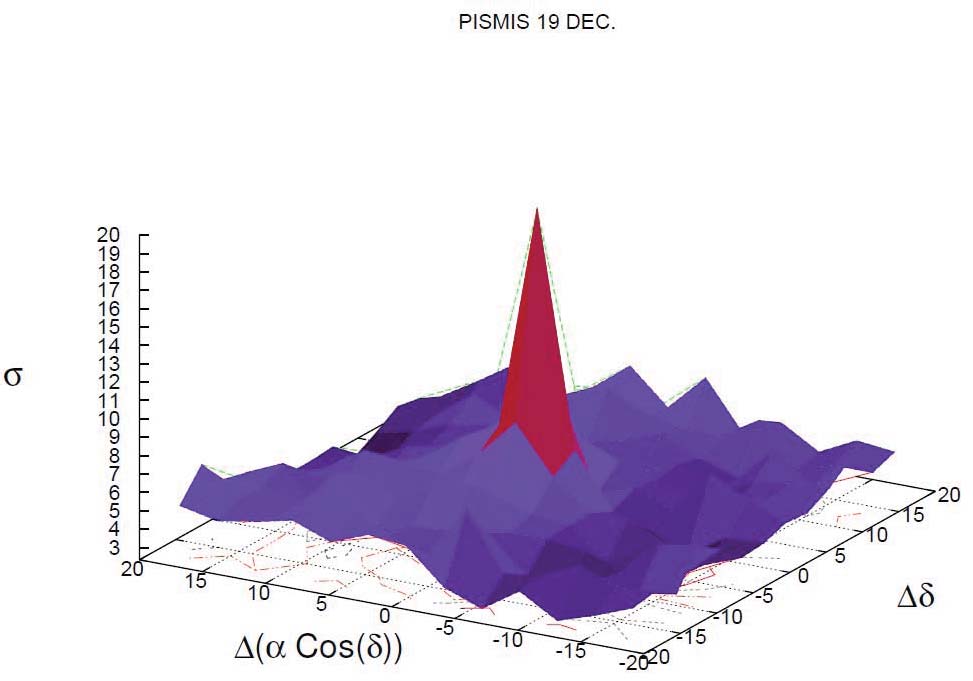}
\includegraphics*[width = 6.5cm, height = 7cm]{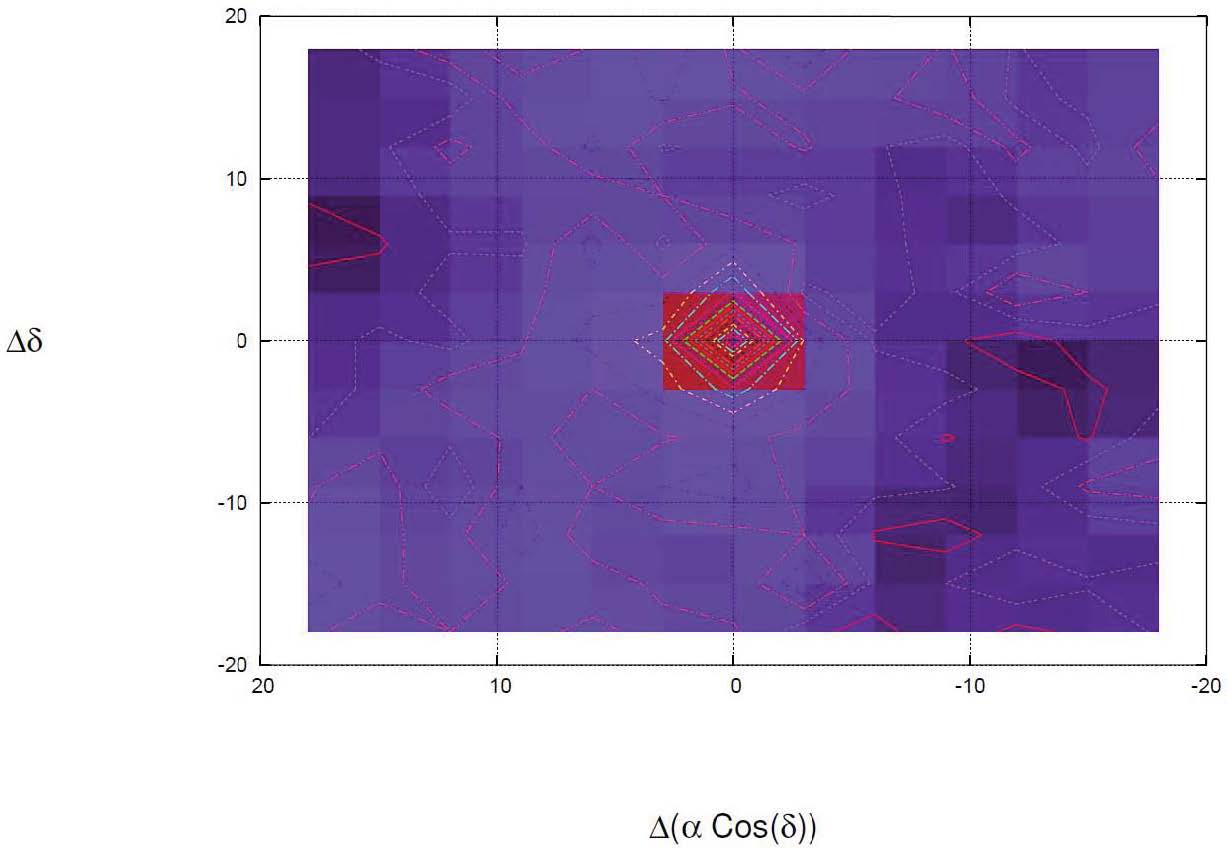}
\caption {For decontaminated photometry, top panel: stellar surface$-$density 
$\sigma (stars\,\rm arcmin^{-2}$) of Pismis~19, computed for a mesh size of 
$3^\prime\times3^\prime$, centred on the coordinates in Table 2. 
Bottom panel : The corresponding isopleth surface.}  
\end{figure}

As \cite{cam10} noted, RDPs of OCs built based on 
the WEBDA coordinates usually show a dip in the inner RDP 
region when a mismatch between the ''true'' and catalogue 
coordinates exists. For this reason, new central coordinates 
of these clusters have been searched to maximise the star 
counts at the innermost RDP bin. Then, the 2MASS photometry 
was extracted again, but now centered on the optimized cluster 
coordinates. As a representative, the optimised central 
coordinate of Pismis~19 is displayed in Fig.~3 as small circle, 
and given in the right section of Table~2. 

To have the intrinsic morphology of the clusters in the CMD, 
as explained above, the statistical field star decontamination 
procedure of \cite{bon07a} is used. This procedure is based on 
the relative stars densities per sky area in a cluster region and 
on a neighboring offset field. It divides the full range of magnitudes 
and colours of a CMD into the cell dimensions of $\Delta{J}=1.0$, 
and $\Delta(J-H)={\Delta(J-K_{s})}=0.15$. These dimensions are 
adequate to allow for sufficient star counts in individual cells 
and preserve the intrinsic morphology of the evolutionary sequences. 
\cite{bon07a} showed that the field star decontamination procedure 
with 2MASS JH${K_{s}}$ photometry is efficient isolating those 
stars with a high probability of being cluster members. More 
details on the algorithm can be found in \cite{bon07a, bon07b}, 
\cite{bon09a, bon09b, bon09c}, and \cite{cam10}. 

By following the field decontamination technique which is briefly 
explained above,the probable cluster members of the 40 OCs have been 
identified for further analysis.

\begin{table*}  
\renewcommand\thetable{3}
\centering
\caption{Derived fundamental astrophysical parameters from 2MASS JH${K_{s}}$ photometry of 40 OCs.}
\renewcommand{\tabcolsep}{1.1mm}
\renewcommand{\arraystretch}{1.1}
\begin{tabular}{lccccccc}  
\hline
 Cluster & Z & {E(J-H)} & {E(B-V)} & {Age(Gyr)} & {(m-M)j} & {d(kpc)} & {R$_{GC}$(kpc)} \\
\hline
NGC 436    & 0.019  & 0.13$\pm$0.03 & 0.42$\pm$0.10 & 0.4$\pm$0.1  & 12.54$\pm$0.31 & 3.22$\pm$0.46 & 9.48$\pm$0.28  \\
King 5     & 0.0105 & 0.26$\pm$0.05 & 0.83$\pm$0.16 & 1.0$\pm$0.2  & 11.53$\pm$0.24 & 2.03$\pm$0.23 & 8.93$\pm$0.18  \\
NGC 1513   & 0.019  & 0.23$\pm$0.02 & 0.74$\pm$0.06 & 0.1$\pm$0.02 & 10.37$\pm$0.28 & 1.18$\pm$0.15 & 8.29$\pm$0.13  \\
Be 15      & 0.019  & 0.27$\pm$0.03 & 0.86$\pm$0.10 & 0.5$\pm$0.1  & 12.45$\pm$0.31 & 3.10$\pm$0.44 & 10.21$\pm$0.42 \\
NGC 1798   & 0.0105 & 0.16$\pm$0.04 & 0.51$\pm$0.13 & 1.5$\pm$0.3  & 13.51$\pm$0.26 & 5.03$\pm$0.59 & 12.07$\pm$0.55 \\
Be 17      & 0.006  & 0.26$\pm$0.04 & 0.83$\pm$0.13 & 5.0$\pm$0.5  & 11.93$\pm$0.29 & 2.43$\pm$0.33 & 9.65$\pm$0.33  \\
NGC 1907   & 0.019  & 0.18$\pm$0.03 & 0.58$\pm$0.10 & 0.4$\pm$0.1  & 11.45$\pm$0.26 & 1.95$\pm$0.24 & 9.16$\pm$0.23  \\
NGC 2112   & 0.019  & 0.20$\pm$0.04 & 0.64$\pm$0.13 & 2.0$\pm$0.3  & 10.15$\pm$0.23 & 1.07$\pm$0.11 & 8.18$\pm$0.10  \\
Koposov 12 & 0.0105 & 0.07$\pm$0.02 & 0.22$\pm$0.06 & 1.8$\pm$0.2  & 11.56$\pm$0.18 & 2.05$\pm$0.17 & 9.26$\pm$0.17  \\
NGC 2158   & 0.019  & 0.05$\pm$0.01 & 0.16$\pm$0.03 & 2.5$\pm$0.3  & 13.21$\pm$0.10 & 4.39$\pm$0.21 & 11.59$\pm$0.21 \\
Koposov 53 & 0.019  & 0.01$\pm$0.00 & 0.03$\pm$0.02 & 1.0$\pm$0.1  & 13.05$\pm$0.18 & 4.08$\pm$0.34 & 11.28$\pm$0.34 \\
NGC 2194   & 0.019  & 0.13$\pm$0.04 & 0.42$\pm$0.13 & 0.8$\pm$0.2  & 11.87$\pm$0.27 & 2.37$\pm$0.30 & 9.51$\pm$0.28  \\
NGC 2192   & 0.019  & 0.01$\pm$0.00 & 0.03$\pm$0.00 & 1.3$\pm$0.1  & 13.12$\pm$0.15 & 4.21$\pm$0.29 & 11.37$\pm$0.28 \\
NGC 2243   & 0.0105 & 0.01$\pm$0.00 & 0.03$\pm$0.00 & 2.0$\pm$0.2  & 13.37$\pm$0.12 & 4.73$\pm$0.26 & 10.36$\pm$0.14 \\
Trumpler 5 & 0.006  & 0.24$\pm$0.05 & 0.77$\pm$0.16 & 3.0$\pm$0.3  & 12.19$\pm$0.29 & 2.74$\pm$0.36 & 9.80$\pm$0.33  \\
Col 110    & 0.019  & 0.06$\pm$0.01 & 0.19$\pm$0.03 & 3.0$\pm$0.2  & 11.93$\pm$0.15 & 2.44$\pm$0.17 & 9.41$\pm$0.15  \\
NGC 2262   & 0.0105 & 0.11$\pm$0.01 & 0.35$\pm$0.03 & 1.3$\pm$0.1  & 12.36$\pm$0.30 & 2.96$\pm$0.41 & 9.88$\pm$0.35  \\
NGC 2286   & 0.019  & 0.03$\pm$0.00 & 0.10$\pm$0.02 & 1.0$\pm$0.2  & 11.82$\pm$0.30 & 2.31$\pm$0.32 & 9.20$\pm$0.26  \\
NGC 2309   & 0.019  & 0.16$\pm$0.02 & 0.51$\pm$0.06 & 0.5$\pm$0.1  & 12.41$\pm$0.21 & 3.03$\pm$0.29 & 9.74$\pm$0.22  \\
Tombaugh 2 & 0.019  & 0.35$\pm$0.05 & 1.12$\pm$0.16 & 3.0$\pm$0.3  & 10.43$\pm$0.24 & 1.22$\pm$0.14 & 8.01$\pm$0.08  \\
Be 36      & 0.019  & 0.12$\pm$0.02 & 0.38$\pm$0.06 & 3.0$\pm$1.0  & 13.67$\pm$0.16 & 5.42$\pm$0.40 & 11.59$\pm$0.27 \\
Haffner 8  & 0.006  & 0.06$\pm$0.02 & 0.19$\pm$0.06 & 1.0$\pm$0.1  & 11.98$\pm$0.16 & 2.49$\pm$0.18 & 9.09$\pm$0.12  \\
Mel 71     & 0.019  & 0.01$\pm$0.00 & 0.03$\pm$0.02 & 1.5$\pm$0.2  & 11.54$\pm$0.15 & 2.03$\pm$0.14 & 8.69$\pm$0.09  \\
NGC 2425   & 0.019  & 0.10$\pm$0.02 & 0.32$\pm$0.06 & 3.2$\pm$0.5  & 12.27$\pm$0.26 & 2.85$\pm$0.34 & 9.26$\pm$0.21  \\
NGC 2506   & 0.006  & 0.03$\pm$0.01 & 0.10$\pm$0.03 & 2.0$\pm$0.3  & 12.27$\pm$0.20 & 2.84$\pm$0.26 & 9.27$\pm$0.17  \\
Pismis 3   & 0.006  & 0.33$\pm$0.02 & 1.06$\pm$0.06 & 3.2$\pm$0.2  & 11.19$\pm$0.11 & 1.73$\pm$0.09 & 7.77$\pm$0.03  \\
NGC 2660   & 0.019  & 0.13$\pm$0.03 & 0.42$\pm$0.10 & 1.5$\pm$0.3  & 11.89$\pm$0.17 & 2.39$\pm$0.19 & 7.76$\pm$0.06  \\
NGC 3680   & 0.019  & 0.05$\pm$0.01 & 0.16$\pm$0.03 & 1.5$\pm$0.2  & 10.16$\pm$0.10 & 1.08$\pm$0.05 & 7.00$\pm$0.02  \\
Ru 96      & 0.019  & 0.07$\pm$0.01 & 0.22$\pm$0.03 & 1.0$\pm$0.1  & 12.01$\pm$0.25 & 2.52$\pm$0.29 & 6.53$\pm$0.15  \\
Ru 105     & 0.019  & 0.05$\pm$0.01 & 0.16$\pm$0.03 & 1.0$\pm$0.4  & 11.56$\pm$0.20 & 2.05$\pm$0.19 & 6.41$\pm$0.10  \\
Trumpler 20& 0.019  & 0.10$\pm$0.03 & 0.32$\pm$0.10 & 1.5$\pm$0.5  & 12.52$\pm$0.31 & 3.20$\pm$0.46 & 6.19$\pm$0.27  \\
Pismis 19  & 0.019  & 0.41$\pm$0.03 & 1.31$\pm$0.10 & 0.8$\pm$0.1  & 11.42$\pm$0.38 & 1.92$\pm$0.34 & 6.02$\pm$0.24  \\
NGC 6134   & 0.019  & 0.10$\pm$0.01 & 0.32$\pm$0.03 & 1.5$\pm$0.1  & 10.22$\pm$0.12 & 1.11$\pm$0.06 & 6.23$\pm$0.06  \\
IC 4651    & 0.019  & 0.02$\pm$0.00 & 0.06$\pm$0.02 & 2.5$\pm$0.3  &  9.64$\pm$0.20 & 0.85$\pm$0.08 & 6.44$\pm$0.07  \\
NGC 6802   & 0.019  & 0.23$\pm$0.03 & 0.74$\pm$0.10 & 0.9$\pm$0.1  & 11.77$\pm$0.31 & 2.25$\pm$0.32 & 6.22$\pm$0.19  \\
NGC 6819   & 0.019  & 0.02$\pm$0.00 & 0.06$\pm$0.02 & 2.5$\pm$0.5  & 11.84$\pm$0.15 & 2.34$\pm$0.16 & 6.96$\pm$0.06  \\
Be 89      & 0.019  & 0.23$\pm$0.02 & 0.74$\pm$0.06 & 2.0$\pm$0.5  & 12.37$\pm$0.21 & 2.97$\pm$0.28 & 7.47$\pm$0.11  \\
NGC 6939   & 0.019  & 0.12$\pm$0.03 & 0.38$\pm$0.10 & 2.0$\pm$0.3  & 11.27$\pm$0.31 & 1.79$\pm$0.26 & 7.61$\pm$0.06  \\
NGC 7142   & 0.019  & 0.13$\pm$0.03 & 0.42$\pm$0.10 & 2.5$\pm$0.3  & 12.04$\pm$0.22 & 2.56$\pm$0.25 & 8.27$\pm$0.10  \\ 
NGC 7789   & 0.0105 & 0.08$\pm$0.02 & 0.26$\pm$0.06 & 1.8$\pm$0.2  & 11.23$\pm$0.21 & 1.76$\pm$0.17 & 8.13$\pm$0.08  \\
\hline
\end{tabular}
\end{table*}

\section{Astrophysical parameters}

We have derived the fundamental parameters of 40 OCs using the 
decontaminated $(J, J-H)$ CMDs (see Figs.~S5-S9 in the 
supplementary material) eye-fitted with Padova isochrones 
\citep[hereafter M08]{mar08}. 
Since the spectroscopic metal abundances [Fe/H]$_{spec}$ are only available for 21 (Col.~7, Table 4) out of 40 OCs,
we have considered the abundances of $Z= +0.019$~([Fe/H]=0), $Z= +0.0105$~([Fe/H]=$-$0.25), and  $Z= 
+0.006$~([Fe/H]=$-$0.50), respectively. In the sense OCs need to be uniformly and homogeneously analysed.
M08 isochrones for three Z abundances were fitted to the $(J, J-H)$ CMDs of 
each of the 40 OCs. The most appropriate $Z$ fit solution on the CMDs 
has been made by eye. Accordingly, the M08 isochrones of $Z= 
+0.019$ for 29 OCs, $Z= +0.0105$ for six OCs, and $Z= +0.006$ 
for five OCs, respectively, have provided us good fits for 
reddening, age and distance modulus. 
As an example, such $(J, J-H)$ CMDs have been 
displayed in Figs.~6(a)-(c)  for Pismis~19, for three Z abundances.
The shaded areas in the panels are the 
colour-magnitude filters which follow the distribution of the 
decontaminated star sequences in the CMDs, or stars comprised 
in the shaded area are considered probable members.
These filters are wide enough to accommodate the colour distributions of main sequence and evolved stars 
of the clusters, allowing 1 $\sigma$ photometric uncertainties. 
The fitted 0.8 Gyr isochrone of $Z= +0.019$ for Pismis~19 in 
panel~(a) provides a good solution.  As can be seen from Fig.~6(a), the 
M08 isochrone fits well the main sequence (MS), turn$-$off 
(TO) and Red Giant/Red Clump (RG/RC) regions on the CMD of Pismis~19. Due to the 
presence of binaries, the M08 isochrones have been shifted to 
the left and below of the main sequence in Figs.~6(a)$-$(c), 
and all CMDs of the 40 OCs are presented in Figs.~S5$-$S9 as 
supplementary material.  
The reddening, distance modulus (i.e. distance), age and the appropriate $Z$ 
abundances were derived this way for all 40 OCs 
of our sample. These astrophysical parameters together with their uncertainties are presented in Table 3.

However, the reddening is degenerate with the metallicity. For this, we have determined E(B-V), 
d~(pc), Age~(Gyr) of 21 OCs (Table 4) for three Z abundances.
The E(B-V) and d~(pc) values (Table 4) of three Z abundances are reasonably 
close to our original ones (Table 3) within the uncertainties.
The age values (Col.~6, Table 4) derived from three Z values are the same. 
As stated by \cite{bon09a}, any metallicity for the range of $+0.006\leq Z 
\leq+0.019$ would produce acceptable solutions for the 
astrophysical parameters, due to the filters of 2MASS.

Our derived ages here are almost robust enough to allow inferences about cluster evolution.
For this, NGC 2286 (Fig.~7) is presented as an example. The 0.8 Gyr (blue line), 1 Gyr (solid black line), 
and 1.2 Gyr (red line) isochrones of $Z= +0.019$ for NGC~2286  
are fitted to CMD of the cluster. As is seen from Fig.~7, 1$\pm$0.2 Gyr  isochrone (solid line) fit well the main sequence (MS), turn$-$off 
(TO) and Red Giant/Red Clump (RG/RC) regions on the CMD of the cluster. 
The uncertainties in our derived ages of 40 OCs are in the level of $\pm$0.02$-$0.5 Gyr (Table 3), except for Be~36 ($\pm$1 Gyr).

JHK photometry is unsensitive to metallicity, in opposition 
to optical photometry, where the blue (B) and principally the 
ultraviolet (U) are sensitive to the photospheric metal lines, 
reaching its maximum blanketing effect by SpT~F5. For later 
than SpT = G2 it becomes too fuzzy to disentagle it from the 
molecular lines. On the other hand, metallicity affects 
significantly the distance and the age of a cluster, i.e. the 
less Z is, the shorter the distance and larger the age.

\begin{figure}       
\centering
\includegraphics*[width = 4.5cm, height = 5.5cm]{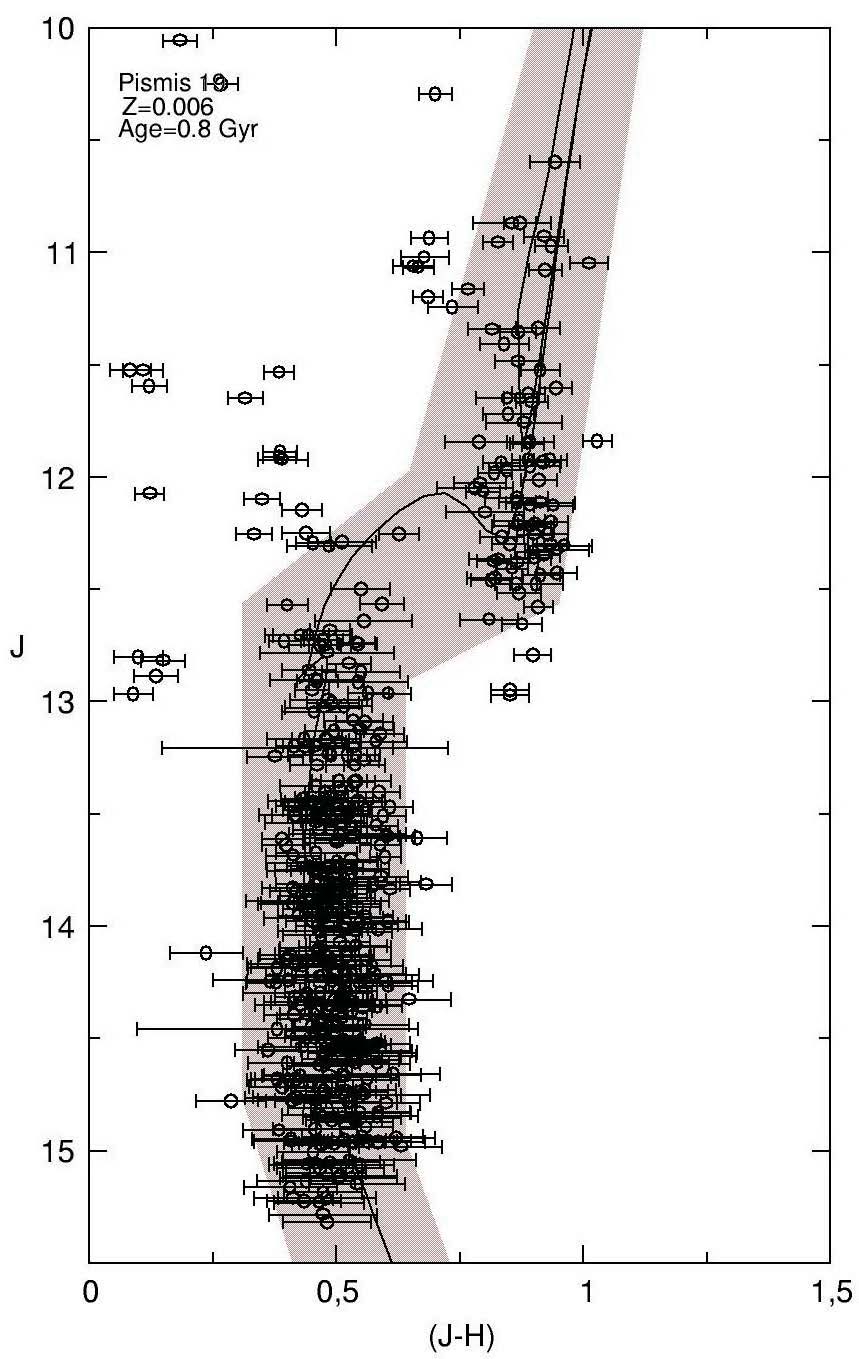}
\includegraphics*[width = 4.5cm, height = 5.5cm]{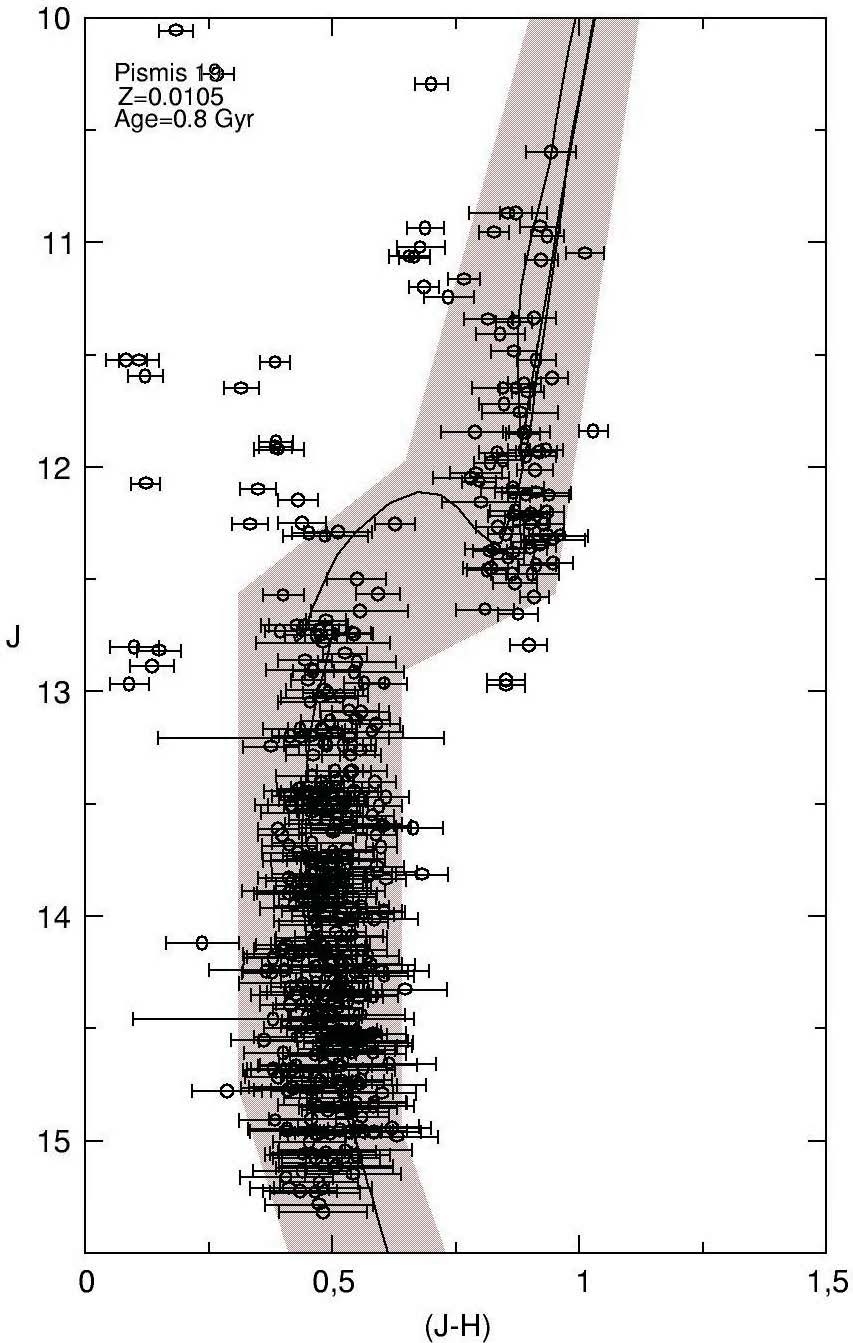}
\includegraphics*[width = 4.5cm, height = 5.5cm]{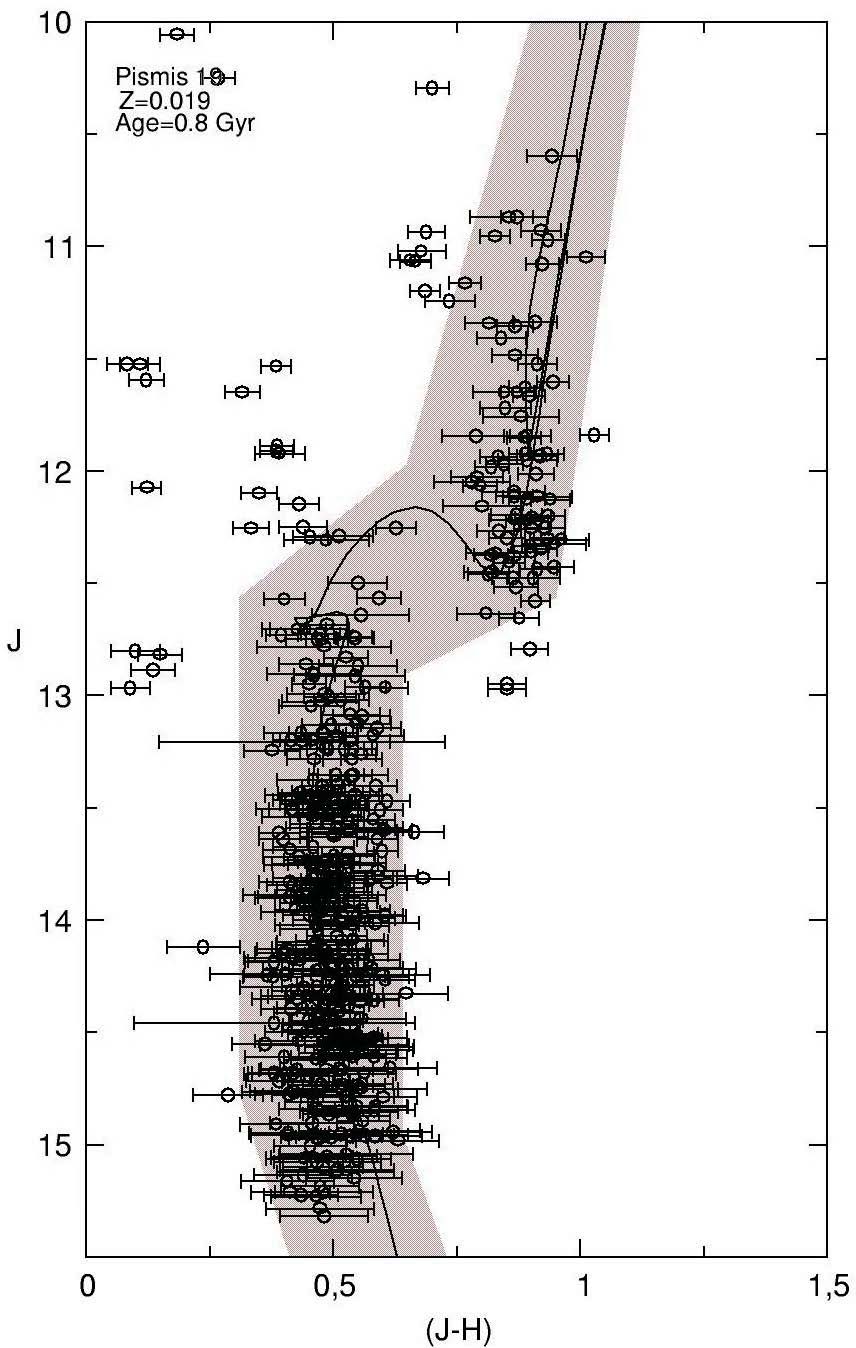}
\caption {Observed decontaminated $J\times(J-H)$ CMDs  
extracted from the region of $R=11'.03$ for Pismis~19.
The solid lines in the panels represent 
the fitted 0.8 Gyr  Padova isochrones 
for Z$=$+0.019 (solar), Z$=$+0.0105, and  Z$=$+0.006, respectively. 
The CMD filter used to isolate cluster 
MS/evolved stars is shown with the shaded area.}  
\end{figure}

\begin{figure}       
\centering
\includegraphics*[width = 4.5cm, height = 5.5cm]{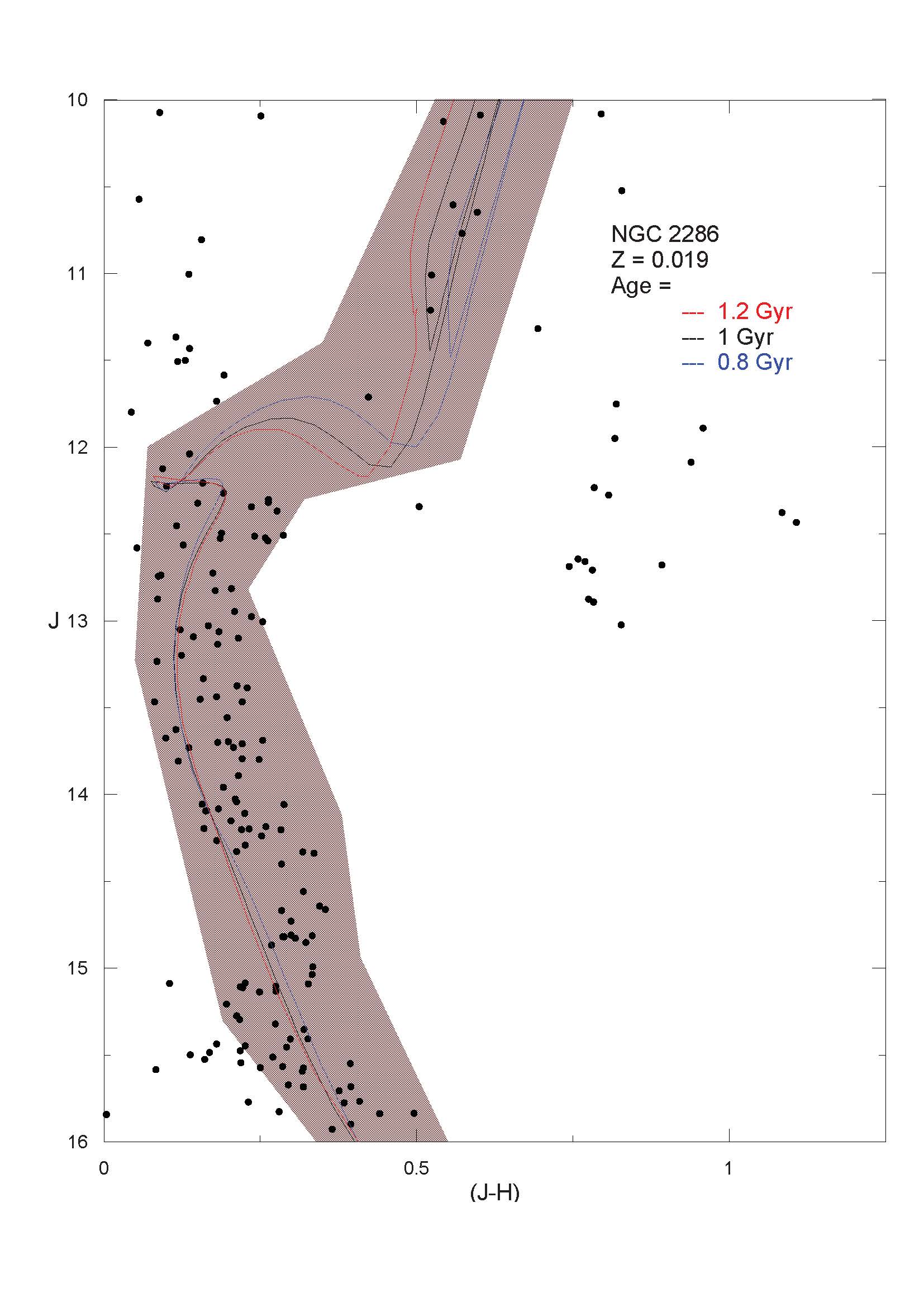}
\caption {Observed decontaminated $J\times(J-H)$ CMD  
of NGC~2286. The solid lines in the panels represent 
the fitted  0.8 Gyr (blue line), 1 Gyr (solid black line), 
and 1.2 Gyr (red line) isochrones of $Z= +0.019$.
The CMD filter used to isolate cluster 
MS/evolved stars is shown with the shaded area.}  
\end{figure}

\begin{table*}                  
\renewcommand\thetable{4}
\tiny
\centering
\tiny
\caption{Z, E(B-V), d~(pc), Age~(Gyr) values of 21 OCs with [Fe/H]$_{spec}$. E(B-V) values are listed in Col.~3 for three Z abundances of 21 OCs in our sample. 
[Fe/H]$_{iso}$ values in Col.~4 are converted from the expression $Z = Z_\odot \cdot 10^{[Fe/H]}$.  
The solar abundance value is taken as $Z_\odot = +0.019$. Ages are given in Col.~6.
[Fe/H]$_{spec}$ values together with literature are listed in Col.~7-8. }
\renewcommand{\tabcolsep}{1.1mm}
\renewcommand{\arraystretch}{1.1}
\tiny
\begin{tabular}{cccccccc}
\hline
Cluster &  Z & E(B-V) & [Fe/H]$_{iso}$ & d (kpc)& Age(Gyr) & [Fe/H]$_{spec}$ & Reference \\
\hline

\multicolumn{ 1}{c}{Trumpler 5} &0.019  &0.54$\pm$0.13  &     &3.13$\pm$0.39&3	&     &                      \\
\multicolumn{ 1}{c}{}           &0.0105 &0.70$\pm$0.16	&     &2.87$\pm$0.38&3	&     &                      \\
\multicolumn{ 1}{c}{}           &0.006  &0.77$\pm$0.16	&-0.50&2.74$\pm$0.36&3	&-0.36& Carrera et al. 2007  \\
\multicolumn{ 1}{c}{NGC 2158}   &0.019  &0.16$\pm$0.03	&   0 &4.39$\pm$0.21&2.5&-0.28& Jacobson et al. 2011 \\
\multicolumn{ 1}{c}{}           &0.0105 &0.29$\pm$0.03	&     &3.98$\pm$0.19&2.5&     &                      \\
\multicolumn{ 1}{c}{}           &0.006  &0.38$\pm$0.03	&     &3.75$\pm$0.18&2.5&     &                      \\
\multicolumn{ 1}{c}{Col 110}    &0.019  &0.19$\pm$0.03	&   0 &2.44$\pm$0.17&3	&-0.01& Carrera et al. 2007  \\
\multicolumn{ 1}{c}{}           &0.0105 &0.29$\pm$0.06	&     &2.29$\pm$0.17&3	&     &                      \\
\multicolumn{ 1}{c}{}           &0.006  &0.42$\pm$0.06	&     &2.03$\pm$0.15&3	&     &                      \\
\multicolumn{ 1}{c}{NGC6134}    &0.019  &0.32$\pm$0.03	&   0 &1.11$\pm$0.06&1.5&0.12 & Smiljanic et al.2009 \\
\multicolumn{ 1}{c}{}           &0.0105 &0.48$\pm$0.06	&     &0.97$\pm$0.07&1.5&     &                      \\
\multicolumn{ 1}{c}{}           &0.006  &0.64$\pm$0.10	&     &0.85$\pm$0.07&1.5&     &                      \\
\multicolumn{ 1}{c}{NGC2425}    &0.019  &0.32$\pm$0.06	&   0 &2.85$\pm$0.34&3.2&-0.15& Jacobson et al. 2011 \\
\multicolumn{ 1}{c}{}           &0.0105 &0.42$\pm$0.10	&     &2.74$\pm$0.33&3.2&     &                      \\
\multicolumn{ 1}{c}{}           &0.006  &0.51$\pm$0.16	&     &2.64$\pm$0.35&3.2&     &                      \\
\multicolumn{ 1}{c}{Trumpler 20}&0.019  &0.32$\pm$0.10	&   0 &3.20$\pm$0.46&1.5& 0.09& Carraro et al. 2014  \\
\multicolumn{ 1}{c}{}           &0.0105 &0.48$\pm$0.10	&     &2.80$\pm$0.40&1.5&     &                      \\
\multicolumn{ 1}{c}{}           &0.006  &0.64$\pm$0.10	&     &2.45$\pm$0.35&1.5&     &           	     \\ 
\multicolumn{ 1}{c}{NGC 2112}   &0.019  &0.64$\pm$0.13	&   0 &1.07$\pm$0.11&2	&-0.10& Brown et al. 1996    \\
\multicolumn{ 1}{c}{}           &0.0105 &0.77$\pm$0.06	&     &0.87$\pm$0.12&2	&     &          	     \\ 
\multicolumn{ 1}{c}{}           &0.006  &0.90$\pm$0.06	&     &0.77$\pm$0.09&2	&     &                      \\
\multicolumn{ 1}{c}{Mel 71}     &0.019  &0.03$\pm$0.02	&   0 &2.03$\pm$0.14&1.5&-0.30& Brown et al. 1996    \\
\multicolumn{ 1}{c}{}           &0.0105 &0.16$\pm$0.06	&     &1.93$\pm$0.14&1.5&     &                      \\
\multicolumn{ 1}{c}{}           &0.006  &0.29$\pm$0.06	&     &1.92$\pm$0.17&1.5&     &                      \\
\multicolumn{ 1}{c}{NGC 7789}   &0.019  &0.19$\pm$0.03	&     &1.81$\pm$0.17&1.8&     &                      \\
\multicolumn{ 1}{c}{}           &0.0105 &0.26$\pm$0.06	&-0.25&1.76$\pm$0.17&1.8& 0.02& Jacobson et al. 2011 \\
\multicolumn{ 1}{c}{}           &0.006  &0.32$\pm$0.06	&     &1.72$\pm$0.19&1.8&     &                      \\
\multicolumn{ 1}{c}{NGC 3680}   &0.019  &0.16$\pm$0.03	&   0 &1.08$\pm$0.05&1.5& 0.04& Smiljanic et al.2009 \\
\multicolumn{ 1}{c}{}           &0.0105 &0.32$\pm$0.06	&     &0.95$\pm$0.06&1.5&     &                      \\
\multicolumn{ 1}{c}{}           &0.006  &0.48$\pm$0.10	&     &0.85$\pm$0.07&1.5&     &                      \\
\multicolumn{ 1}{c}{IC 4651}    &0.019  &0.06$\pm$0.02	&   0 &0.85$\pm$0.08&2.5& 0.10& Pasquini et al. 2004 \\
\multicolumn{ 1}{c}{}           &0.0105 &0.16$\pm$0.03	&     &0.78$\pm$0.08&2.5&     &                      \\
\multicolumn{ 1}{c}{}           &0.006  &0.26$\pm$0.06	&     &0.72$\pm$0.08&2.5&     &                      \\
\multicolumn{ 1}{c}{NGC 6819}   &0.019  &0.06$\pm$0.02	&   0 &2.34$\pm$0.16&2.5& 0.09& Bragaglia et al.2001 \\
\multicolumn{ 1}{c}{}           &0.0105 &0.16$\pm$0.03	&     &2.15$\pm$0.18&2.5&     &                      \\
\multicolumn{ 1}{c}{}           &0.006  &0.32$\pm$0.10	&     &1.84$\pm$0.15&2.5&     &                      \\
\multicolumn{ 1}{c}{NGC 1798}   &0.019  &0.32$\pm$0.06	&     &5.69$\pm$0.54&1.5&     &                      \\
\multicolumn{ 1}{c}{}           &0.0105 &0.51$\pm$0.13	&-0.25&5.03$\pm$0.59&1.5&-0.12& Carrera 2012         \\
\multicolumn{ 1}{c}{}           &0.006  &0.7$\pm$0.16	&     &4.45$\pm$0.58&1.5&     &                      \\
\multicolumn{ 1}{c}{NGC 2243}   &0.019  &0.005$\pm$0.005&     &5.01$\pm$0.23&2	&     &                      \\
\multicolumn{ 1}{c}{}           &0.0105 &0.03$\pm$0.005	&-0.25&4.73$\pm$0.26&2	&-0.48& Gratton et al. 1994  \\
\multicolumn{ 1}{c}{}           &0.006  &0.16$\pm$0.03	&     &4.49$\pm$0.25&2	&     &                      \\
\multicolumn{ 1}{c}{NGC 6939}   &0.019  &0.38$\pm$0.10	&   0 &1.79$\pm$0.26&2	&    0& Jacobson et al. 2007 \\
\multicolumn{ 1}{c}{}           &0.0105 &0.45$\pm$0.13	&     &1.64$\pm$0.26&2	&     &                      \\
\multicolumn{ 1}{c}{}           &0.006  &0.58$\pm$0.19	&     &1.50$\pm$0.26&2	&     &                      \\
\multicolumn{ 1}{c}{NGC 7142}   &0.019  &0.42$\pm$0.10	&   0 &2.56$\pm$0.25&2.5& 0.08& Jacobson et al. 2008 \\
\multicolumn{ 1}{c}{}           &0.0105 &0.54$\pm$0.13	&     &2.32$\pm$0.26&2.5&     &                      \\
\multicolumn{ 1}{c}{}           &0.006  &0.67$\pm$0.16	&     &2.11$\pm$0.25&2.5&     &                      \\
\multicolumn{ 1}{c}{NGC 2194}   &0.019  &0.42$\pm$0.13	&   0 &2.37$\pm$0.30&0.8&-0.08& Jacobson et al. 2011 \\
\multicolumn{ 1}{c}{}           &0.0105 &0.51$\pm$0.16	&     &2.15$\pm$0.26&0.8&     &                      \\
\multicolumn{ 1}{c}{}           &0.006  &0.61$\pm$0.16	&     &1.97$\pm$0.24&0.8&     &                      \\
\multicolumn{ 1}{c}{NGC 2660}   &0.019  &0.42$\pm$0.10	&   0 &2.39$\pm$0.19&1.5& 0.04& Bragaglia et al.2008 \\
\multicolumn{ 1}{c}{}           &0.0105 &0.51$\pm$0.13	&     &2.20$\pm$0.19&1.5&     &                      \\
\multicolumn{ 1}{c}{}           &0.006  &0.67$\pm$0.16	&     &1.97$\pm$0.21&1.5&     &                      \\
\multicolumn{ 1}{c}{Be 17}      &0.019  &0.64$\pm$0.13	&     &3.02$\pm$0.38&5	&     &                      \\
\multicolumn{ 1}{c}{}           &0.0105 &0.74$\pm$0.13	&     &2.71$\pm$0.40&5	&     &                      \\
\multicolumn{ 1}{c}{}           &0.006  &0.83$\pm$0.13	&-0.50&2.43$\pm$0.33&5	&-0.10& Friel et al. 2005    \\
\multicolumn{ 1}{c}{Tombaugh 2} &0.019  &1.12$\pm$0.16	&   0 &1.22$\pm$0.14&3	&-0.45& Brown et al. 1996    \\
\multicolumn{ 1}{c}{}           &0.0105 &1.22$\pm$0.19	&     &1.18$\pm$0.14&3	&     &                      \\
\multicolumn{ 1}{c}{}           &0.006  &1.31$\pm$0.19	&     &1.13$\pm$0.18&3	&     &                      \\
\multicolumn{ 1}{c}{NGC 2506}   &0.019  &0.02$\pm$0.005	&     &3.07$\pm$0.28&2	&     &                      \\
\multicolumn{ 1}{c}{}           &0.0105 &0.06$\pm$0.03	&     &2.94$\pm$0.27&2	&     &                      \\
\multicolumn{ 1}{c}{}           &0.006  &0.10$\pm$0.03	&-0.50&2.84$\pm$0.26&2	&-0.20& Carretta et al. 2004 \\
\hline													     
\end{tabular}												     
\end{table*}

The reddenings $E(J-H)$ (Col.~3 in Table 3) of the 40 OCs 
were derived from the CMD diagrams. These are converted to  $E(B-V)$ (Col.~4 in Table 3) with  
the extinction law  $A_{J}/{A_{V}}=0.276,\, A_{H}/{A_{V}}=0.176,\, 
A_{K_{s}}/{A_{V}}=0.118, A_{J}=2.76\times{E(J-H)}$, and $E(J-H)=0.33
\times{ E(B-V)}$ \citep{dut02a}, assuming a constant total-to-selective 
absorption ratio $R_{V}=3.1$.  The distance 
moduli of the clusters have been derived and listed in Col.~6 of 
Table 3. The estimated heliocentric $d~(kpc)$ and its corresponding 
galactocentric $R_{GC}$ (kpc) distances are given in 
Cols.~7$-$8, respectively. When estimating the $R_{GC}$ distances, 
we adopted the galactocentric distance of the Sun as $R_{\odot}=7.2\pm 
0.3$ kpc of \cite{bic06b}. 

The errors in E(J-H), hence in colour excess E(B-V), distance moduli 
and ages, given in Table~3 have been estimated as follows: 

\begin{enumerate}
 
\item The uncertainties of E(J-H) were estimated moving the M08 
isochrones up and down, back and forward and in direction of the 
reddening vector in the colour-magnitude diagram $(J,J-H)$ until a 
good fit with the observed MS, TO, the subgiant branch (SG), RG/RC  sequences were 
achieved. 

\item The uncertainties of distance moduli in Table 3 stem to a lesser 
degree from the photometric errors and fitting the appropriate isochrone 
to the observational data points in the CMDs.  A larger uncertainty, up 
to 2 mag in the distance moduli, originates by the assumption of the 
metallicity: for a larger Z the OCs are more distant and metal poor 
stars are nearer.
 
\item For the uncertainties in the age estimates, see those of the 
distance moduli. Again, metal-rich stars are younger than the metal-
poor ones.  

\end{enumerate}

The precision of the parameters depends on the scatter 
of the data points in the CMDs. The uncertainties of distance moduli 
in Table 3 stem from fitting the appropriate isochrone to the observation 
in the CMDs, by taking into the uncertainties of the photometric data. 
The uncertainties of distance moduli of 40 OCs are at the level of 
0.10$-$0.31. The uncertainty of age is obtained from fitting the M08 
isochrone with appropriate heavy element to the CMDs. In this regard, 
the uncertainty of the age depends on the uncertainties of E(J-H) and 
distance moduli of 40 OCs.  The uncertainties of the ages of 40 OCs in 
Table~3 fall in the range of 0.02$-$1.0 Gyr.

The relations of $E(B$--$V)$ versus Galactic longitude $\textit 
l^{\circ}$ and $E(B$--$V)$ versus Galactic latitude $\it b^{\circ}$ 
as a function of the cluster distances, are displayed in Figs.~8(a) 
and (b), respectively. In Figs.~8 open and filled circles show the 
$d=[0,~2.1]$ kpc  and  $d=(2.1,~5.42]$ kpc subsets, respectively. 
The reddenings of the OCs in the anticentre directions have   
$0.03 \leq E(B-V) \leq 1.31$. From panel~(a), the bulk of the 40 
clusters lies within $|b|\leq 5^{\circ}$ and $0.03\leq E(B-V) \leq 1.31$. 
There are two OCs with $E(B-V)>0.50$ in the Galactic centre directions. 

\begin{figure}    
\centering
\includegraphics*[width = 7cm, height = 10cm]{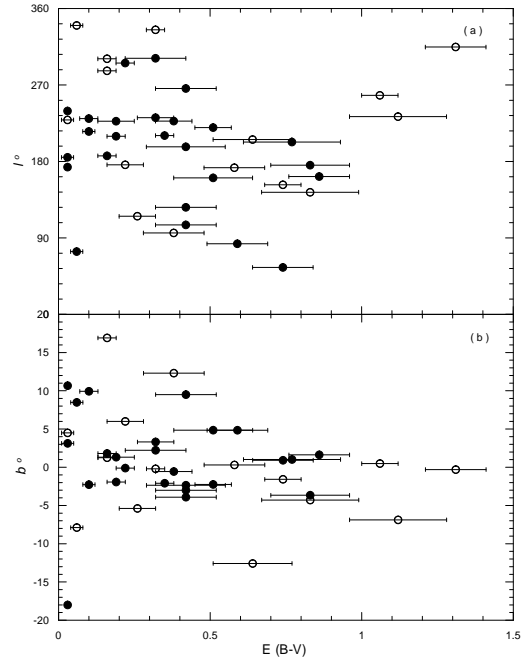}
\caption {$E(B$-$V)$ versus $\textit l^{\circ}$ (panel a) and versus 
$b^{\circ}$ (panel b) for the 40 OCs.  Open and filled circles 
show clusters with $d=[0, 2.1]$ kpc and $d=(2.1, 5.42]$
kpc, respectively.}
\end{figure}

The reddenings of 40 OCs have been compared to those of the dust maps 
of \cite[hereafter SFD]{sch98}, which are based on the COBE/DIRBE and 
IRAS/ISSA maps. These maps take into account the dust absorption 
$E(B-V)_{\infty}$ all the way to infinity.  

The relations of $E(B$--$V)_{\rm SFD,\infty}$ versus $E(B$--$V)$, and 
$E(B$--$V)_{\rm SFD}$versus $E(B$--$V)$ of the 40 OCs are displayed in Figs.~9(a) 
and (b), respectively. As is seen from Fig.~9(a), the values of $E(B$--$V)_{\rm 
SFD,\infty}$ are at the level of $0.07 \leq E(B$--$V)_{\rm SFD,\infty}\leq25.81$. 
For seven clusters, differences in between both reddenings are $\Delta E(B-V)\leq0.10$, 
while the differences of 33 OCs are larger than 0.10 mag. The equation given by 
\cite{bon} has been adopted to correct the SFD reddening estimates.  Then the final 
reddening, $E(B$--$V)_{\rm SFD}$, for a given star is reduced compared to the total 
reddening $E(B$--$V)(\ell, b)_\infty$by a factor $\lbrace1-\exp[-d \sin |b|/H]\rbrace$, 
given by \cite{bs80}, where $b$, $d$, and $H$ are the Galactic latitude (Col.~9 of 
Table~2), the distance from the observer to the object (Col.~7 of Table~3), and the 
scale height of the dust layer in the Galaxy, respectively. The value of $H=125$ pc is 
adopted \citep{bon}. The reduced final reddenings have been compared with the ones of 40 
OCs in Fig.~9(b).  The reduced $E(B$--$V)$ values fall in the range of $0.07 \leq E(B-V) 
\leq 1.261$. 

\begin{figure}    
\centering
\includegraphics*[width = 7cm, height = 10cm]{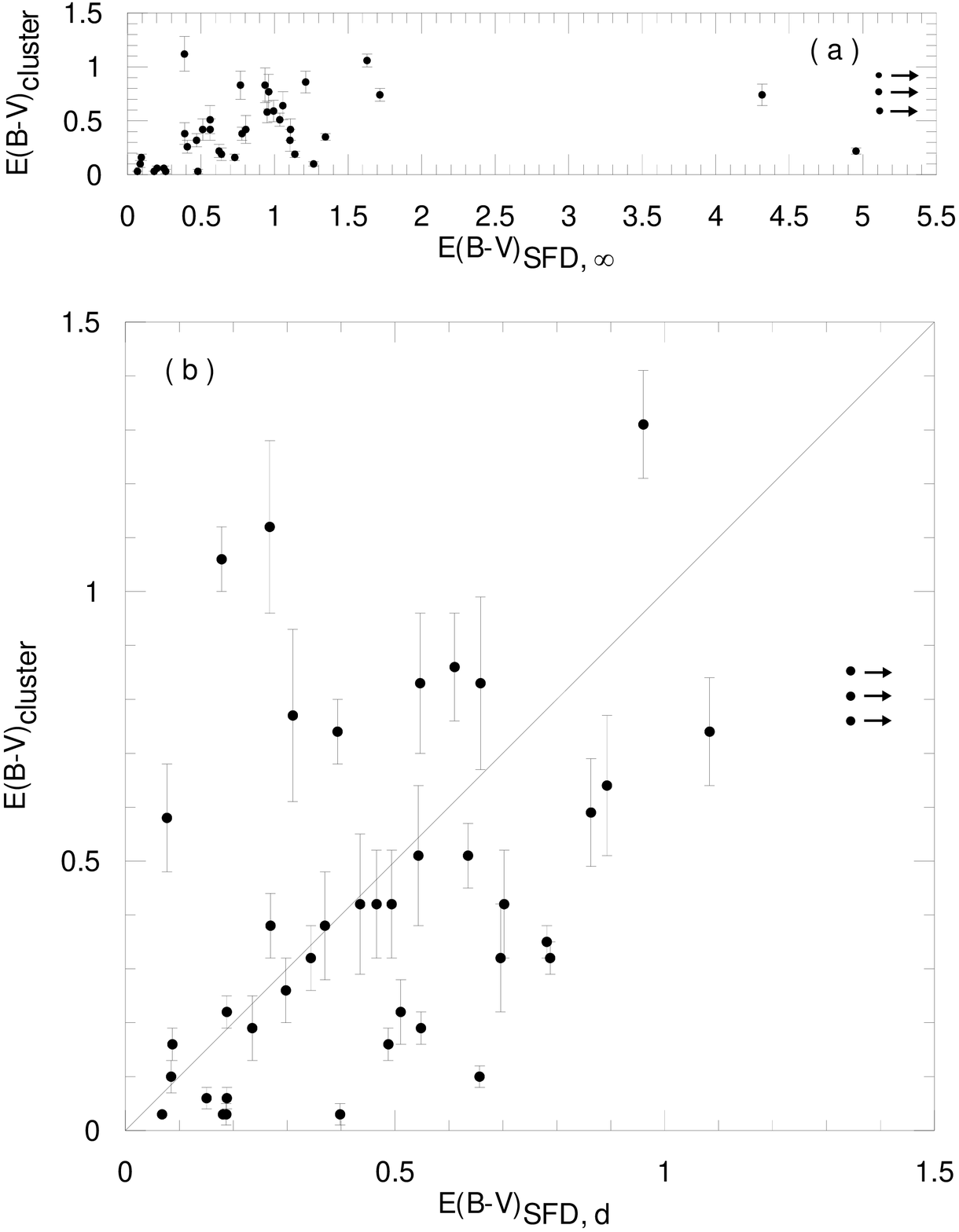}
\caption {Relations of E(B-V)$_{cluster}$-E(B-V)$_{SFD, \infty}$ (panel a), 
E(B-V)$_{cluster}$-E(B-V)$_{SFD, d}$ (panel b), respectively.} 
\end{figure}

There are significant differences for 27 OCs between both  $E(B$--$V)$ color excess 
values. For the rest, the $E(B$--$V)$ values of 13 OCs are quite close to the ones of 
SFD.  Note that  SFD maps are not reliable at regions  $|b|<5^{\circ}$ due to 
contaminating sources and uncertainties in the dust temperatures \citep{gon12}. 
Therefore, the SFD values resulted from line-of-sight integral through the Milky Way and 
with low spatial resolution, it is quite a normal to have different reddening values for 
these relatively close ($\sim 1$~kpc) star clusters.

\begin{table*}   
\renewcommand\thetable{5}
\centering
\caption{Structural parameters of 40 OCs. Col.~2 represents arcmin to parsec scale.  
$\sigma_0K$ in Col.~3 and 7 is the central density of stars. 
$\sigma_{bg}$ in Col.~4 and 8 is the residual background density. 
R$_{core}$ in Col.~5 and 8 and R$_{RDP}$ in Col.~6 and 10 are the core and cluster radii, respectively.
The symbols $* pc^{-2}$ and $*^{-2}$ in cols.~3, 4, 7 and 8 mean $stars~pc^{-2}$ and $stars~arcmin^{-2}$, respectively.
$\Delta$ R($'$) in Col.~11 denotes comparison field ring. Col.~12 represents the correlation coefficient.}
\tiny
\begin{tabular}{lccccccccccc} 
\hline 
Cluster & (1$'$) pc & $\sigma_{0K}$ (*pc$^{-2}$) & $\sigma_{bg}$ (*pc$^{-2}$) & R$_{core}$(pc) 
& R$_{RDP}$ (pc) & $\sigma_{0K}$ (*'$^{-2}$) & $\sigma_{bg}$ (*'$^{-2}$) & R$_{core}$ ($'$) 
& R$_{RDP}$ ($'$) & $\Delta$ R($'$) & C.C. \\
\hline  
NGC 436     &0.94&10.94$\pm$2.97&0.71$\pm$0.03&1.04$\pm$0.20&6.97$\pm$0.26& 9.60$\pm$2.60& 0.62$\pm$0.03&1.11$\pm$0.22&7.44$\pm$0.27&22-32&0.93\\    
King 5      &0.59&24.79$\pm$5.28&2.70$\pm$0.06&0.95$\pm$0.15&5.62$\pm$0.18&8.65$\pm$1.84&0.94$\pm$0.02&1.60$\pm$0.25&9.52$\pm$0.30&20-30&0.94\\	     
NGC 1513    &0.34&30.60$\pm$4.73&17.44$\pm$0.44&1.65$\pm$0.26&6.51$\pm$0.20&3.61$\pm$0.55&2.05$\pm$0.05&4.80$\pm$0.75&16.99$\pm$0.58&42-57&0.94\\    
Be 15       &0.90&25.70$\pm$11.33&0.97$\pm$0.02&0.35$\pm$0.10&5.04$\pm$0.30&20.89$\pm$9.20&0.79$\pm$0.02&0.39$\pm$0.11&5.59$\pm$0.33&30-40&0.86\\    
NGC 1798    &1.46&6.64$\pm$1.90&0.58$\pm$0.01&1.10$\pm$0.22&9.11$\pm$0.48&18.20$\pm$5.22&1.59$\pm$0.03&0.67$\pm$0.13&5.51$\pm$0.29&50-60&0.92\\	     
Be 17       &0.71&7.25$\pm$1.38&3.26$\pm$0.10&2.10$\pm$0.39&5.29$\pm$0.20&3.62$\pm$0.69&1.63$\pm$0.05&2.98$\pm$0.55&7.48$\pm$0.29&42-52&0.94\\	     
NGC 1907    &0.57&17.03$\pm$4.16&4.32$\pm$0.19&1.28$\pm$0.27&4.26$\pm$0.16&5.47$\pm$1.34&1.39$\pm$0.06&2.26$\pm$0.47&7.50$\pm$0.28&50-60&0.91\\	     
NGC 2112    &0.31&24.69$\pm$3.51&6.96$\pm$0.21&1.64$\pm$0.21&5.92$\pm$0.19&2.39$\pm$0.34&0.67$\pm$0.02&5.28$\pm$0.68&19.01$\pm$0.61&50-60&0.95\\     
Koposov 12  &0.60&8.96$\pm$3.74&1.27$\pm$0.05&0.87$\pm$0.27&3.82$\pm$0.20&3.18$\pm$1.33&0.45$\pm$0.02&1.46$\pm$0.46&6.41$\pm$0.33&15-25&0.83\\	     
NGC 2158    &1.27&28.85$\pm$4.67&1.45$\pm$0.06&1.74$\pm$0.20&14.03$\pm$0.71&47.05$\pm$7.61&2.37$\pm$0.10&1.36$\pm$0.16&10.99$\pm$0.56&45-60&0.97\\   
Koposov 53  &1.18&7.12$\pm$0.23&0.48$\pm$0.05&0.66$\pm$0.04&4.18$\pm$0.33&10.04$\pm$0.33&0.67$\pm$0.08&0.56$\pm$0.03&3.52$\pm$0.28&25-35&0.99\\	     
NGC 2194    &0.69&21.00$\pm$3.31&3.30$\pm$0.15&1.66$\pm$0.22&6.55$\pm$0.21&9.98$\pm$1.58&1.57$\pm$0.07&2.41$\pm$0.32&9.5$\pm$0.31&40-50&0.96\\	     
NGC 2192    &1.22&5.26$\pm$1.59&0.41$\pm$0.01&1.11$\pm$0.24&5.47$\pm$0.34&7.87$\pm$2.42&0.62$\pm$0.02&0.91$\pm$0.19&4.47$\pm$0.28&45-55&0.90\\	     
NGC 2243    &1.38&13.35$\pm$4.18&0.22$\pm$0.01&0.89$\pm$0.18&12.94$\pm$0.37&25.28$\pm$7.95&0.42$\pm$0.02&0.65$\pm$0.13&9.40$\pm$0.27&20-30&0.93\\    
Trumpler 5  &0.79&13.62$\pm$1.72&3.37$\pm$0.09&3.86$\pm$0.43&15.18$\pm$0.47&8.65$\pm$1.09&2.14$\pm$0.06&4.85$\pm$0.54&19.05$\pm$0.58&27-37&0.97\\    
Col 110     &0.71&5.32$\pm$0.51&2.68$\pm$0.05&6.25$\pm$0.63&12.12$\pm$0.40&2.68$\pm$0.26&1.35$\pm$0.02&8.79$\pm$0.88&17.07$\pm$0.57&40-50&0.97\\     
NGC 2262    &0.86&22.03$\pm$5.17&1.93$\pm$0.05&0.85$\pm$0.14&6.37$\pm$0.24&16.32$\pm$3.81&1.43$\pm$0.04&0.99$\pm$0.17&7.40$\pm$0.28&30-45&0.94\\     
NGC 2286    &0.67&6.81$\pm$2.29&2.66$\pm$0.12&1.59$\pm$0.48&6.39$\pm$0.19&3.07$\pm$1.03&1.20$\pm$0.05&2.37$\pm$0.72&9.51$\pm$0.29&27-37&0.85\\	     
NGC 2309    &0.88&12.41$\pm$5.96&0.64$\pm$0.05&0.84$\pm$0.28&7.50$\pm$0.25&9.64$\pm$4.63&0.50$\pm$0.04&0.95$\pm$0.32&8.51$\pm$0.29&50-60&0.84\\	     
Tombaugh 2  &0.35&134.29$\pm$83.62&3.67$\pm$0.25&0.17$\pm$0.07&1.92$\pm$0.11&16.91$\pm$10.5&0.46$\pm$0.03&0.47$\pm$0.18&5.42$\pm$0.31&20-25&0.98\\   
Be 36 	    &1.57&3.30$\pm$1.76&0.51$\pm$0.02&1.32$\pm$0.51&10.23$\pm$0.40&8.21$\pm$4.39&1.27$\pm$0.04&0.83$\pm$0.32&6.50$\pm$0.25&25-40&0.79\\	     
Haffner 8   &0.72&5.31$\pm$2.85&3.89$\pm$0.08&1.47$\pm$0.68&6.93$\pm$0.21&2.79$\pm$1.50&2.04$\pm$0.04&2.03$\pm$0.94&9.56$\pm$0.29&45-60&0.69\\	     
Mel 71      &0.59&22.61$\pm$3.97&4.39$\pm$0.09&1.27$\pm$0.18&5.00$\pm$0.17&7.89$\pm$1.39&1.53$\pm$0.03&2.16$\pm$0.30&8.46$\pm$0.29&45-60&0.95\\	     
NGC 2425    &0.83&11.00$\pm$2.04&2.28$\pm$0.04&1.16$\pm$0.17&5.43$\pm$0.22&7.55$\pm$1.41&1.57$\pm$0.03&1.40$\pm$0.20&6.54$\pm$0.26&42-47&0.95\\	     
NGC 2506    &0.82&18.56$\pm$3.13&1.07$\pm$0.04&1.65$\pm$0.20&10.76$\pm$0.48&12.67$\pm$2.14&0.73$\pm$0.03&2.00$\pm$0.24&13.02$\pm$0.58&40-50&0.96\\   
Pismis 3    &0.50&26.09$\pm$2.91&6.53$\pm$0.11&2.10$\pm$0.20&8.58$\pm$0.29&6.61$\pm$0.74&1.65$\pm$0.03&4.17$\pm$0.40&17.05$\pm$0.57&37-47&0.98\\     
NGC 2660    &0.69&90.44$\pm$24.32&3.84$\pm$0.09&0.39$\pm$0.07&5.27$\pm$0.17&43.71$\pm$11.74&1.86$\pm$0.04&0.55$\pm$0.10&7.58$\pm$0.25&25-35&0.94\\   
NGC 3680    &0.31&19.37$\pm$7.61&1.84$\pm$0.05&0.47$\pm$0.13&2.98$\pm$0.10&1.90$\pm$0.75&0.18$\pm$0.005&1.49$\pm$0.41&9.49$\pm$0.32&40-50&0.85\\     
Ru 96 	    &0.73&6.52$\pm$2.75&5.95$\pm$0.13&1.43$\pm$0.53&2.60$\pm$0.21&3.54$\pm$1.45&3.20$\pm$0.07&1.94$\pm$0.73&3.54$\pm$0.29&50-60&0.77\\	     
Ru 105 	    &0.56&2.63$\pm$1.81&1.52$\pm$0.05&1.35$\pm$0.78&3.83$\pm$0.21&0.93$\pm$0.64&0.54$\pm$0.02&2.27$\pm$1.30&6.42$\pm$0.34&47-57&0.64\\	     
Trumpler 20 &0.93&10.22$\pm$1.22&4.73$\pm$0.12&3.12$\pm$0.38&13.98$\pm$0.55&8.86$\pm$1.06&4.10$\pm$0.10&3.36$\pm$0.41&15.02$\pm$0.59&40-50&0.97\\    
Pismis 19   &0.56&104.06$\pm$15.47&13.68$\pm$0.18&0.54$\pm$0.06&6.16$\pm$0.33&32.46$\pm$4.82&4.27$\pm$0.05& 0.96$\pm$0.10&11.03$\pm$0.59&47-57&0.97\\
NGC 6134    &0.32&72.12$\pm$24.74&7.82$\pm$0.18&0.45$\pm$0.11&3.08$\pm$0.10&7.52$\pm$2.59&0.81$\pm$0.02&1.39$\pm$0.34&9.53$\pm$0.30&13-22&0.86\\     
IC 4651     &0.25&38.57$\pm$7.88&12.50$\pm$0.99&1.02$\pm$0.22&2.35$\pm$0.08&2.36$\pm$0.48&0.76$\pm$0.06&4.13$\pm$0.91&9.49$\pm$0.31&50-60&0.92\\     
NGC 6802    &0.65&34.48$\pm$9.18&7.90$\pm$0.27&1.03$\pm$0.18&4.24$\pm$0.18&14.72$\pm$3.63&3.38$\pm$0.12&1.58$\pm$0.28&6.49$\pm$0.38&45-60&0.92\\     
NGC 6819    &0.68&38.22$\pm$4.18&3.97$\pm$0.08&1.50$\pm$0.12&12.92$\pm$0.40&17.15$\pm$1.95&1.84$\pm$0.04&2.20$\pm$0.18&18.98$\pm$0.59&40-50&0.98\\   
Be 89       &0.88&7.02$\pm$1.35&3.80$\pm$0.12&2.75$\pm$0.53&7.48$\pm$0.26&5.42$\pm$1.04& 2.90$\pm$0.09&3.10$\pm$0.60&8.50$\pm$0.30&15-20&0.93\\	     
NGC 6939    &0.52&33.06$\pm$5.19&3.35$\pm$0.17&1.16$\pm$0.15&4.92$\pm$0.18&8.96$\pm$1.41&0.91$\pm$0.05&2.24$\pm$0.28&9.46$\pm$0.28&35-45&0.97\\	     
NGC 7142    &0.74&10.15$\pm$1.87&1.72$\pm$0.10&1.98$\pm$0.32&11.19$\pm$0.44&5.63$\pm$1.04&0.95$\pm$0.05&2.65$\pm$0.43&15.02$\pm$0.59&50-60&0.94\\    
NGC 7789    &0.51&31.23$\pm$2.22&3.88$\pm$0.04&2.32$\pm$0.13&26.88$\pm$0.74&8.18$\pm$0.58&1.02$\pm$0.01&4.52$\pm$0.25&52.5$\pm$1.44&55-70&0.99\\     
\hline    
\end{tabular}
\end{table*}

\section{Structural parameters}

We derived the structural parameters of 40 OCs from the stellar radial density profiles (RDPs). 
Usually, the RDPs of star clusters can be described by an analytical profile, like the 
empirical, single mass, modified isothermal spheres of 
\cite{kin66} and \cite{wil75}, and the power law with a core
of \cite{els87}. These functions are characterized by different 
sets of parameters that are related to the cluster structure. 
Here we adopted the two-parameter function $\sigma(R) = 
\sigma_{bg} + \sigma_0/(1+(R/R_c)^2)$, where $\sigma_{bg}$ 
is the residual background density, $\sigma_0$ the central 
density of stars, and R$_{core}$ the core radius. Applied to 
star counts, this function is similar to that used by 
\cite{kin62} to describe the surface brightness profiles in 
the central parts of globular clusters. To minimize degrees of 
freedom in RDP fits with the King-like profile, $\sigma_{bg}$ 
was kept fixed (measured in the respective comparison fields) 
while $\sigma_{0}$ and $R_{core}$ were determined by the best 
profile fit to the data. As a representative of the OCs sample, 
the RDP of Pismis~19 fitted with King's profile is shown in  
Fig.~10, where the solid line shows the best profile fit. The 
horizontal red bar in the figure denotes the stellar background 
level measured in the comparison field, and the $1\sigma$ profile 
fit uncertainty is shown by the 
shaded domain. The stellar RDPs fitted profiles of the 40 OCs have 
been given in Figs.~S10$-$S13 as supplementary material. The cluster 
radius (R$_{RDP}$) is also obtained from the measured distance from 
the cluster centre where the RDP and residual background are 
statistically indistinguishable \citep{bon07a}. The R$_{RDP}$ can 
be taken as an observational truncation radius, whose value depends 
on the radial distribution of member stars and the stellar field 
density. $\Delta R$  means the wide external ring of the stellar 
comparison field (see also Sect.~3). These structural parameters and 
their meaning are listed in Table~5.  

\begin{figure}     
\centering
\includegraphics*[width = 7cm, height = 7cm]{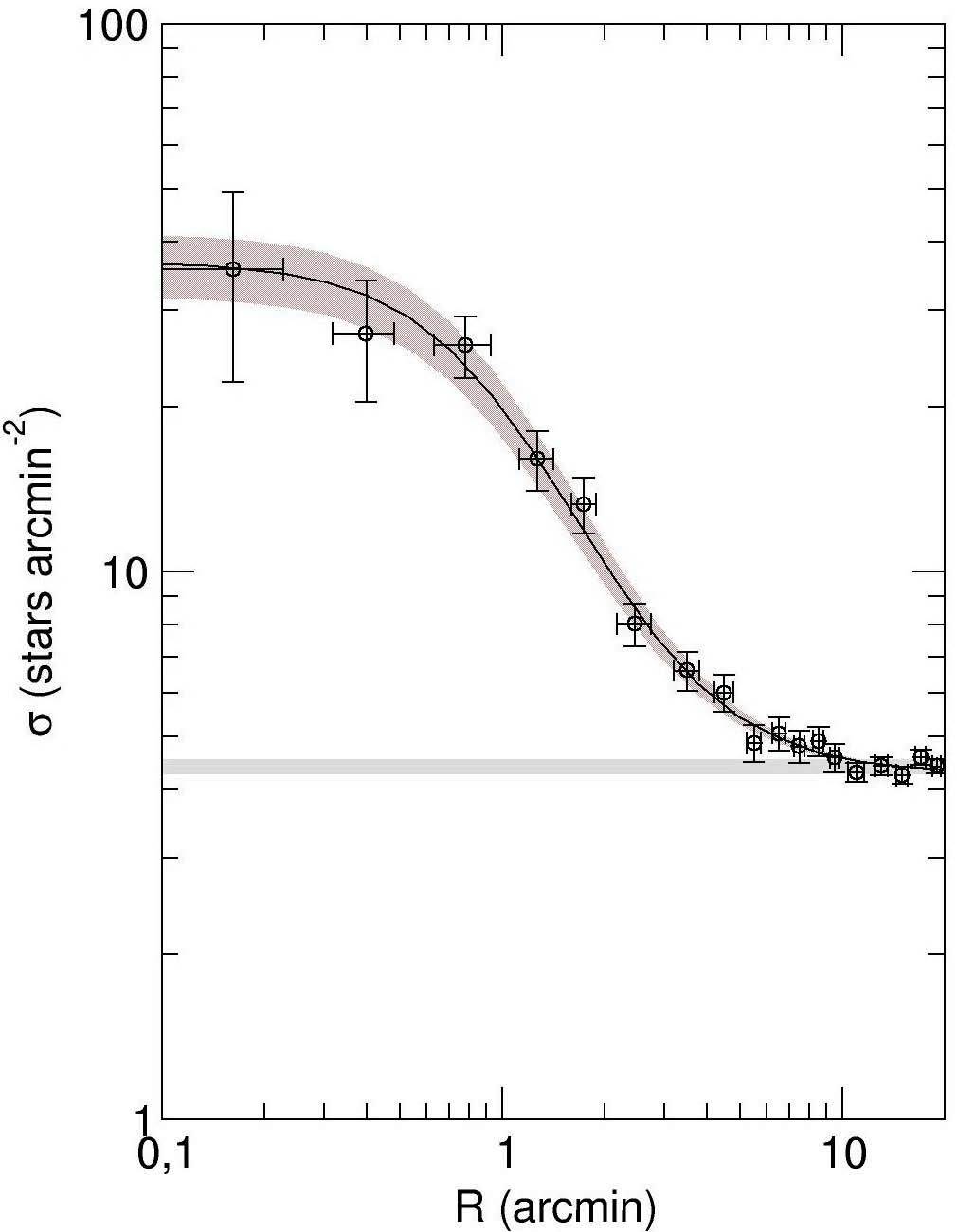}
\caption {Stellar RDP (open circles) of Pismis~19 built with 
CMD filtered photometry. Solid line shows the best-fit King profile. 
Horizontal red bar: stellar background level measured 
in the comparison field. Shaded region: $1\sigma$ King fit uncertainty.}
\end{figure}

From the distributions of R$_{core}$ and R$_{RDP}$, given in 
Fig.~11(a) and (b), there seems to be two groupings at R$_{RDP}
$$=$7 pc and R$_{core}$$=$1.5 pc, respectively, which are  
close to the values of 10 pc and 1.5 pc of  \cite{buk11}.

\begin{figure}      
\centering
\includegraphics*[width = 7cm, height = 11cm]{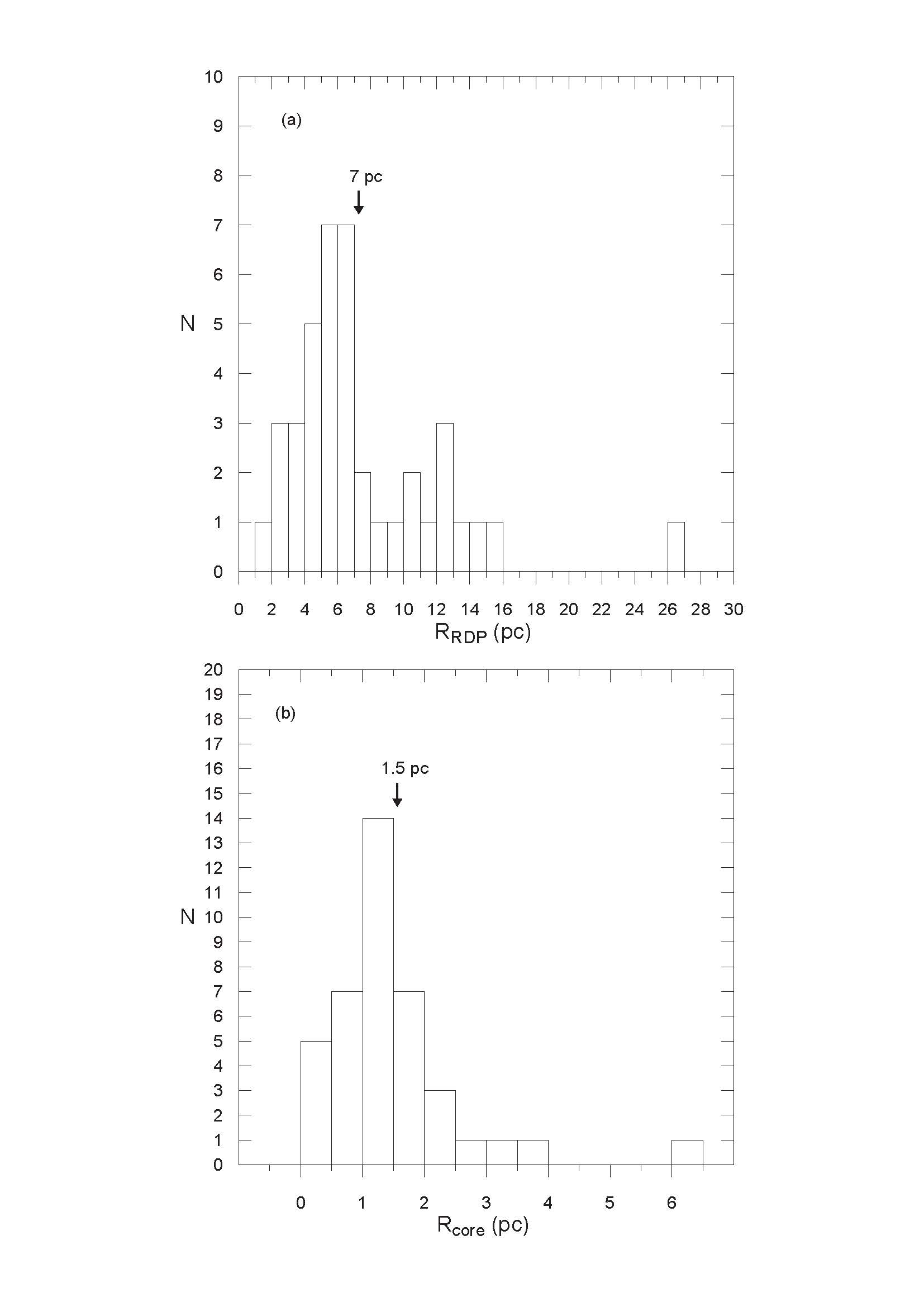}
\caption {Distributions of R$_{RDP}$ 
(panel a) and R$_{core}$ (panel b) of 40 OCs, respectively.} 
\end{figure}

\section{Mass and Mass functions}

The stellar masses stored in the OCs of our sample have been 
determined by means of their mass functions (MFs), built for 
the observed MS mass range, according to \cite{bic06a}. 
By following the algorithm, which is basically defined by 
\cite{bon05}, luminosity functions from the decontaminated 
$(J, J-H)$ diagrams of the OCs have been transformed into MFs 
through the corresponding mass-luminosity relations derived 
from the M08 isochrones which correspond to the ages in Col.~5 
of Table 3. We determined the overall masses of 26 OCs and the 
core masses of 24 OCs in our sample. The total mass locked up 
in stars of these OCs was obtained by considering all stars from 
the turnoff to the H-burning mass limit. We do this by directly 
extrapolating the low-mass MFs down to $0.08M_{\odot}$. Here we 
have based our results on the CMD filtered photometry of open 
cluster and offset field stars. The filtering process contemplates 
most of the background, leaving a residual contamination. Due 
to the relatively large sizes of the OCs and the brightness 
limitation of the 2MASS photometry, we do not have access to 
the whole stellar mass range of the OCs. 
Here, we stress that the values we derive should be taken as approximations.

\begin{figure}    
\centering
\includegraphics*[width = 7cm, height = 7cm]{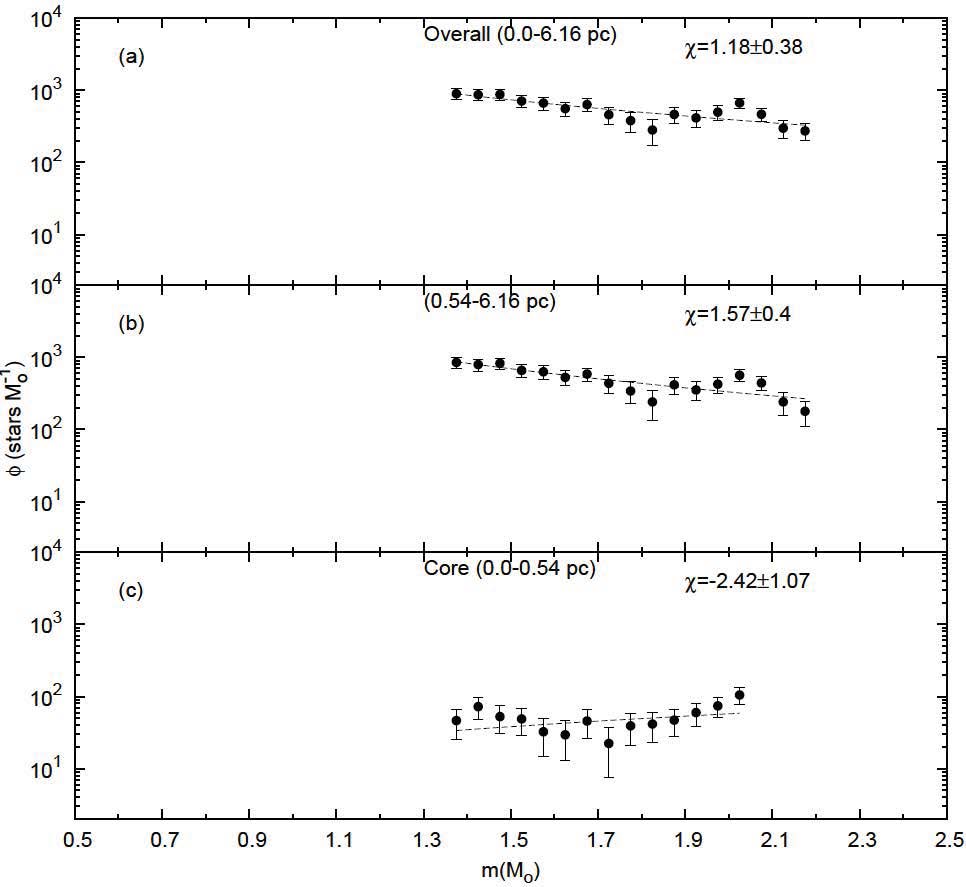}
\caption {$\phi(m)(stars ~m_\odot^{-1})$ versus $m_\odot$ of Pismis~19 cluster,
as a function of distance from the core.} 
\end{figure}

The relation of $\phi(m)(stars ~m_{\odot}^{-1})$ versus $m_{\odot}$ 
of our representative open cluster Pismis~19 is shown in 
Figs.~12(a)$-$(c) for different cluster regions. 
The main sequence mass functions (MFs) in the panels~(a)$-$(c) of 
Fig.~12 are fitted with the function $\phi(m)\propto{m}^{-(1+\chi)}$, 
and the MF slopes ($\chi$) have been determined for the different 
segments of the mass function MF in Col.~1 of Table~6. More details 
of this approach are given in Table~6, where we also show the number 
and mass of the evolved stars (m$_{evol}$). The MF slopes of the core 
(29 OCs) and the overall (31 OCs) regions of OCs are presented in 
Cols.~2 and 5 of Table 7. Since the lower MS is not accessible on the 
$(J, J-H)$ diagrams of the OCs sample, we assumed that the low-mass 
content is still present, and use Kroupa's MF\footnote{$\chi=0.3\pm0.5$ 
\cite{kro01} for $0.08<M_{\odot}<0.5$, $\chi=1.3\pm0.3$ for $0.5 < 
M_{\odot}<1.0$, and $\chi=1.3\pm 0.7$ for $1.0<M_{\odot}$} to estimate 
the total stellar mass, down to the H-burning mass limit. The results: 
number of stars, MS and evolved star contents (m$_{obs}$), MF 
slope ($\chi$), and mass extrapolated (m$_{tot}$) to 0.08~$M_{\odot}$) 
for each cluster region are given in Table 6. The mass densities of 
$\rho$ in unit of $M_{\odot}\: pc^{-3}$ are also estimated and given 
in Cols.~8 and 11 of Table 6 (See also sect.~7.7).

When deriving the mass functions, the part of the steep, that is  
observed in the core may come from crowding and completeness. 2MASS 
is not very photometrically deep and has just a moderate spatial 
resolution. So, in crowded regions (such as the core of most 
clusters) many stars are not detected, especially the faint ones. 
This, in turn may mimic mass segregation.

The relaxation time $t_{rlx}$ (Myr) is the characteristic 
time-scale for a cluster to reach some level of energy 
equipartition \citep{bin98}. As discussed in \cite{bon05}, 
\cite{bon06a}, and  \cite{bon07a}, the evolutionary parameter 
($\tau = Age/t_{rlx}$) appears to be a good indicator of dynamical 
state. Following \cite{bon06a}, we parameterize $t_{rlx}$ as 
$t_{rlx}\approx0.04\left(\frac{N}{lnN}\right)\left(\frac{R}{1pc}\right)$, 
where N is the number of stars located inside the region of radius R. 
The relaxation time and evolutionary parameter for both core and the 
overall regions are listed in Table 7. The uncertainties in the 
evolutionary parameters ($\tau$) of OCs have been estimated by 
propagating the errors in Age (Table 3), Radii (Table 5) and N 
(Table 6) into $t_{rlx}$ and $\tau$.  When propagated, the latter 
two errors produce a large uncertainty in $t_{rlx}$ (Table 7) and, 
consequently, a large uncertainty in the evolutionary parameter. 
In this sense, both $t_{rlx}$ and $\tau$ should be taken simply as 
an order of magnitude estimate.

From the overall mass distribution ($m_{overall}$) of 26 OCs 
displayed in Fig.~13, 2000~$M_\odot$ value is considered as a 
criteria in classifying the clusters as less massive and 
massive. 

\begin{figure}         
\centering
\includegraphics*[width = 6.5cm, height = 7cm]{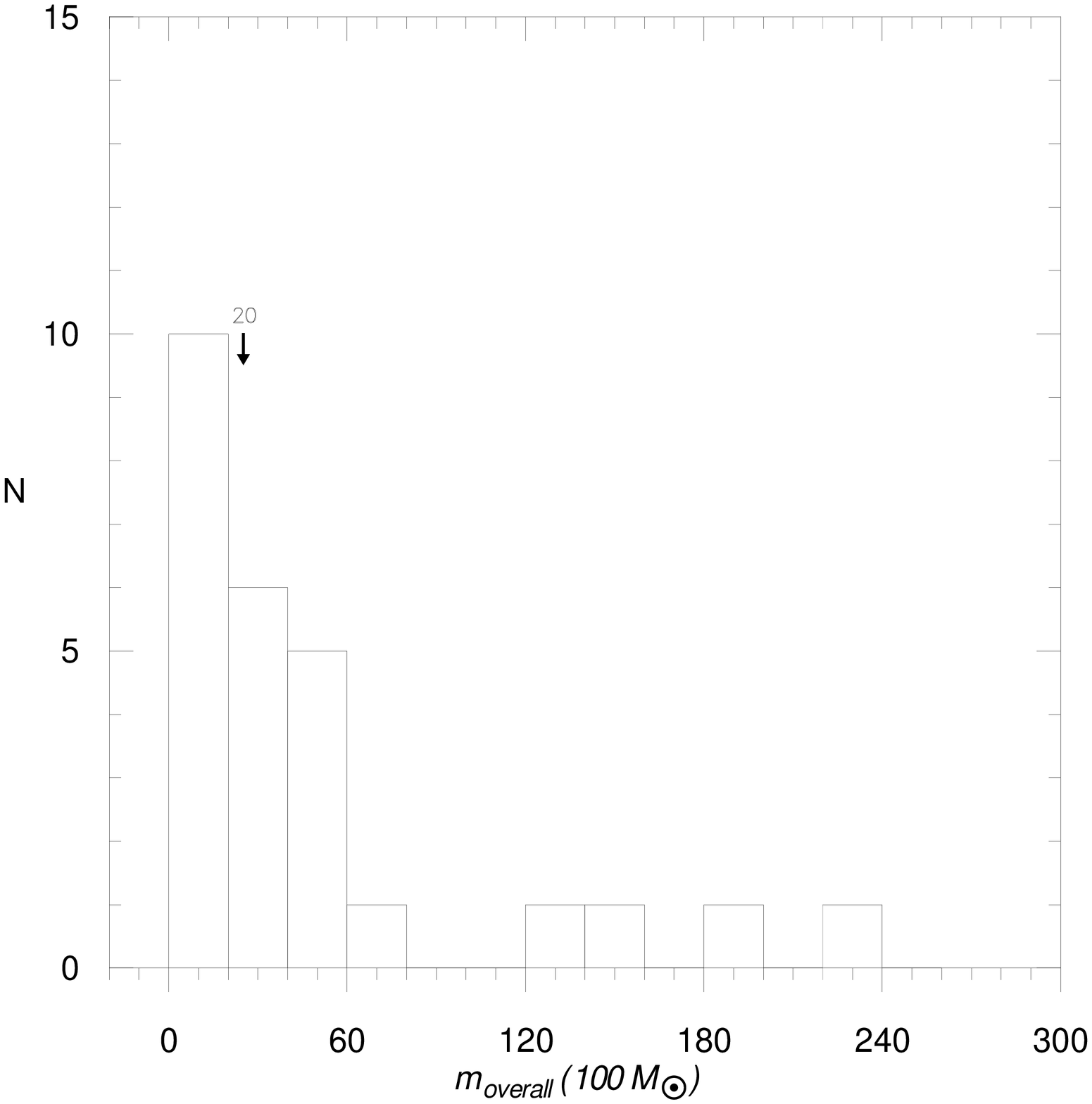}
\caption {The overall mass distribution of 26 OCs.}
\end{figure}
\clearpage

\begin{table*}   
\renewcommand\thetable{6}
\centering
\caption{The number of stars, mass information, mass function slope, mass density, 
which correspond to cluster regions of available clusters for the cases of 
Evolved, Observed+Evolved, and Extrapolated+Evolved. 
The full version is available in the online version of this manuscript in the supplementary material section (Table~S6).}
\tiny
\begin{tabular}{c|c|c|c|c|c|c|c|c|c|c}
\hline
\hline
\multicolumn{11}{c}{NGC 436}\\\cline {5-7} \\
      & \multicolumn{2}{c}{Evolved} & 
         \multicolumn{2}{c} {$\chi$} &  
          \multicolumn{3}{c}{Observed+Evolved} &    
           \multicolumn{3}{c}{Extrapolated+Evolved} \\\cline {2-11} \\
   
 {Region}  & {N*}    & {m$_{evol}$}        & {1.38-2.78} & {-} & {N*}             & {m$_{obs}$}         & {$\rho$} & {N*}& {m$_{tot}$} & {$\rho$} \\
  (pc)     & (Stars) & ($10^1 M_{\odot}$)&             &     &  ($10^2 Stars$)   & ($10^2 M_{\odot}$)  &  $M_{\odot} pc^{-3}$  & ($10^2 Stars$) & ($10^2 M_{\odot}$ & $M_{\odot} pc^{-3}$ \\
\hline
0.0-1.04 &  1$\pm$1   &  0.4$\pm$0.4 &  -1.46$\pm$0.47 &  {-} &  0.25$\pm$0.03 &  0.56$\pm$0.28 &  11.9$\pm$5.97 &  0.4$\pm$0.1 &  0.7$\pm$0.03 &  15.2$\pm$6.11 \\
1.04-6.97 &  12$\pm$6  &  3.5$\pm$1.8 &  1.74$\pm$0.36 &  {-} &  1.01$\pm$0.1 &  2$\pm$0.57 &  0.14$\pm$0.04 &  25.6$\pm$19.6 &  9.8$\pm$3.8 &  0.69$\pm$0.27 \\
0.0-6.97 &  14$\pm$6  &  3.9$\pm$1.9 &  0.86$\pm$0.29 &  {-} &  1.12$\pm$0.09 &  2.55$\pm$0.63 &  0.18$\pm$0.04 &  17.1$\pm$11.9 &  7.9$\pm$2.4 &  0.56$\pm$0.17 \\
$\cdot \cdot \cdot$&$\cdot \cdot \cdot$&$\cdot \cdot \cdot$&$\cdot \cdot \cdot$&$\cdot \cdot \cdot$
&$\cdot \cdot \cdot$&$\cdot \cdot \cdot$&$\cdot \cdot \cdot$&$\cdot \cdot \cdot$&$\cdot \cdot \cdot$&$\cdot \cdot \cdot$ \\
\hline
\end{tabular}
\\
Col.~1: the distance from the core. Cols.~2,6,9 : cluster stars for the regions in Col.~1.
Col.~4 gives the MF slopes ($\chi$),  derived for the low-mass and high-mass ranges. 
The masses of $m_{evol}$, $m_{obs}$, and $m_{tot}$ are listed in Cols.~3, 7 and 10, respectively.
The mass densities are given in Cols.~8 and 11.
\\
\end{table*}

\begin{table*}   
\renewcommand\thetable{7}
\centering
\caption{Mass function slopes ($\chi$), relaxation time (t$_{rlx}$(Myr)) 
and evolutionary parameter ($\tau$) of core and overall regions of the available clusters.}
\scriptsize
\begin{tabular}{lcccccc}
\hline
& \multicolumn{3}{c}{Core}& \multicolumn{3}{c}{Overall} \\\cline{2-7}
Cluster & $\chi$     & t$_{rlx}$(Myr) & $\tau$$_{core}$ & $\chi$& t$_{rlx}$(Myr) & $\tau$$_{overall}$ \\
\hline
NGC 436  &$-$1.46$\pm$0.47   &0.46$\pm$0.12      & 869.57$\pm$314.19   & 0.86$\pm$0.29   & 63.95$\pm$38.60     & 6.25$\pm$4.08 \\   
King 5    &$-$3.06$\pm$0.96  &0.58$\pm$0.11      & 1724.14$\pm$475.22  & 1.80$\pm$0.49   & 215.00$\pm$148.09   & 4.65$\pm$3.34 \\   
NGC 1513 &1.12$\pm$0.24      &10.21$\pm$6.40     & 9.79$\pm$6.44       & 1.90$\pm$0.12   & 175.88$\pm$118.13   & 0.57$\pm$0.40 \\   
Be 15    & -                 & -                 & -                   &$-$1.54$\pm$1.15 & 5.05$\pm$2.78       & 99.01$\pm$57.99 \\ 
NGC 1907 &$-$0.76$\pm$0.40   &1.71$\pm$0.43      & 233.92$\pm$82.95    & 0.00$\pm$0.23   & -                   & - \\		    
NGC 2112 &$-$1.28$\pm$0.51   &3.28$\pm$3.59      & 609.76$\pm$673.63    & 0.50$\pm$0.42   & 126.22$\pm$78.13   & 15.85$\pm$10.10 \\ 
NGC 2158 &$-$4.24$\pm$1.00   &5.05$\pm$0.16      & 495.05$\pm$61.44    &$-$1.55$\pm$0.71 & -                   & - \\		    
Koposov 53&$-$3.96$\pm$3.40  & -                 & -                   & 0.93$\pm$0.81   & 15.51$\pm$11.41     & 64.47$\pm$47.86 \\ 
NGC 2194  & 0.38$\pm$0.42    &13.15$\pm$7.87     & 60.84$\pm$39.46     & 2.52$\pm$0.37   & 456.55$\pm$311.22   & 1.75$\pm$1.27 \\   
NGC 2192 &$-$2.78$\pm$0.96   &0.36$\pm$0.10      & 3611.11$\pm$1040.84 &$-$3.12$\pm$0.43 & 7.23$\pm$0.85       & 179.81$\pm$25.26 \\
NGC 2243 & -                 & -                 & -                   & 2.09$\pm$1.01   & 826.05$\pm$654.27   & 2.42$\pm$1.93 \\   
Trumpler 5&0.42$\pm$0.75     &1195.27$\pm$869.82 & 2.51$\pm$1.84       & 1.32$\pm$1.18   & 3804.93$\pm$2817.13 & 0.79$\pm$0.59 \\   
Col 110  &$-$2.58$\pm$0.21   &29.78$\pm$4.37     & 100.74$\pm$16.24     &$-$2.84$\pm$0.57 & 100.01$\pm$22.83   & 30.00$\pm$7.13 \\  
NGC 2262 &$-$1.49$\pm$0.80   &0.43$\pm$0.15      & 3023.26$\pm$1079.96 & 1.01$\pm$0.44   & 186.81$\pm$126.13   & 6.96$\pm$4.73 \\   
NGC 2286 &1.30$\pm$0.50      &5.97$\pm$4.26      & 167.50$\pm$124.13   & 1.45$\pm$0.30   & 99.95$\pm$65.65     & 10.01$\pm$6.87 \\   
NGC 2309 &$-$1.52$\pm$1.03   &0.35$\pm$0.07      & 1428.57$\pm$404.06  &$-$0.89$\pm$0.60 & -                   & - \\		    
Haffner 8&1.28$\pm$0.77      &5.50$\pm$4.79      & 181.82$\pm$159.39   & 1.82$\pm$0.59   & 101.14$\pm$68.47    & 9.89$\pm$6.77 \\   
Mel 71   &0.30$\pm$1.04      & -                 & -                   & 1.29$\pm$0.40   & 146.55$\pm$99.37    & 10.24$\pm$7.08 \\   
NGC 2506  & 4.11$\pm$1.63    &56.00$\pm$49.28    & 35.71$\pm$31.88     & 0.97$\pm$0.63   & 822.94$\pm$594.95   & 2.43$\pm$1.79 \\   
Pismis 3 &$-$1.60$\pm$0.84   &10.42$\pm$2.03     & 307.10$\pm$62.83    & 1.71$\pm$0.49   & 1348.20$\pm$939.15  & 2.37$\pm$1.66 \\   
Ru 96    &4.58$\pm$0.98      &17.70$\pm$15.06    & 56.50$\pm$48.40     & 4.55$\pm$0.65   & 57.49$\pm$40.92     & 17.39$\pm$12.50 \\ 
Trumpler 20&$-$1.07$\pm$0.50 &12.85$\pm$17.60    & 116.73$\pm$164.55   & 2.06$\pm$0.68   & 2272.78$\pm$1663.64 & 0.66$\pm$0.53 \\   
Pismis 19&$-$2.42$\pm$1.07   &0.45$\pm$0.10      & 1777.78$\pm$453.27  & 1.18$\pm$0.38   & 402.04$\pm$272.33   & 1.99$\pm$1.37 \\   
NGC 6134  &$-$0.95$\pm$0.90  & -	         & -		       &$-$0.83$\pm$1.11 & -                   & -	\\	    
IC 4651  &$-$2.78$\pm$0.75   &1.09$\pm$0.06      & 2293.58$\pm$302.81  &$-$0.60$\pm$0.41 & -                   & - \\		    
NGC 6802 &$-$0.46$\pm$0.84   & -                 & -                   & 1.66$\pm$0.24   & 232.20$\pm$156.40   & 3.88$\pm$2.65 \\   
NGC 6819 &$-$1.07$\pm$0.55   & -                 & -                   & 0.47$\pm$0.40   & 680.42$\pm$429.71   & 3.67$\pm$2.43 \\   
Be 89    &  0.18$\pm$0.66    & -                 & -                   & 1.64$\pm$0.93   & 312.83$\pm$220.34   & 6.39$\pm$4.78 \\   
NGC 6939 &$-$2.17$\pm$0.66   &1.42$\pm$0.07      & 1408.45$\pm$222.38  & 0.84$\pm$0.46   & -                   & - \\		    
NGC 7142  &$-$1.97$\pm$1.44  & -	         & -		       &$-$1.06$\pm$0.56 & -                   & -		\\  
NGC 7789 &$-$0.42$\pm$0.45   & -                 & -                   & 0.79$\pm$0.65   & 5471.49$\pm$3972.27 & 0.33$\pm$0.24 \\   
\hline
\end{tabular}
\end{table*}

\section{Results}

\subsection{Relation of R$_{RDP}$--R$_{core}$}

The cluster and core radii (R$_{RDP}$, R$_{core}$) of 40 OCs, given 
in Fig.~14 are  related by the following relation,    
R$_{RDP}$$=(4.69\pm0.35)R_{core}^{(0.56\pm0.11)}$ 
with a mild  correlation coefficient (CC, hereafter) of 0.61. 
This relation of Fig.~14 is almost linear between $log\: (R_{RDP})$ and 
$log\: (R_{core})$, where the axes are in a log-log scale.  
Their core and cluster sizes  are $0.17\leq R_{core}~(pc) \leq 6.25$ 
and $1.92\leq R_{RDP}~(pc)\leq 26.88$, respectively. The OCs in our 
sample which do not follow the relation above are either intrinsically 
small or have been suffering significant evaporation effects. Our 
coefficient value (4.69) of Fig.~14 falls in the 
range of $3.1-8.9$ of the literature (Table 8). However the coefficients 
in Table~8 are affected by the sample size. The relation between R$_{RDP}$ 
and R$_{core}$ found by us is reasonably similar to that given by \cite{cam10}. 
However analogue functions were found by other authors, \cite{nil02, bic05, sha06, mn07, buk11}.

\begin{figure}        
\centering
\includegraphics*[width = 7cm, height = 7cm]{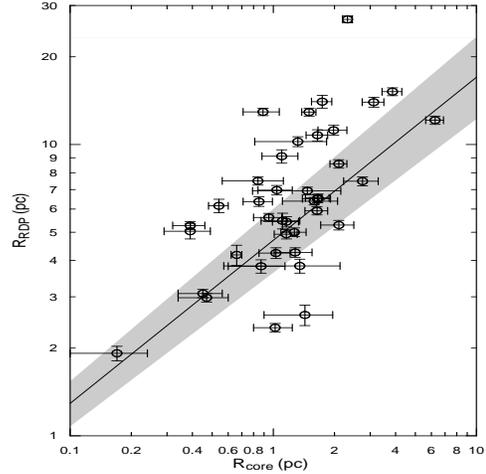}
\caption {Relation of R$_{RDP}$ - R$_{core}$ of 40 OCs. 
Empty circles show 40 OCs. Solid line and shaded area show 
the best fit and $1\sigma$ uncertainty, respectively.}
\end{figure}

\begin{table}   
\tiny
\renewcommand\thetable{8}
\centering    
\tiny
\caption{The coefficients of the relation, R$_{RDP}$$=$a+bR$_{core}$, given   
in the literature between  R$_{core}$ and R$_{RDP}$. 
The form of the relation of Camargo et al.~(2010) is $R_{RDP}=bR_{core}^{a}$.
CC and N in last two columns mean the correlation coefficient and data number, respectively.}
\begin{tabular}{lcccc}
\hline 
Author &  a &  b & CC & N \\
\hline 
\cite{nil02} &   - &6    &  -  & 38 \\
\cite{bic05} &1.05 &7.73 &0.95 & 16 \\
\cite{sha06} &   - &3.1  &  -  &  9 \\
\cite{mn07}  &   - &3.1  &0.74 & 42 \\
\cite{cam10} & 0.3 &8.9  &  -  & 50 \\
\cite{buk11} &0.58 &6.98 &0.93 &140 \\
\hline    
\end{tabular}
\end{table}

\subsection{Relations of Cluster Dimensions to the Distance and Age}

The relations of R$_{RDP}$ and R$_{core}$ with d(kpc) are apparently 
linear, and are displayed in Figs.~15(a)$-$(b). The linear best fit to the 
data (solid lines) are the following, R$_{RDP}=(2.67\pm0.27)\: d(kpc)$ 
(CC$=$0.84) and R$_{core}= (0.50 \pm 0.07)$~d(kpc) 
(CC$=$0.76), respectively. Given a couple of deviants, the sizes (R$_{RDP}$ and 
R$_{core}$) increase on the average with the distance from the Sun.
Similar trends were also obtained by \cite{lyn82}, 
\cite{van91}, \cite{tad02}, \cite{bon10} and \cite{buk11}.

\begin{figure}     
\centering
\includegraphics*[width = 6.5cm, height = 11cm]{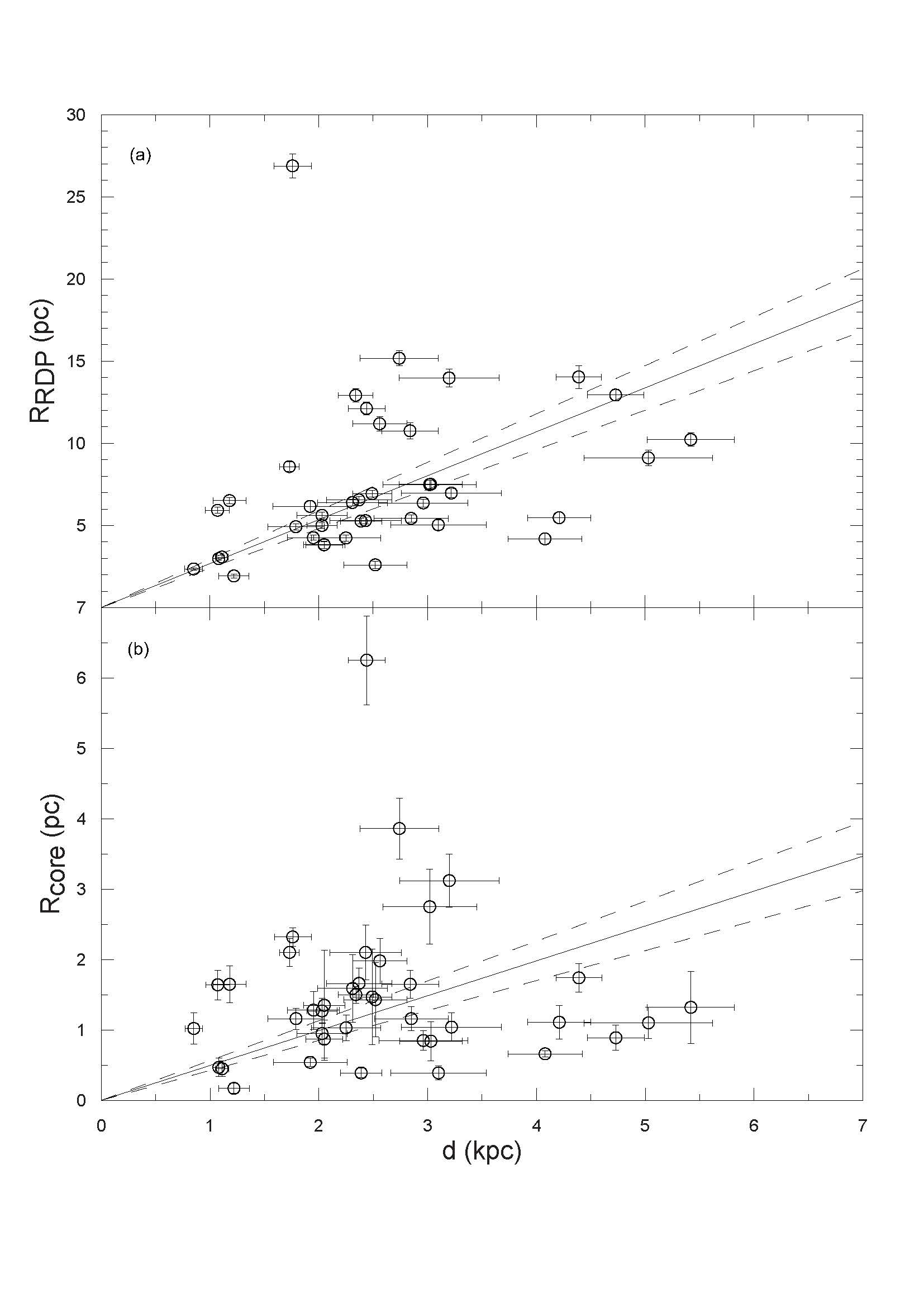}
\caption {Relations of R$_{RDP}$ - d(kpc)(panel a), 
R$_{core}$ - d(kpc)(panel b), respectively. Solid and dashed lines 
show the best fit and $1\sigma$ uncertainty, respectively.} 
\end{figure}

The relations of $|z|$ and R$_{GC}$ as a function of Age and R$_{RDP}$, 
respectively, are presented  in Figs.~16(a)$-$(b). Younger and older 
clusters than 1 Gyr in panels~(a)$-$(b) lie inside/outside the Solar 
circle.  No cluster with R$_{RDP} > 8$~pc is seen in panel~(a).  
The OCs with Age$\geq$1~Gyr in panel~(b) 
do not show any dependence of R$_{GC}$ and R$_{RDP}$. The OCs, NGC 2243 
and NGC 2192 with $|z| > 800$ pc outside the Solar circle in panels~(b)$-$(c),
where GMCs are scarce, might have been moved to outer 
parts of the Galactic radii via tidal interactions with the disc and the 
Galactic bulge, and collisions with GMCs. Alternatively, they may have been 
formed from molecular clouds at these distances. Note that 
\cite{sch06} have also detected large and small sized clusters outside 
the Solar circle. From panel~(b) we note that most of large/small sized 
OCs inside or outside the Solar circle are located near the galactic 
plane ($|z| < 300$ pc) and the OCs inside the Solar circle seem to 
survive four or more rotations around the Galactic centre. Their 
survival can be explained by which they survived against external shocks 
\citep{jan94}.

\begin{figure}    
\centering
\includegraphics*[width = 7cm, height = 13cm]{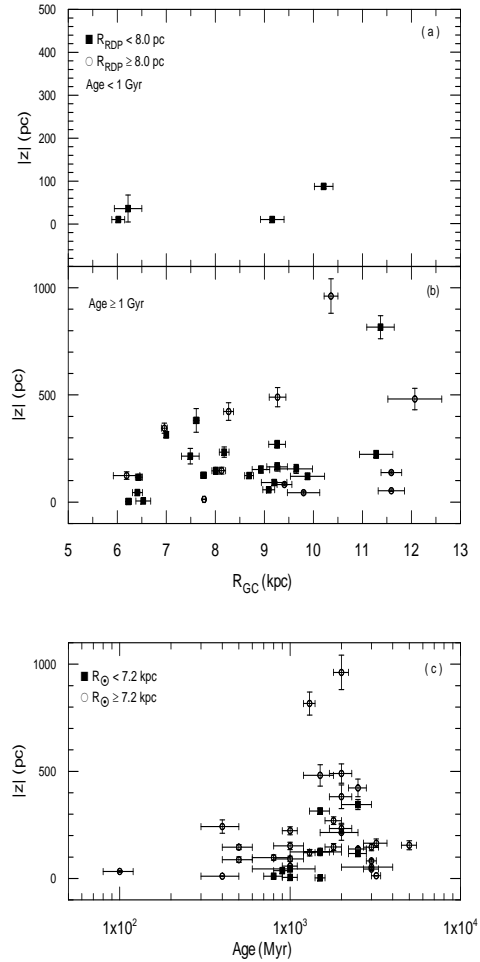}
\caption {Relations of $|z|$- R$_{GC}$ in terms of  
R$_{RDP}$ (panel a) and Age (Myr) (panel b).
Filled squares and empty circles show the OCs with  
R$_{RDP}<8$~pc and R$_{RDP}\ge8$~pc, respectively. 
Relation of $|z|$-Age as function of R$_{\odot}$ (panel c).}
\end{figure}

Old clusters with large dimensions inside the Solar circle in panel~(b) 
may have a primordial origin, or their sizes may have been increased 
via expansion due to stellar mass black hole couples. 
For the relation, $|z|$- Age as a function of R$_{\odot}$ 
in Fig.~16(c), the OCs with Age $\geq 1$ Gyr reach higher z distances, 
whereas those with Age $<1 $ Gyr  have $|z|<300$~pc.

\subsection{Relations of R$_{RDP}$-Age and R$_{core}$-Age}

The relations of R$_{core}$$-$Age and R$_{RDP}$$-$Age have been 
displayed in Figs.~17(a)$-$(b). In Fig.~17, filled circles and 
empty triangles show 16 OCs with $m_{overall} \ge 2000~M_{\odot}$ 
and 10 OCs with $m_{overall}<2000~M_{\odot}$, respectively. 14 
OCs which have no mass determinations are marked by open squares. 
The relation in Figs.~17(a)$-$(b) suggests a bifurcation which is seen 
at an age $\approx$ 1 Gyr. In the sense, in Fig.~17 some clusters 
appear to expand (`A' arrow), while others contract (`B' arrow) 
with a bifurcation occuring at about 1 Gyr.  \cite{mac08} observed 
the bifurcation at $\approx$~500-600~Myr (shown with `C' in the 
panels of Fig.~17). This kind of relations in the panels were 
also observed by  \cite{bon07a}, \cite{mn07}, and  \cite{cam10} 
from their OC samples.

\begin{figure}     
\includegraphics*[width = 7cm, height = 9.5cm]{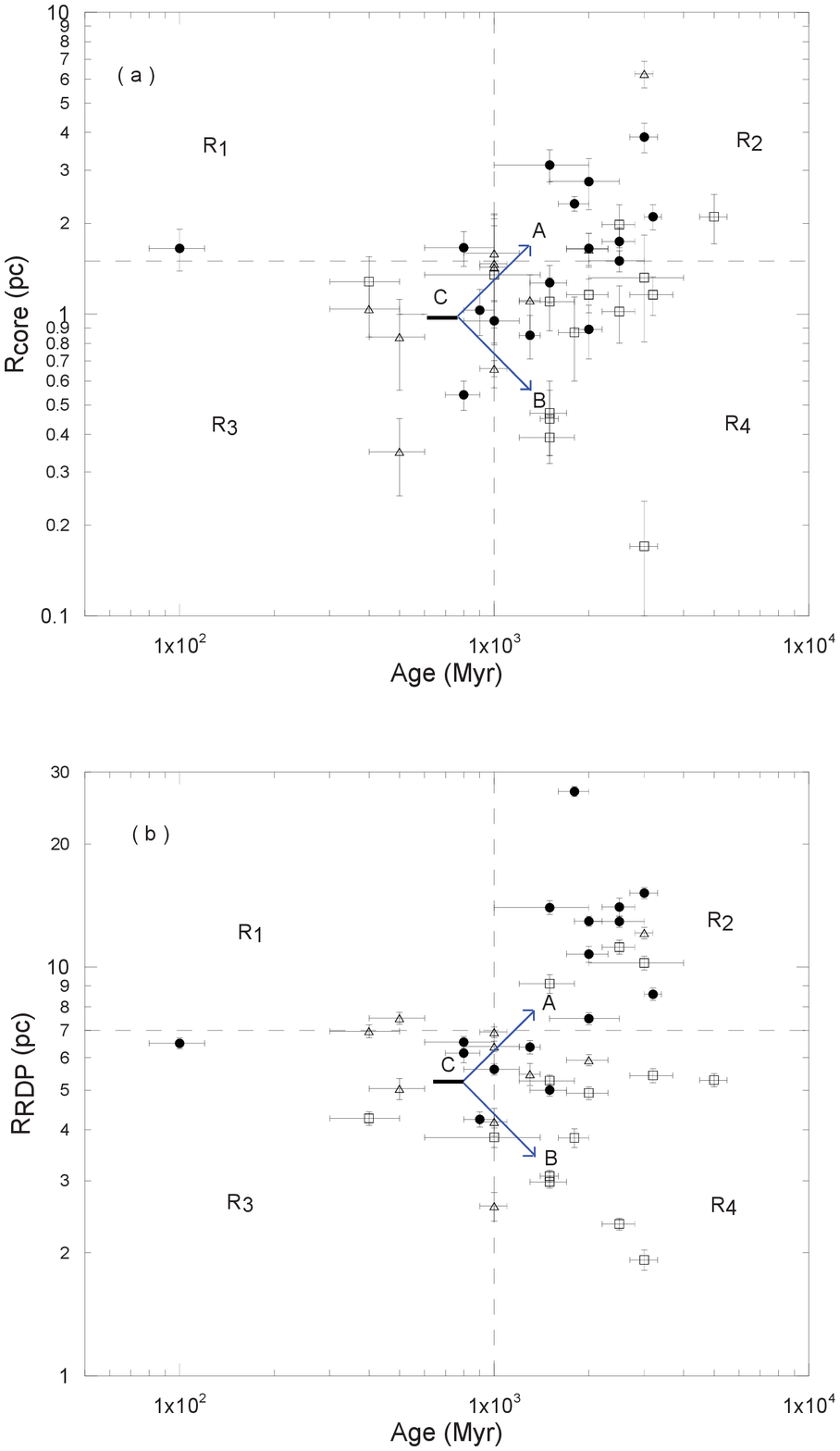}
\caption {Relations of Age--R$_{RDP}$ (panel a) and 
Age--R$_{core}$ (panel b), respectively. 
Filled circles and empty triangles show 
16 OCs with $m_{overall}\ge20~M_{\odot}$ and  10 OCs with $m_{overall}<20~M_{\odot}$, respectively.
14 OCs which have no mass determinations are marked by open squares. R1, R2, R3 and R4 mean the regions.} 
\end{figure} 

\cite{mac08} argue that some clusters show the expanded cores due to 
stellar mass black holes (hereafter BHs), and others contract due to 
dynamical relaxation and core collapse. To be able to see the effect 
of BHs in our core radius-age relation, the information of 
the OCs in regions of R2 and R4 in Fig.~17(a) is given in Tables~9$-$10.
We call the regions in Fig.~17 as R1, R2, R3 and R4.
N$_{bh}$ in Tables~9$-$10 means the estimated number of stellar mass 
black holes (BHs). This value is estimated from a relation N$_{bh}=6 
\times 10^{-4}N_{star}$, given by \cite{pm00}. Here, $N_{star}$ is the 
extrapolated number of stars in the OCs, and is given in Col.~9 of 
Table 6 for the overall regions of OCs. Because the extrapolated stellar number for NGC 2158 is 
not available (Col.~9 of Table 6; supplementary material), the number 
of BHs could be estimated from the relation of $N_{bh} \approx 0.002 \: 
M_{cluster}$, given by  \cite{pm00}. The BH numbers of seven OCs in the regions 
R2 and R4 in Fig.~17(a) cannot be estimated, because their extrapolated 
star numbers or overall  masses are not available (see Col.~9 of 
Table 6; supplementary material). 

\begin{table}   
\tiny
\renewcommand\thetable{9}
\centering
\tiny
\caption{Age, dimensions and mass (Cols.~2$-$5) for OCs, 
which show the core expansion in Fig.~17(a).
The number of black holes (N$_{bh}$) is listed in last column.}
\begin{tabular}{lccccc}
\hline
Cluster &Age &R$_{RDP}$&R$_{core}$ 
&\textit{m}$_{overall}$& N$_{bh}$ \\
& (Myr) & (pc) & (pc)& (100m$_{\odot})$ & \\
\hline
NGC 2112   & 2000  & 5.92  & 1.64  & 18.10  & 4 \\
NGC 2158   & 2500  & 14.03 & 1.74  & 33.40  & 7 \\
Trumpler 5 & 3000  & 15.18 & 3.86  & 223.0 & 45 \\
Col 110    & 3000  & 12.12 & 6.25  & 16.50  & 3 \\
NGC 2286   & 1000  & 6.39  & 1.59  & 11.10  & 2 \\
NGC 2506   & 2000  & 10.76 & 1.65  & 65.80  & 13 \\
Pismis 3   & 3200  & 8.58  & 2.10  & 133.00 & 27 \\
Trumpler 20& 1500  & 13.98 & 3.12  & 150.00 & 30 \\
NGC 6819   & 2500  & 12.92 & 1.50  & 49.10  & 10 \\
Be 89      & 2000  & 7.48  & 2.75  & 32.00 &  6 \\
NGC 7789   & 1800  & 26.88 & 2.32  & 194.00 & 39 \\
\hline
\end{tabular}
\end{table}

\begin{table}     
\tiny
\renewcommand\thetable{10}
\small
\centering
\tiny
\caption{Age, dimensions and mass (Cols.~2$-$5) for OCs, 
which show the core shrinkage in Fig.~17(a).
The number of black holes (N$_{bh}$) is listed in last column.}
\begin{tabular}{lccccc}
\hline
Cluster & Age & R$_{RDP}$&  R$_{core}$  
& \textit{m}$_{overall}$& N$_{bh}$ \\
& (Myr) & (pc) & (pc)& (100m$_{\odot})$ & \\
\hline
King 5     & 1000  & 5.62  & 0.95  & 30.00  & 6 \\
Koposov 53 & 1000  & 4.18  & 0.66  & 2.37  & 0 \\
NGC 2192   & 1300  & 5.47  & 1.11  & 2.27  & 0 \\
NGC 2243   & 2000  & 12.94 & 0.89  & 51.70  & 10 \\
NGC 2262   & 1300  & 6.37  & 0.85  & 24.10  & 5 \\
Haffner 8  & 1000  & 6.93  & 1.47  & 9.85   & 2 \\
Mel 71     & 1500  & 5.00  & 1.27  & 21.80  & 4 \\
Ru 96      & 1000  & 2.60  & 1.43  & 15.80  & 3 \\
\hline
\end{tabular}
\end{table}

\subsection{Relations of R$_{RDP}$  and  R$_{core}$ with R$_{GC}$}

The dependence of the structural parameters (R$_{RDP}$ \& R$_{core}$) 
with their galactocentric distance R$_{GC}$ of the 40 OCs and as a 
function of the ages are plotted in Figs.~18(a)$-$(b). The large and 
small-sized clusters in Fig.~18(a) occupy the inner- and the outer- 
Galactic radii. Two OCs with R$_{RDP}$ $ < $ 7 pc and  Age $ < $ 1 Gyr 
in Fig.~18(a) are locate inner Galactic radius. Such OCs with these 
sizes and ages are also seen in \citet[their fig.~3(c)]{sch06}. 
The relation between R$_{RDP}$ and R$_{GC}$ is the following, 
R$_{RDP}$$=(0.98\pm0.25)R_{GC}+(-3.07\pm2.03)$, with a 
correlation coefficient of 0.53 (see Fig.~18a), Our result shows that 
there is no strong dependence of R$_{RDP}$ on R$_{GC}$. However, 
\cite{lyn82}, \cite{tad02} and  \cite{cam09, cam10} 
mention a correlation from their OC samples.

\begin{figure}    
\centering
\includegraphics*[width = 7cm, height = 7cm]{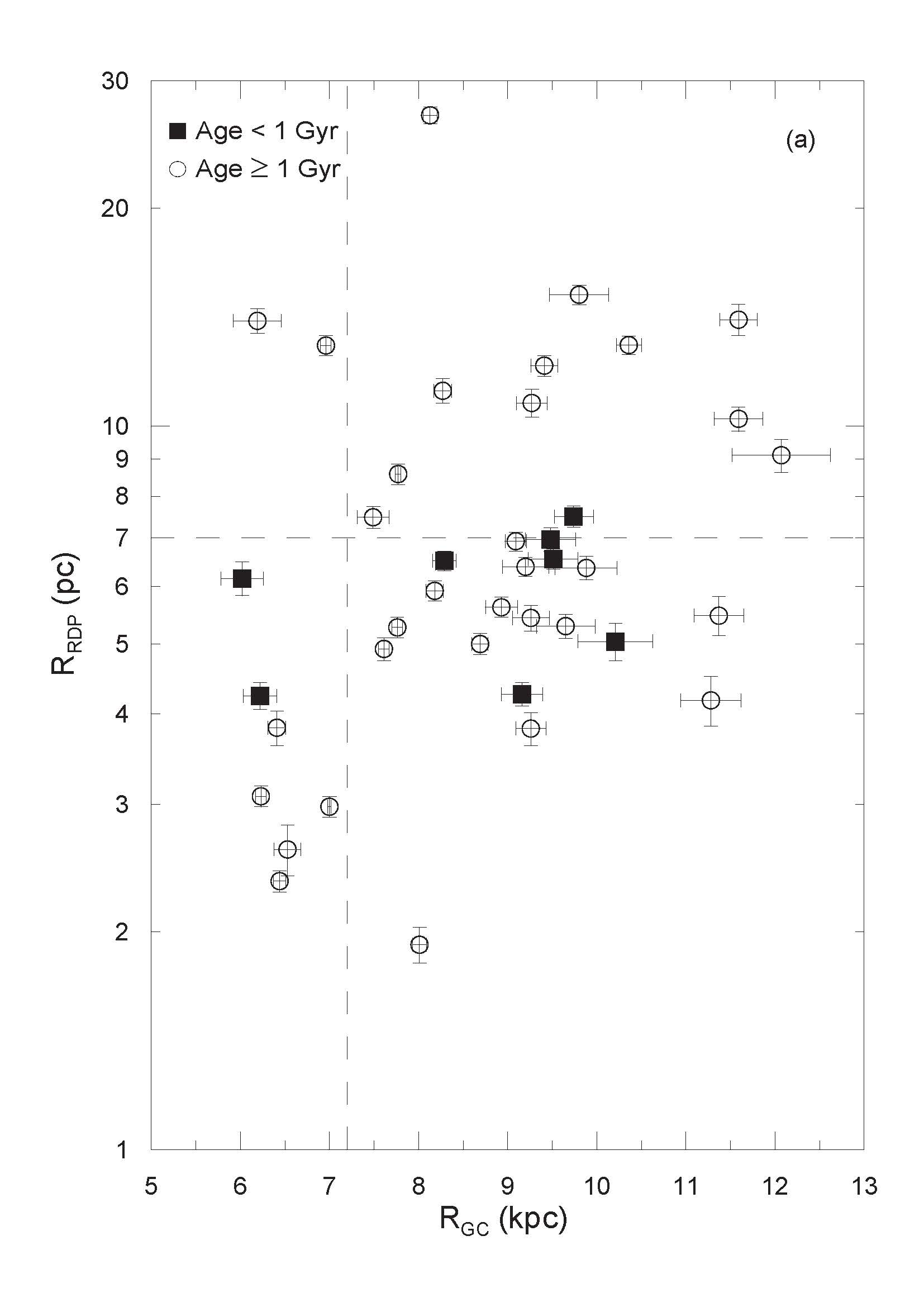}
\includegraphics*[width = 7cm, height = 7cm]{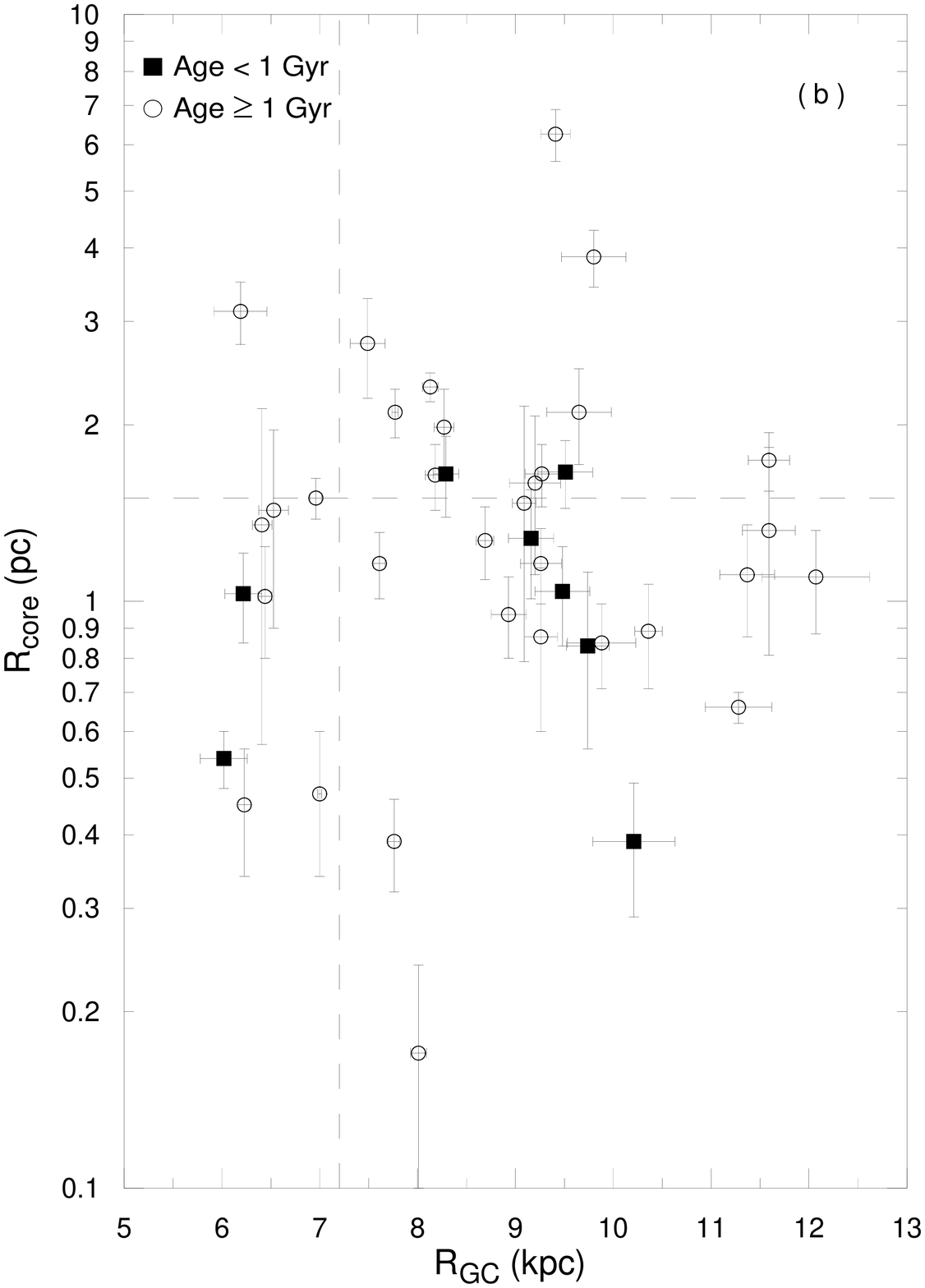}
\caption {Relations of R$_{GC}$--R$_{RDP}$ (panel a) and 
R$_{GC}$--R$_{core}$ (panel b), respectively. 
Filled squares and open circles denote the OCs with Age $<$ 1 Gyr 
and Age $\ge$ 1 Gyr, respectively.} 
\end{figure}

\subsection{Relations of m$_{overall}$ with R$_{RDP}$, R$_{core}$, Age and R$_{GC}$}

Figs.~19(a) and (b) show the relations of m$_{overall}$ versus R$_{RDP}$ 
and  m$_{overall}$ versus R$_{core}$ as a function of Age of 26  
of our 40 OCs. The relations to fit  m$_{overall}$ with R$_{RDP}$  
and with R$_{core}$ are $\ln m_{overall} = (1.57\pm0.42)\: \ln R_{RDP}+(0.01\pm0.02)$ 
(CC$=$0.60) and  $\ln m_{overall} = (1.14\pm0.37)\\ln R_{core}+(2.81\pm0.26)$ 
(CC$=$0.53), respectively. These correlations between size and mass of the clusters 
are in concordance with the mass-radius relation for massive OCs with 
Age $ > $ 100 Myr \citep{por10,cam10}. 
In Figs.~20(a) and (b) the relations of m$_{overall}$ with R$_{GC}$ and of 
m$_{overall}$ with Age of 26 of our 40 OCs are shown. 
As is seen from Fig.~20(a), 
massive and less massive OCs than m$_{overall} =2000\: 
M_{\odot}$ are located indistinctly in- or outwards of the Solar circle.

\begin{figure}       
\centering
\includegraphics*[width = 7cm, height = 7cm]{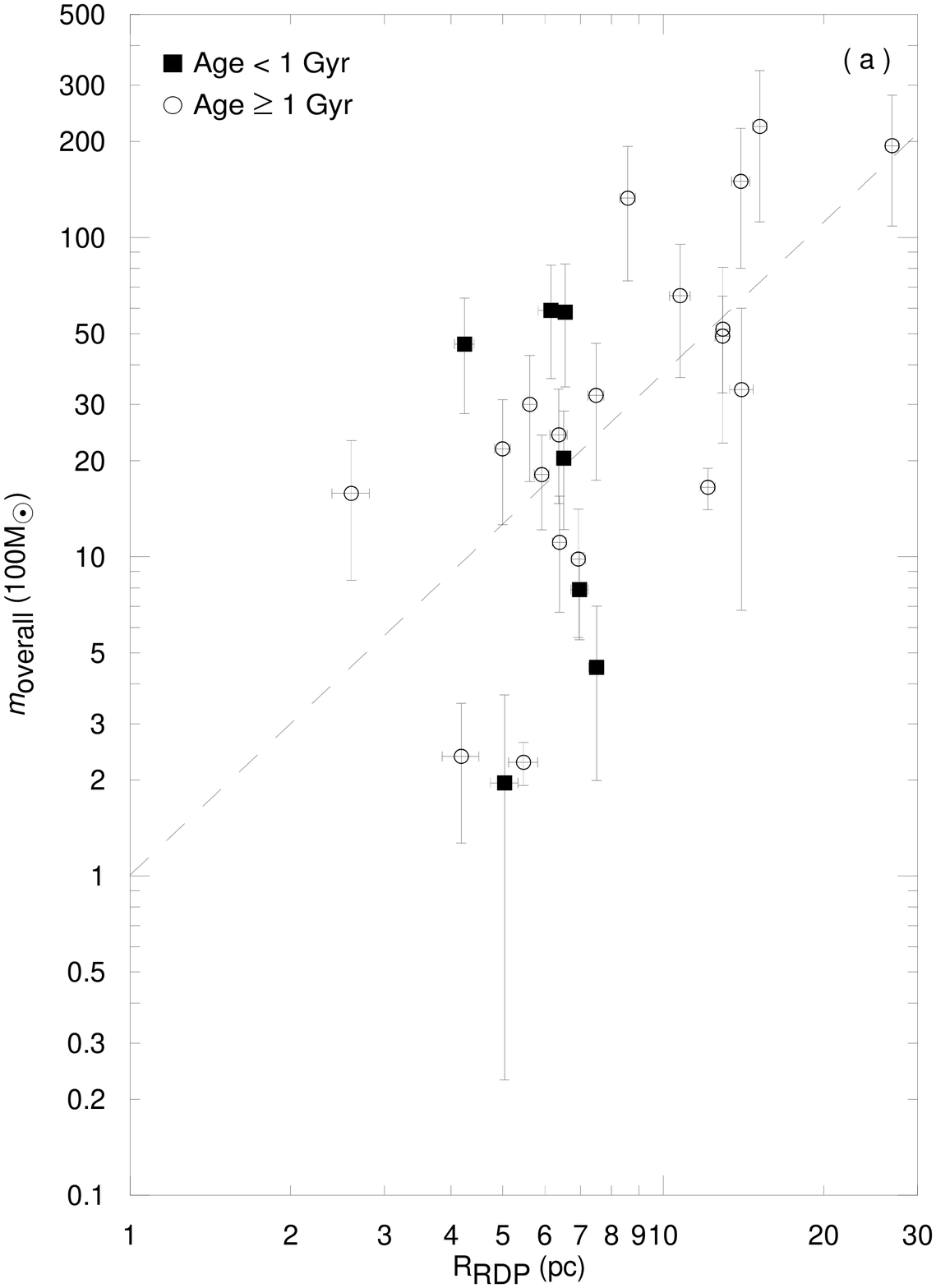}
\includegraphics*[width = 7cm, height = 7cm]{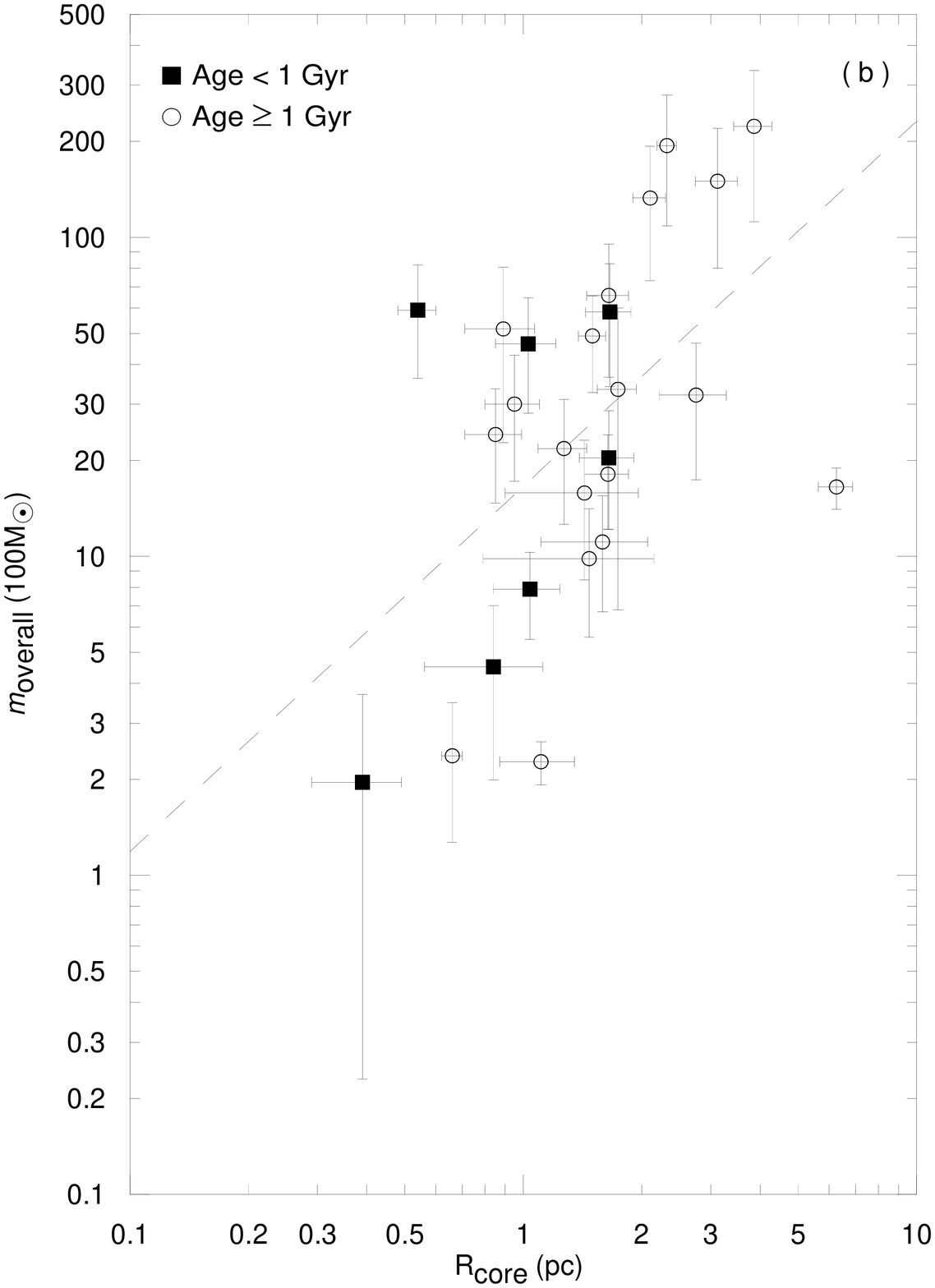}
\caption {Relations of R$_{RDP}$ - m$_{overall}$ 
(panel a) and R$_{core}$ - m$_{overall}$(panel b) of 26 OCs. 
Filled squares and open circles represent the OCs with Age $<$ 1 Gyr 
and Age $\ge$ 1 Gyr. Dashed lines denote the best fits.}
\end{figure}

\begin{figure}      
\centering
\includegraphics*[width = 7cm, height = 7cm]{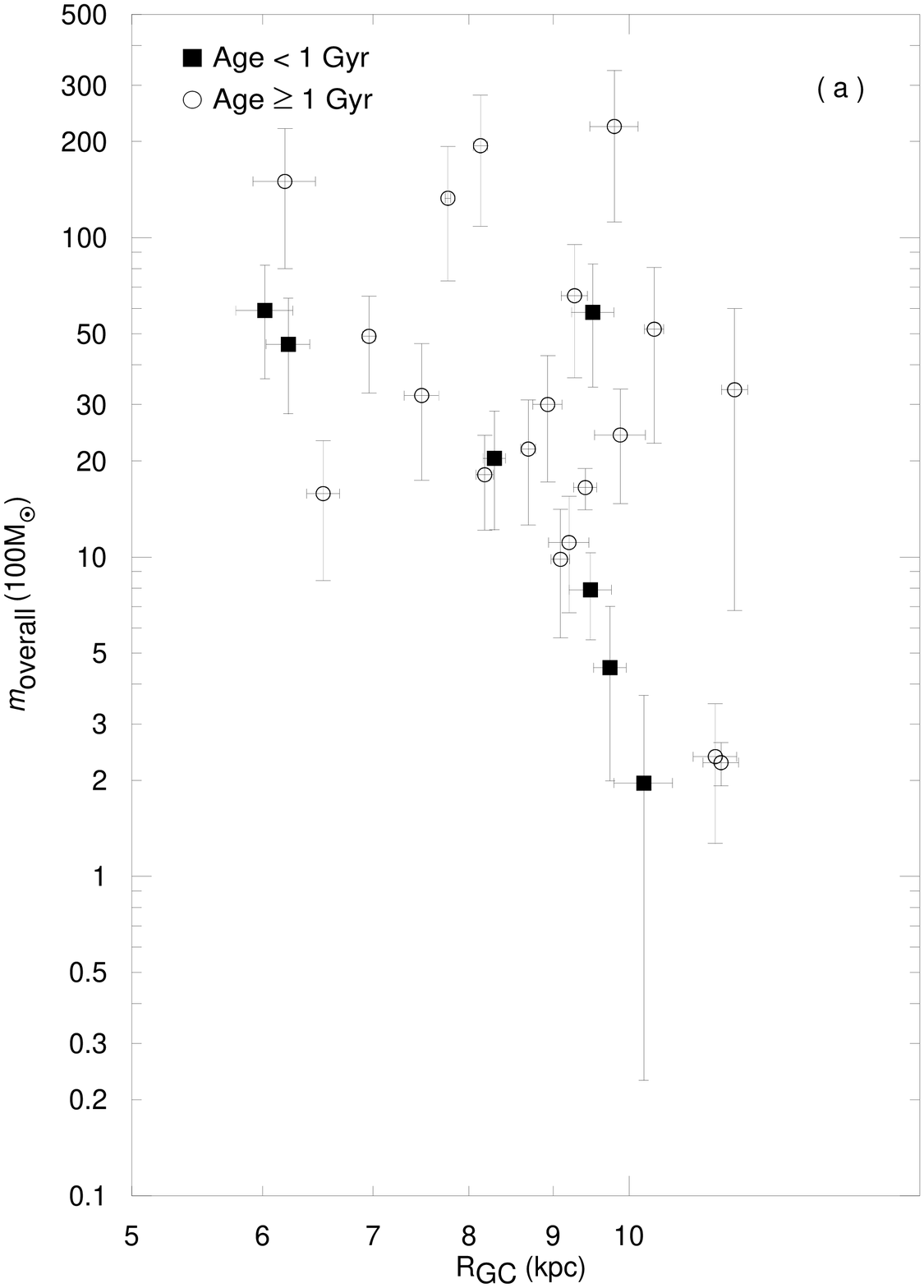}
\includegraphics*[width = 7cm, height = 7cm]{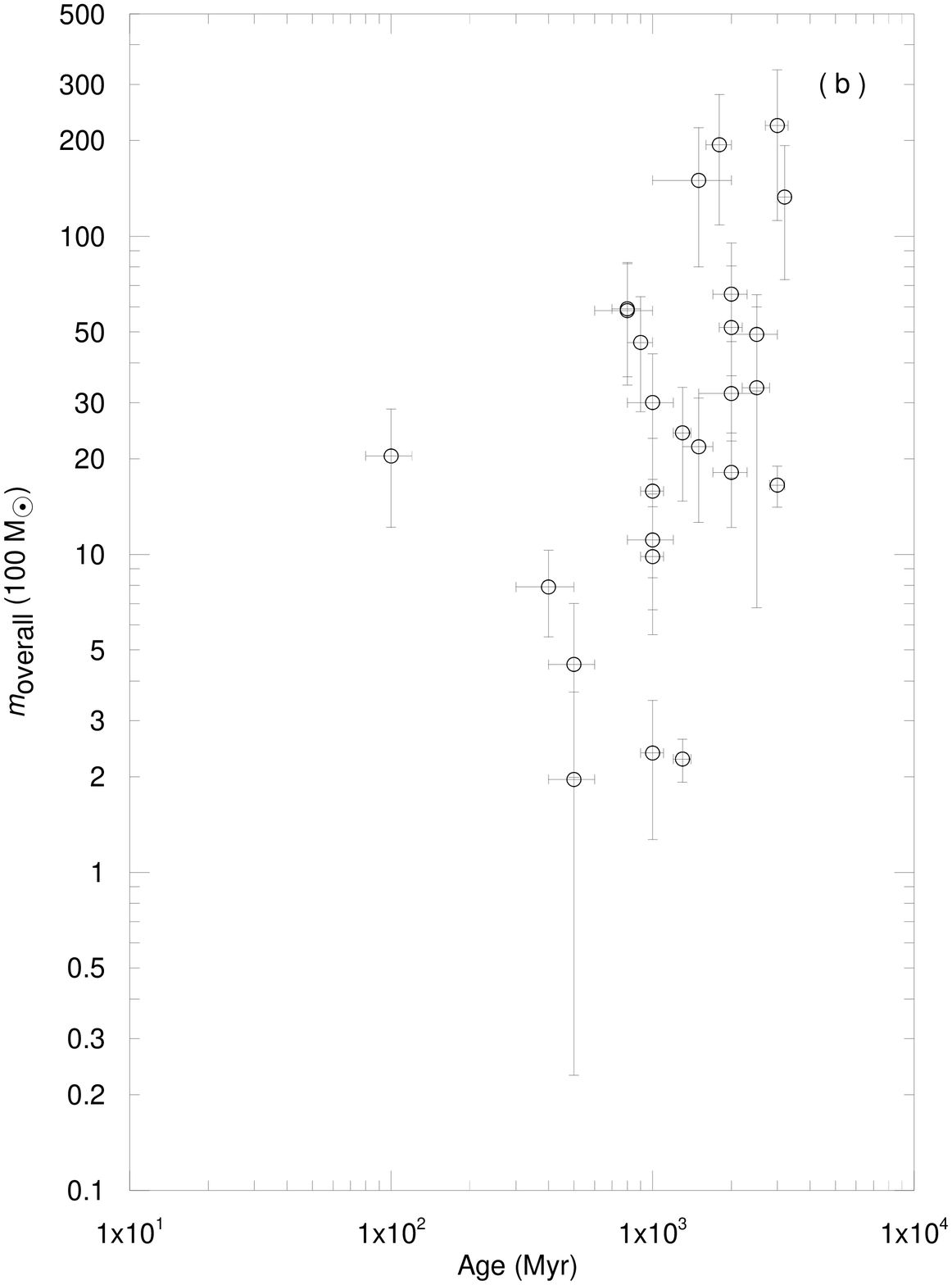}
\caption {Relations of R$_{GC}$ - m$_{overall}$ (panel a) 
and Age - m$_{overall}$ (panel b) of 26 OCs.}
\end{figure}

\subsection{Relations between MF slopes, Age, R$_{RDP}$, R$_{GC}$, and the mass density}

The relation of $\chi_{overall}$ with $\chi_{core}$ of 29 OCs is 
presented in Fig.~21. The fit which is applied to the data is given as 
following, $\chi_{overall}=(0.47\pm0.12)\chi_{core}+(1.10\pm0.26)$, 
with a moderate CC$=$0.60. The OCs with flat/steep 
positive overall MF slopes for $\chi_{core}<0$ in Fig.~21 show signs of 
a mild to large scale mass segregation, whereas the OCs with negative 
overall MF slopes for ~$\chi_{core} < 0$ ~indicate an advanced dynamical 
evolution. These MF slopes of ~$\chi_{core}<0$ ~in Fig.~21 can be explained 
by the external dynamical effects such as tidal stripping by tidal 
interactions (in the form of shocks) due to disc and bulge 
crossings, as well as encounters with GMCs.

\begin{figure}     
\centering
\includegraphics*[width = 7cm, height = 7cm]{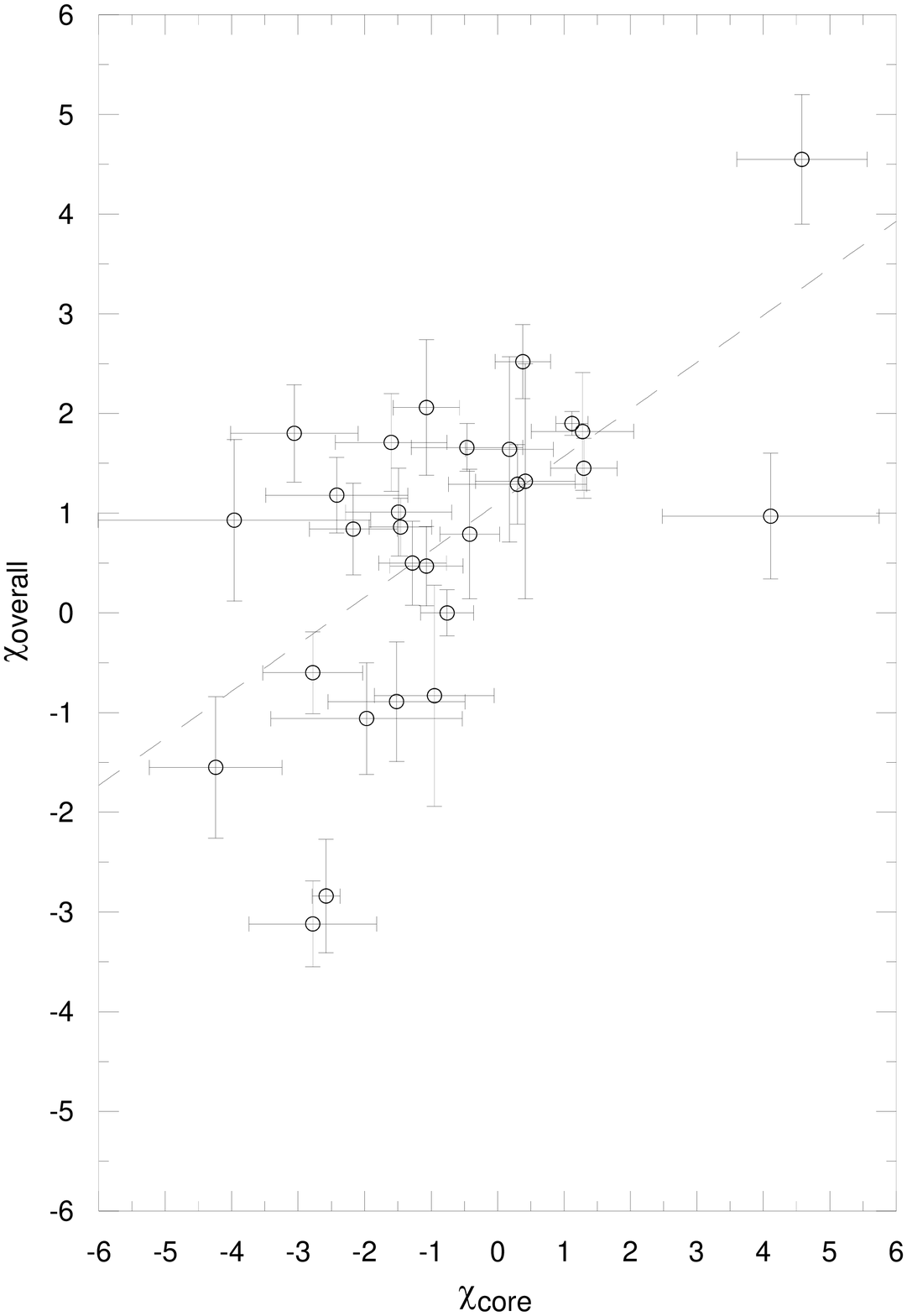}
\caption {Relation of $\chi_{core}$ - $\chi_{overall}$ of 29 OCs. 
Dashed line shows the best fit.}  
\end{figure}

The relations of Age vs. $\chi_{overall}$ of 31 OCs and Age vs. $\chi_{core}$ 
of 29 OCs of our sample are displayed in Figs.~22(a) and (b).  
The age dependence of the overall and core MF slopes 
has been parameterised by the linear-decay function 
(shown as dashed curve) $\chi (t) = \chi_\circ - t/t_{f}$, 
where  $\chi_\circ$ represents the MF slope in the early phases and 
$t_{f}$ is  the flattening time scale. For the overall MF we derive 
$\chi_\circ = 1.68\pm0.30$ and  $t_{f}=1569\pm600$~Myr (CC$=$0.44); 
the core values $\chi_\circ = 0.74\pm0.39$ and  
$t_{f}=1006\pm206$~Myr (CC$=$0.68). Within the 
expected uncertainties the overall MF values are quite close to 
$\chi_\circ=1.30\pm0.30$ of \cite{kro01} and $\chi_\circ=1.35$ of  
\cite{sal55}. 

\begin{figure}     
\centering
\includegraphics*[width = 7cm, height = 7cm]{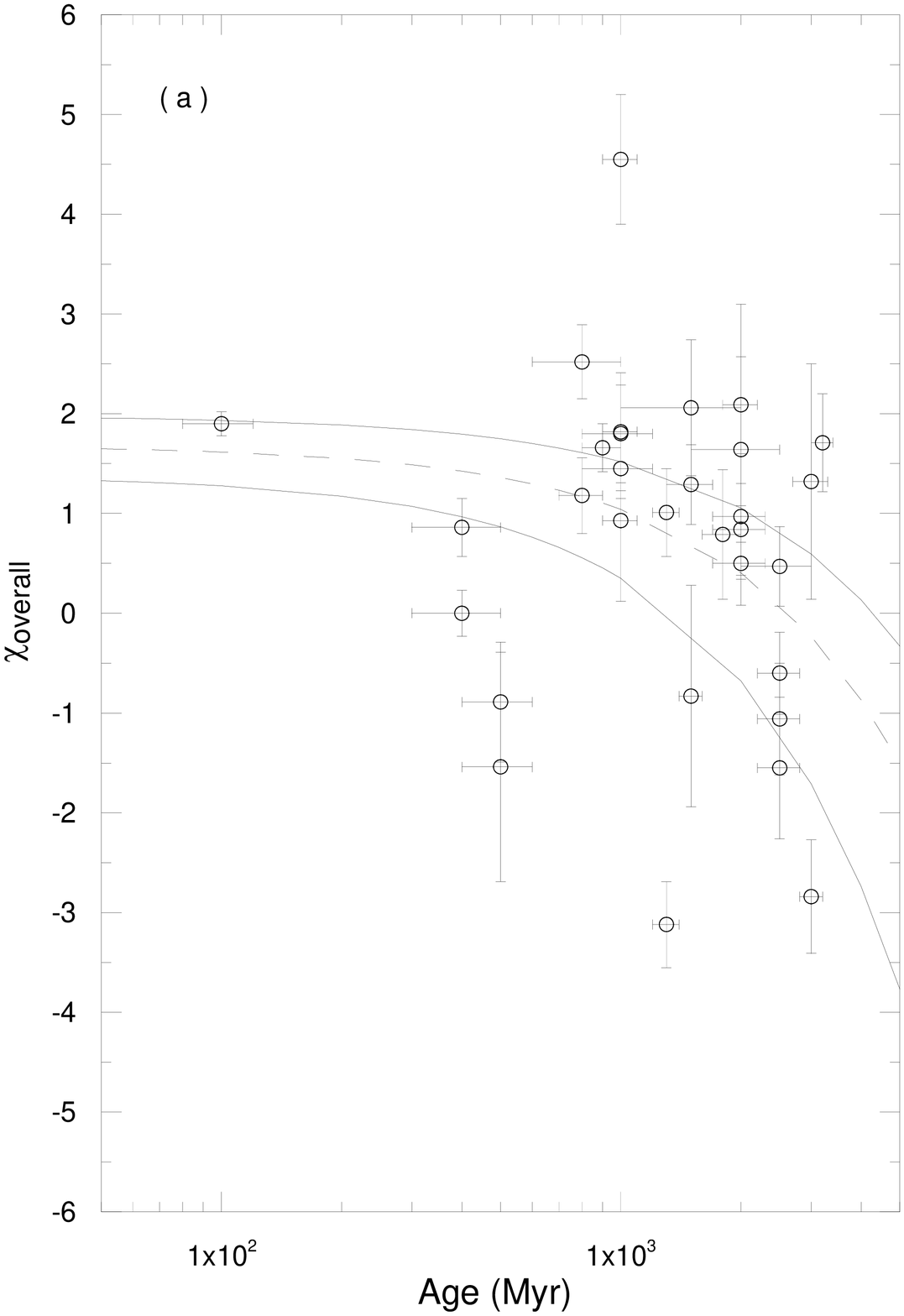}
\includegraphics*[width = 7cm, height = 7cm]{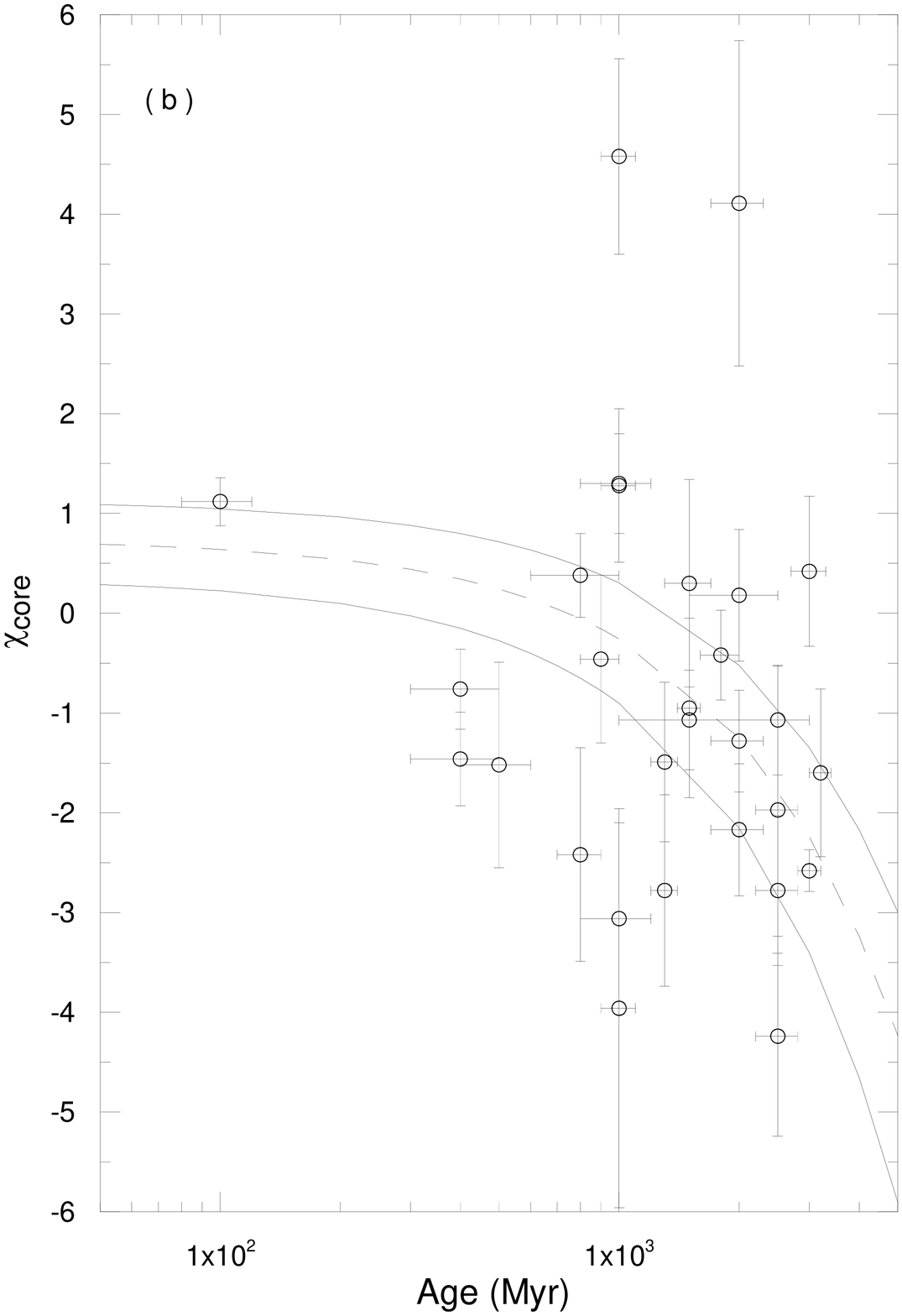}
\caption {Relations of Age - $\chi_{overall}$ (31 OCs,~panel a) and 
Age - $\chi_{core}$ (29 OCs,~ panel b). Dashed and solid lines show
the best fit and $1\sigma$ uncertainty, respectively.} 
\end{figure}

The relations of m$_{overall}$ with the slope $\chi_{overall}$ of 26 
OCs and  m$_{core}$ with $\chi_{core}$ of 24 OCs have been presented 
in Figs.~23(a) and (b).  In Fig.~23(a), most of the OCs of 
with positive overall slopes are mass-rich and present little or no 
signs of mass segregation. 
For the relation of m$_{core}$ with $\chi_{core}$ dispayed in Fig.~23(b), 
most of OCs with m$_{core}<1000\: m_{\odot}$ have negative core MF 
slopes implicating  mass segregation effects at a larger scale. 

\begin{figure}     
\centering
\includegraphics*[width = 7cm, height = 7cm]{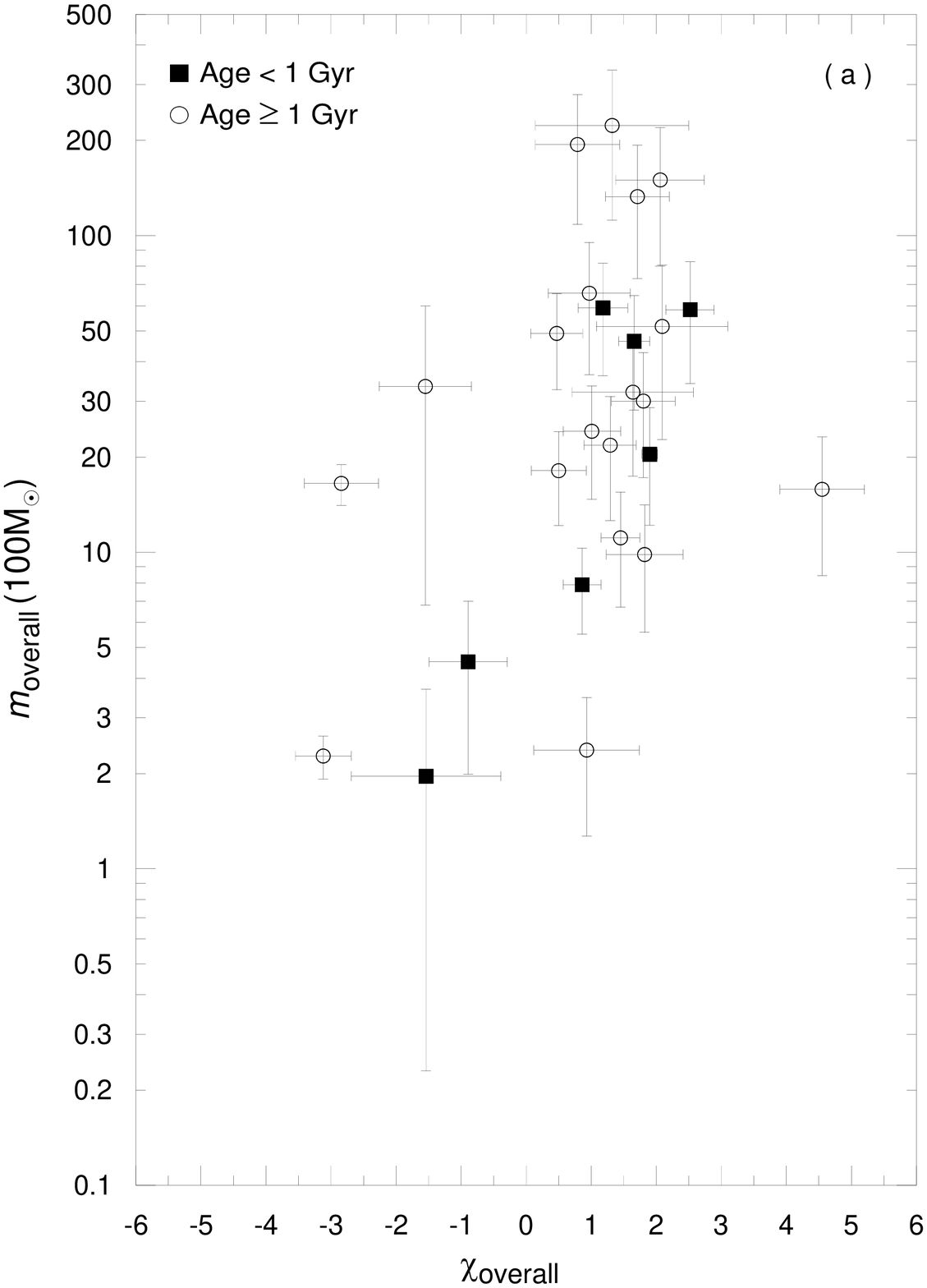}
\includegraphics*[width = 7cm, height = 7cm]{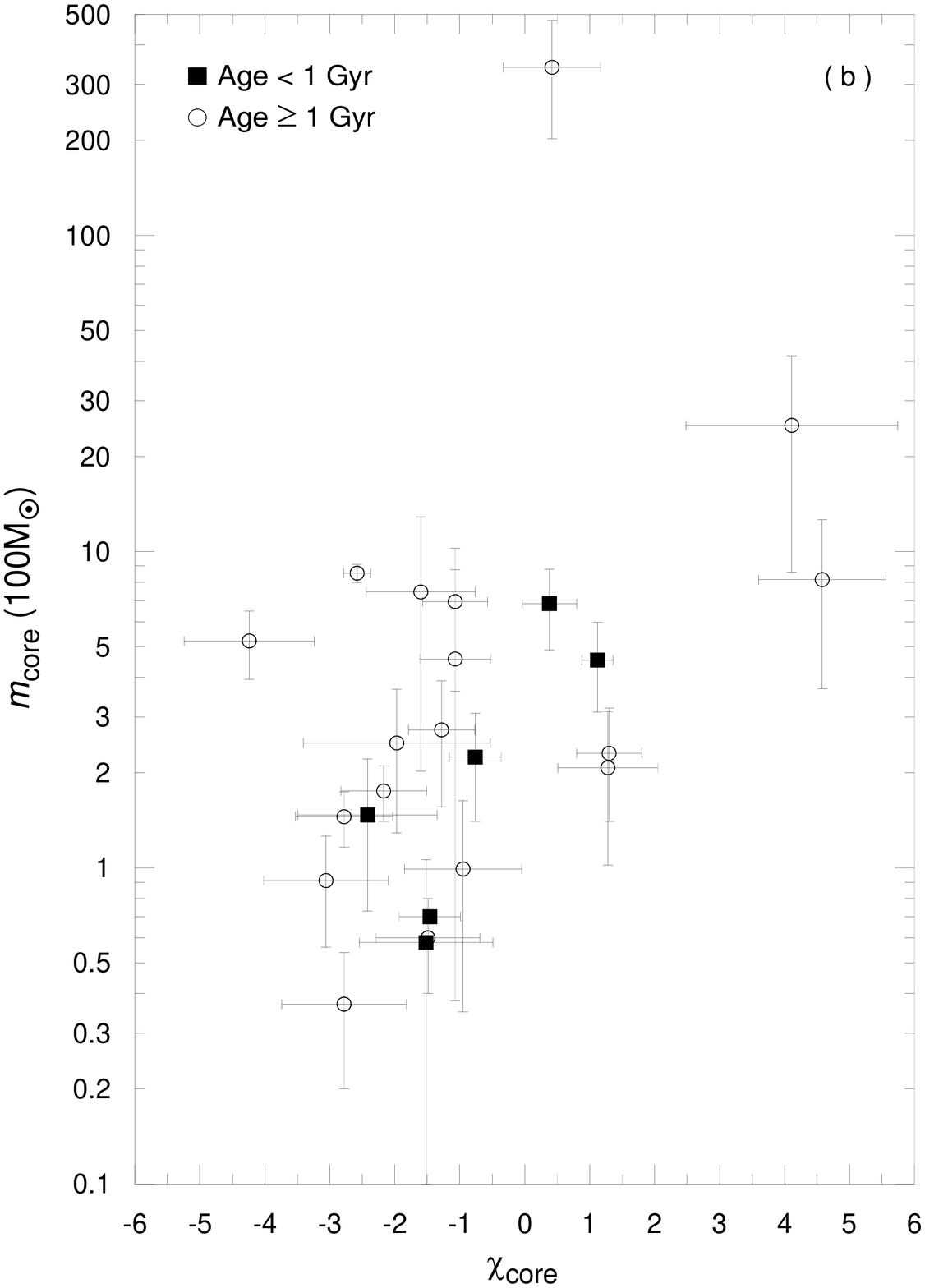}
\caption {Relations of m$_{overall}$ - $\chi_{overall}$ 
(panel a) of 26 OCs and m$_{core}$ - $\chi_{core}$ (panel b) of 24 OCs, respectively.} 
\end{figure}

In the relations of R$_{RDP}$ with $\chi_{overall}$ and R$_{RDP}$ with 
$\chi_{core}$ for 31 and 29 OCs of our sample, respectively,  given in 
Figs.~24(a) and (b), the OCs with larger or smaller dimensions than R$_{RDP} = 
7\:pc$  have a positive or negative sloped overall- and core-MFs, 
repectively.

\begin{figure}     
\centering
\includegraphics*[width = 7cm, height = 7cm]{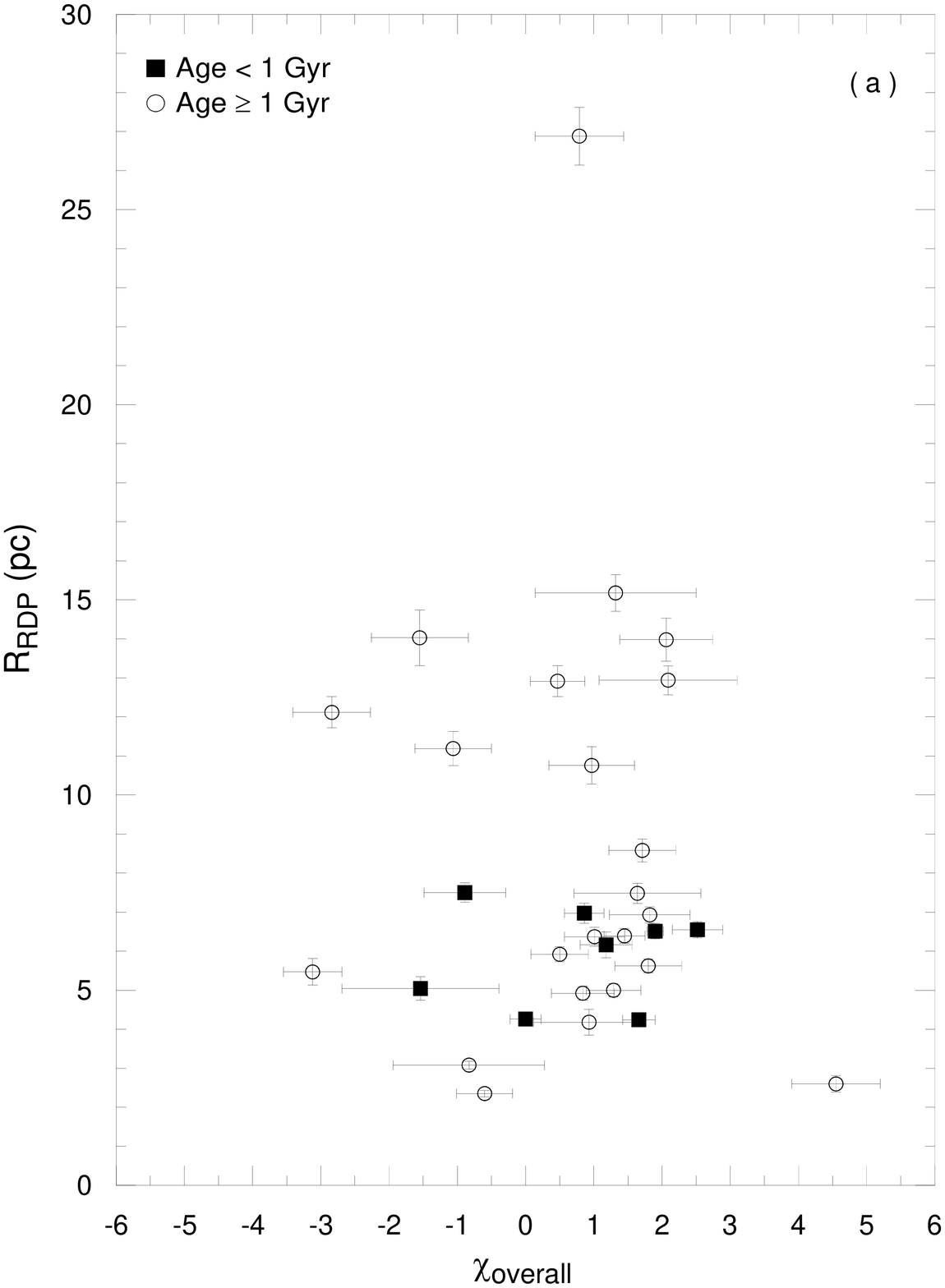}
\includegraphics*[width = 7cm, height = 7cm]{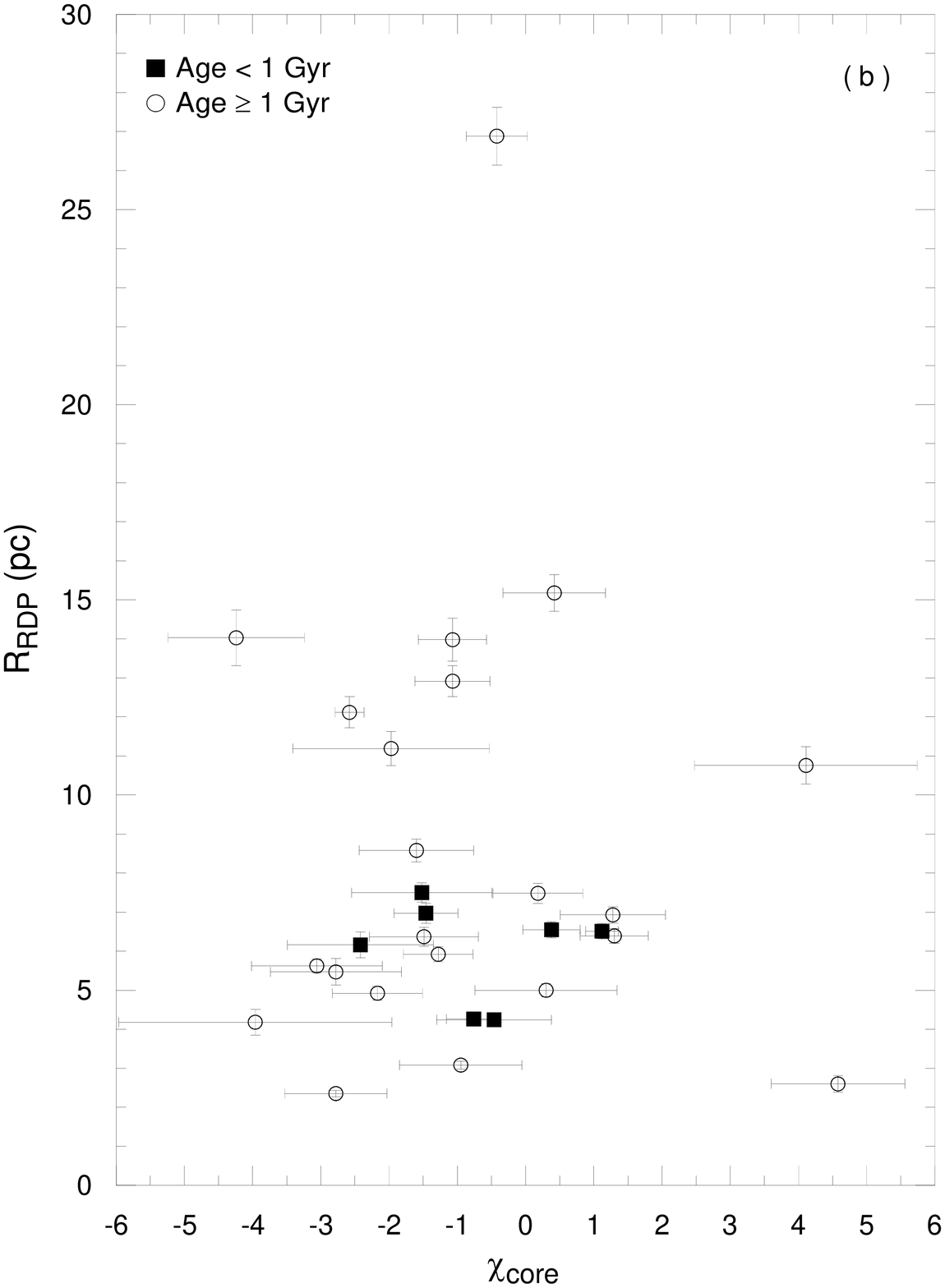}
\caption {Relations of $\chi_{overall}$ - R$_{RDP}$ 
(panel a) of 31 OCs and $\chi_{core}$ - R$_{RDP}$ (panel b) of 29 OCs. 
Filled squares and open circles denote clusters with Age $<$ 1 Gyr 
and Age $\ge$ 1Gyr, respectively.} 
\end{figure}

From the relations between R$_{GC}$ and $\chi_{overall}$ of 31 OCs, and  
between R$_{GC}$ and $\chi_{core}$ of 29 OCs shown in Figs.~25(a) and (b), 
apparently MF slopes are not correlated with R$_{GC}$ or Age. 

\begin{figure}     
\centering
\includegraphics*[width = 7cm, height = 7cm]{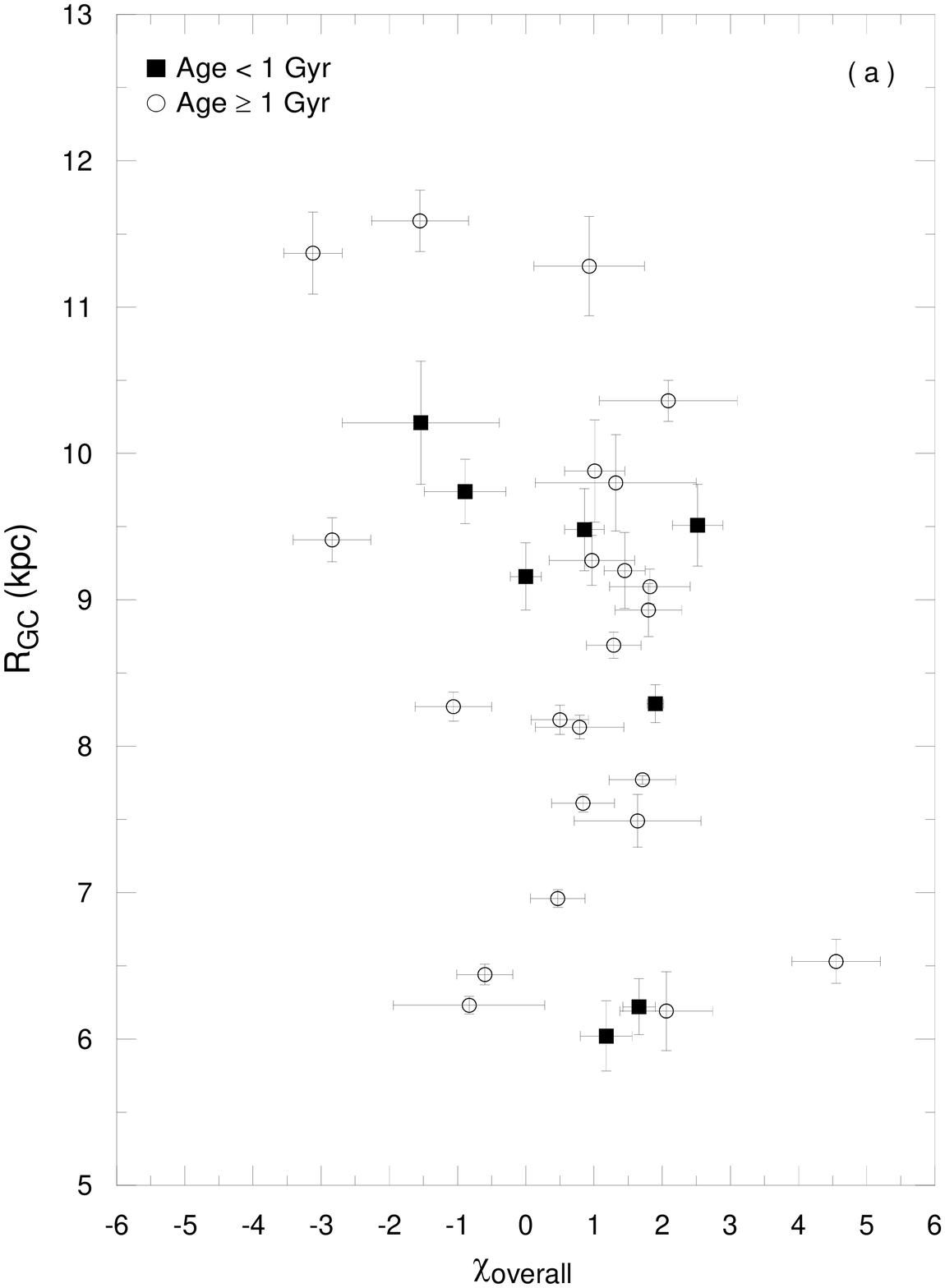}
\includegraphics*[width = 7cm, height = 7cm]{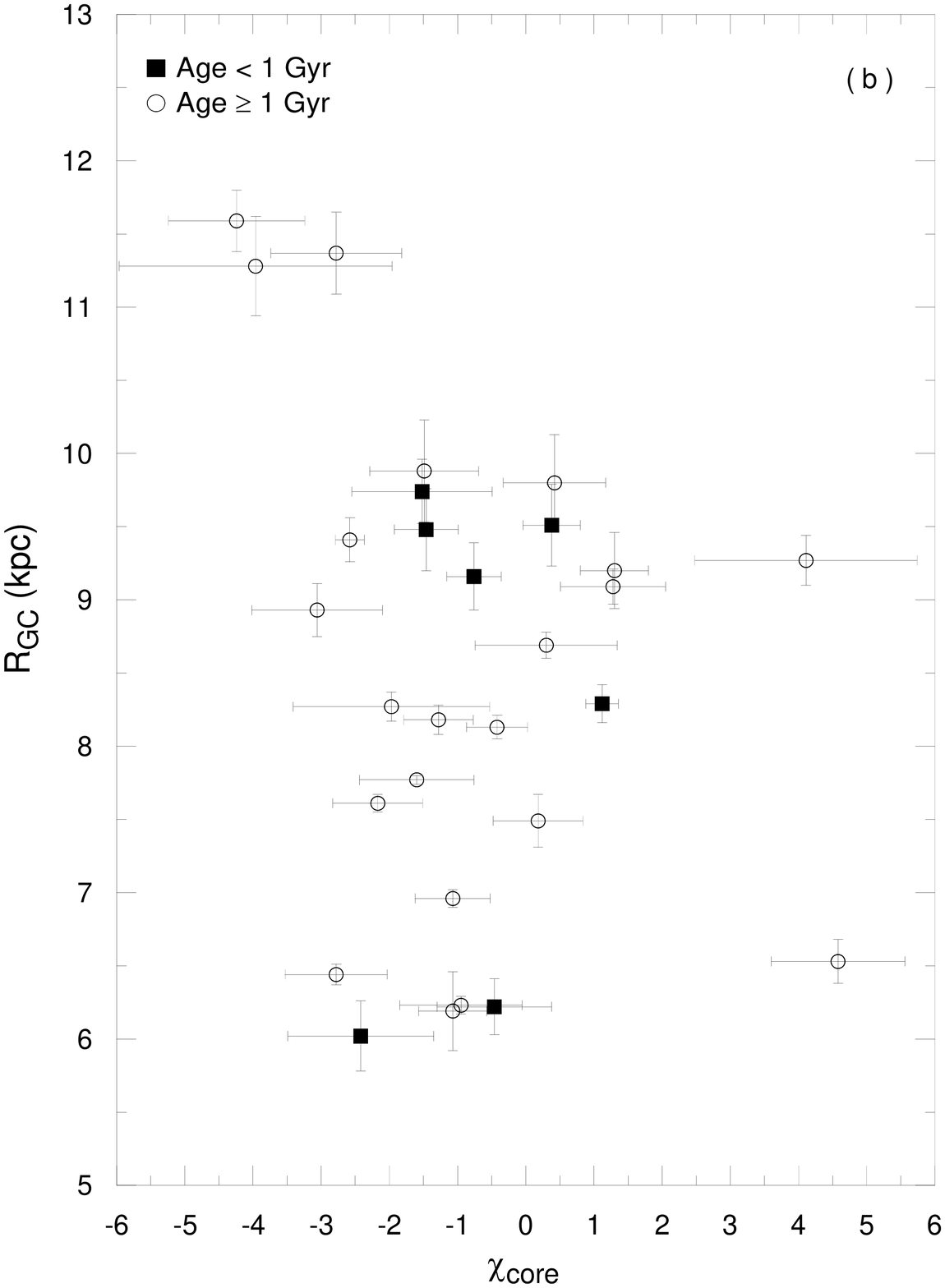}
\caption {Relations of $\chi_{overall}$ - R$_{GC}$ (panel a) of 31 OCs 
and $\chi_{core}$ - R$_{GC}$ (panel b) of 29 OCs, respectively.}
\end{figure}

The cluster mass density $\rho(m_\odot\: pc^{-3})$ is plotted  in 
Fig.~26(a) and (b) as a function of $\chi_{overall}$ for 26 OCs and 
$\chi_{core}$ for 24 OCs, respectively. In panel~(a) the mass 
densities of the OCs having $\chi_{overall} < 0$ are low, as 
compared to the ones of the OCs with $\chi_{overall} > 0$. This 
indicates that low mass stars of OCs with negative MF slopes are 
significantly lost due to external dynamical processes. From 
panel~(b) one can see that $\chi_{core}$ and $\rho_{core}$ are not 
correlated. 

\begin{figure}   
\centering
\includegraphics*[width = 7cm, height = 7cm]{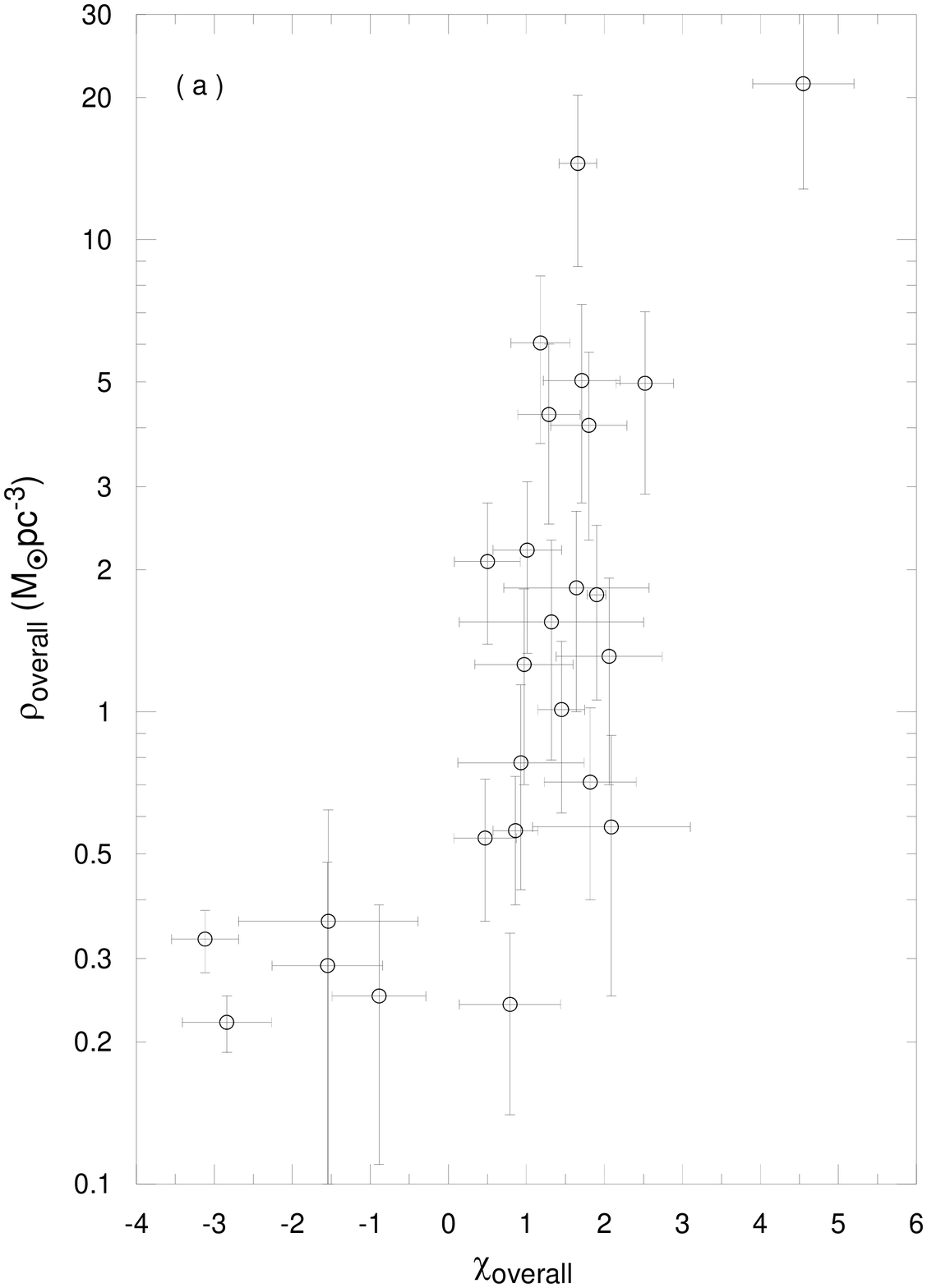}
\includegraphics*[width = 7cm, height = 7cm]{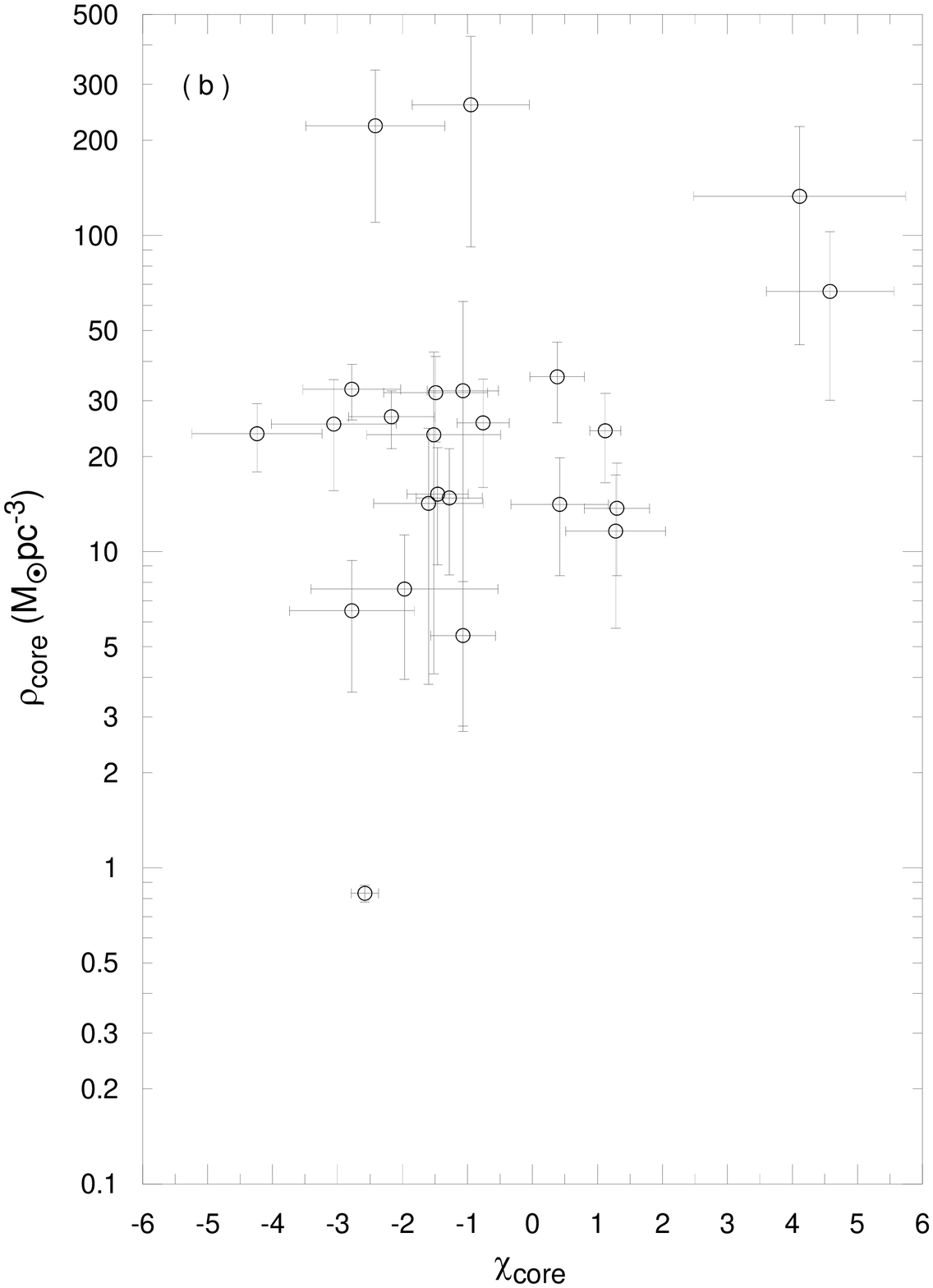}
\caption {Relations of $\chi_{overall}$ - $\rho_{overall}$ (panel a) of 26 OCs
and $\chi_{core}$ - $\rho_{core}$ (panel b) of 24 OCs, respectively.} 
\end{figure}

\subsection{Relation between the MF slope and the evolutionary 
parameter and a comparison to Kroupa's IMF}

From MF slopes and evolutionary parameters of the overall and the 
core of the OCs given in Table~7, the relations of $\tau_{overall}$ 
with  $ \chi_{overall} $ of 24 OCs and similarly, of $\tau_{core}$ 
with $\chi_{core}$ of 21 OCs have been plotted in Figs.~27(a)$-$(b). 
The dashed curve on the figure shows the fit, $\chi (\tau) = \chi_\circ 
- \chi_{1}e^{-(\frac{\tau_{o}}{\tau})}$. As seen in panels~ (a) and 
(b) of Figure~27, the overall- and core-MF slopes undergo an exponential 
decay with $\tau$. Here, $\chi_\circ$ and $\chi_{1}$ mean MF slopes at 
birth and in the advanced stage, respectively. For the overall MF slope, 
we derive $\chi_\circ = 1.67\pm 0.18$ and $\tau_{o}=29.92\pm 12.29$ 
(CC$=$0.77); the core values $\chi_\circ = 
1.19\pm 0.89$ and  $\tau_{o} = 31.62\pm 34.79$ (CC$=$0.64). 
Similar relations were obtained by  \citet[see their Fig.~8(a) and (b)]{bic06a} 
and  \citet[see their Figs.~7(b), (d), (e)]{mn07}.

\begin{figure}   
\centering
\includegraphics*[width = 7cm, height = 7cm]{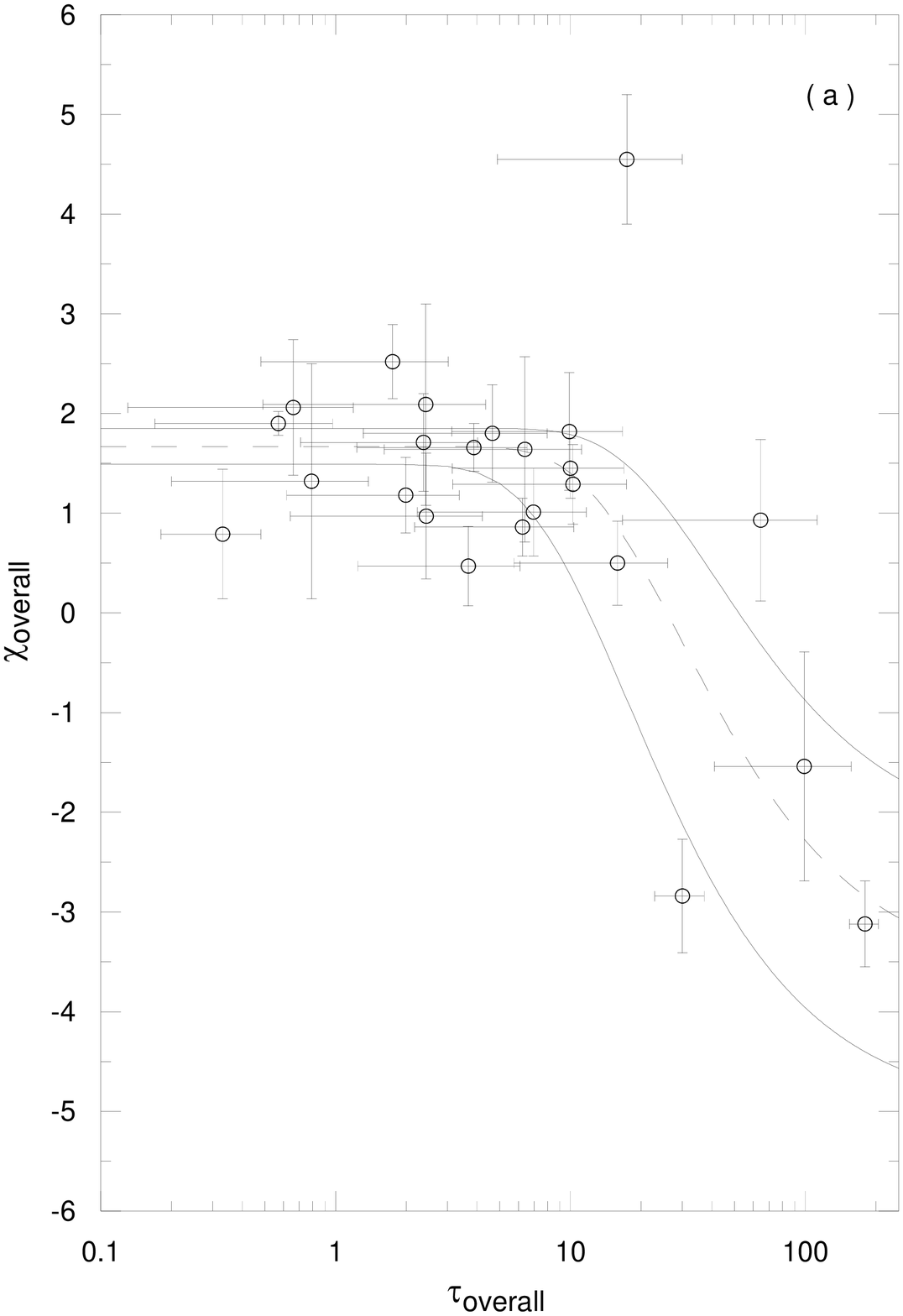}
\includegraphics*[width = 7cm, height = 7cm]{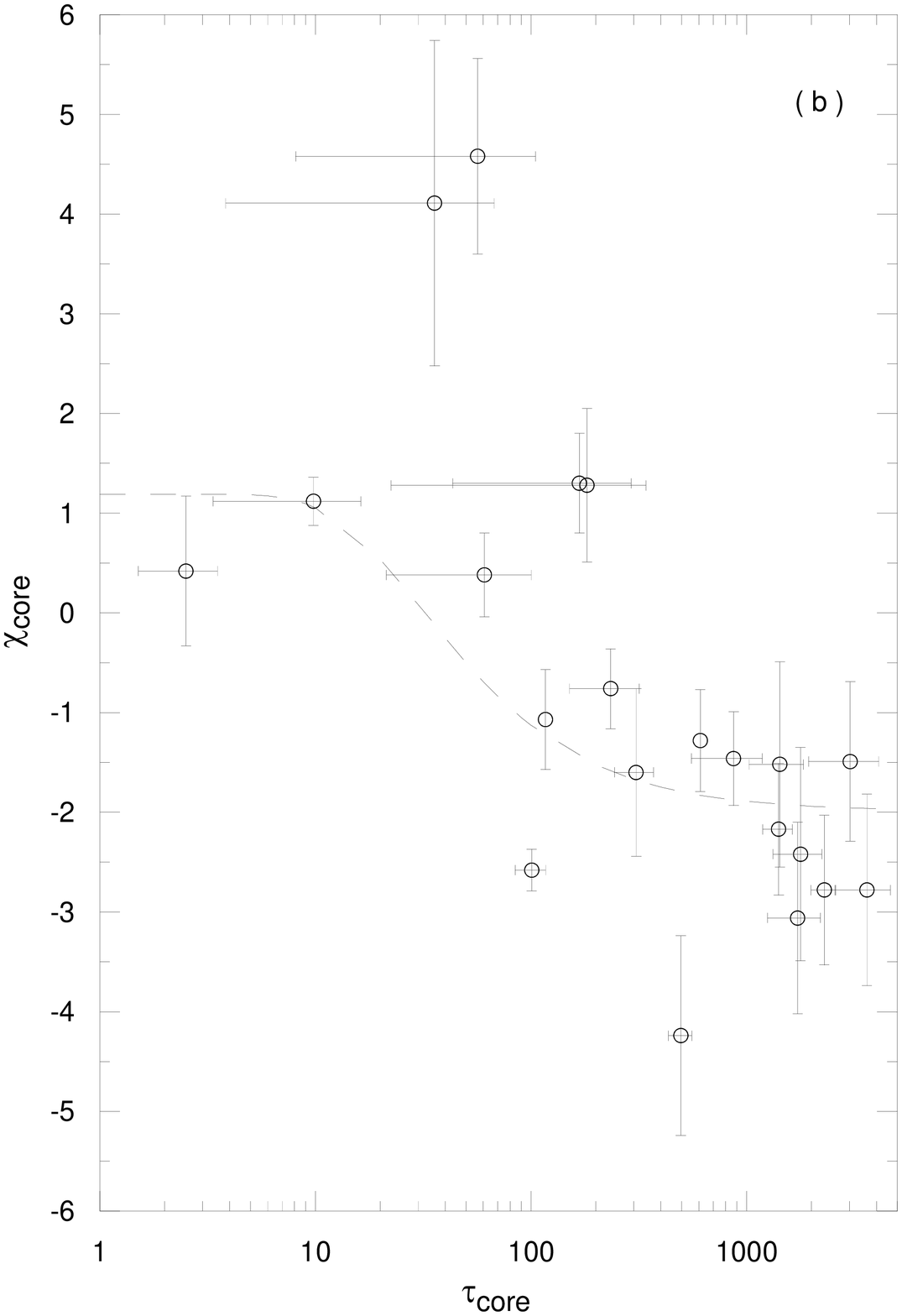}
\caption {Relations of $\tau_{overall}$ - $\chi_{overall}$ (panel a) of 24 OCs and 
$\tau_{core}$ - $\chi_{core}$ (panel b) of 21 OCs. Dashed and solid lines show the best fit
and $1\sigma$ uncertainty, respectively. In panel~(b), $1\sigma$ uncertainty is 
not shown due to too large error of $\tau_{core}$.}
\end{figure}

\section{Conclusions}

Our main conclusions are summarized as follows:

\begin{enumerate}
  
\item The astrophysical and structural parameteres of 40 OCs  
have been derived from the filtered 2MASS $(J, J-H)$ CMDs, and
the stellar RDPs. The field star decontamination technique is utilised 
for separating the cluster members. The astrophysical parameters (Age,~d,~E(B-V)) 
of 40 OCs comprise with ages in the range of 0.1 Gyr  to 5.0 Gyr, 
at heliocentric distances, 0.85 kpc to 5.42 kpc, and  with reddenings, $0.03 \leq E(B-V) 
\leq 1.31$ (Table 3). 
Having combined the derived structural, mass and mass functions, relaxation and evolutionary 
parameters with astrophysical parameters of 40 OCs, dynamical evolution of these OCs have been studied.
The reduced final reddenings from the dust maps of SFD  have been compared with the ones of 40 
OCs (Fig.~9(b)). There are significant differences for 27 OCs between both  $E(B$--$V)$ color excess 
values. For the rest, the $E(B$--$V)$ values of 13 OCs are quite close to the ones of 
SFD.  Note that  SFD maps are not reliable at regions  $|b|<5^{\circ}$ due to 
contaminating sources and uncertainties in the dust temperatures \citep{gon12}. 
Therefore, the SFD values resulted from line-of-sight integral through the Milky Way and 
with low spatial resolution, it is quite a normal to have different reddening values for 
these relatively close ($\sim 1$~kpc) star clusters.

\item The relation between R$_{RDP}$ and R$_{core}$ in Fig.~14 found by us is reasonably 
similar to that given by \cite{cam10}. The OCs in our sample which do not follow the relation
are either intrinsically small or have been suffering significant evaporation effects.
The dimensions  (R$_{RDP}$ and R$_{core}$) in Figs.~15(a) and (b) increase on 
the average with the distance from the Sun.

\item From Fig.~17(a) and Tables 9-10, apparently the sizes of core radii 
of the OCs are related with their respective BH numbers. The black 
hole numbers of the OCs in region R2 of Fig.~17(a) are 
generally larger than the ones of the OCs in the region of IV. Note 
that the BH numbers of six, out of 10 OCs in Table 10 are almost 
close to the ones of the OCs in Table 9. However, with almost similar BH numbers, these six OCs show 
shrinkage, whereas those in Table 9 indicate the expanded cores. For 
example, If the statement of \cite{mac08} is correct, NGC~2243 with 
its 10 BHs would develope a large core. However, NGC~2243 has small 
core value, R$_{core}$ = 0.89 pc. Col.~110 with 
a few BHs shows an expanding core with R$_{core}$ = 6.25 pc. The core 
sizes of NGC~7789 (R$_{core}$ = 2.32 pc) and Be~89 (R$_{core}$ = 2.75 
pc), respectively are quite close together. But their BH numbers of 
the two OCs are very  different. Therefore, the presence of BHs is not the only possible
explanation for the bifurcation seen in Figs.~17(a) and (b).
Alternatively, one should also consider the effect of the mass 
range of OCs. In other words, for clusters older than ~1 Gyr, 
Fig. 17 shows that massive OCs (filled circles) can be found in 
both regions R2 and R4. There are two low-mass OCs (open triangles) 
in the region R4, (see Fig. 17(a)). In this sense, the distribution
of OCs in Fig. 17 can be partly attributed to clusters with large radii 
retaining larger masses.

\cite{mac03, mac08} also argue that the expanded cores are the cause of  
growth of the limiting radii and the shrinking cores lead to the 
contraction of the limiting radii. There are 32 OCs of our sample in the 
regions R2 and R4 in Fig.~17(a); 16 out of 19 OCs with R$_{core} < 1.5$ 
pc in the region R4 of Fig.~17(a) have R$_{RDP} < 7$ pc; three of 19 OCs 
have R$_{RDP} > 7$ pc.  Similarly, 10 out of 13 OCs in the region R2 of 
Fig.~17(a) with R$_{core} > 1.5$ pc  have R$_{RDP} > 7$ pc; three of them  
have R$_{RDP} < 7$ pc. Here, R$_{RDP} = 7$ pc means the separation into 
two groups of our sample (Fig.11a). These findings imply that the OCs 
with their core expanding could have small cluster limiting radii, in a 
similar manner, the OCs with shrinking cores could have large limiting 
cluster radii. Note that there are six OCs with incompatible cores and 
limiting radii in the regions R2 and R4 of Fig.~17(b). These six OCs are 
inconsistent with the arguments of \cite{mac03, mac08}. 
 
\item  For this paper, we do not make an effort to determine the binary fractions of our sample OCs.
However, the OCs Binaries widen the main sequence of the OCs  by as much as 0.75 mag, so
theoretical isochrones are fitted to the mid-points of CMDs of the OCs, 
rather than the faint or blue sides, as emphasized by \cite{car01}.
Binaries are indeed an effective way of storing energy in a cluster. Non-primordial
binary formation, especially the close ones, requires many encounters of at least 3
stars, 2 of which end up having orbits around each other and the 3rd one gets
"ejected". So, depending on the binary fraction, a cluster can get dynamically
swollen. As a consequence of the dynamical evolution in OCs, multiple systems
tend to concentrate in central regions \cite{tak00}. As indicated by \cite{bont05}, the main effect of 
a significant fraction of  binaries in central parts of OCs
is that the number of low-mass stars is underestimated with respect to the higher mass stars.
\cite{sol10} give the complete fractions of binaries as 35 $\%$ to 70 $\%$. 
They also give a minimum binary fraction, which is larger than  11 $\%$ within the core of OCs.

\item It is seen from Figs.~18(a) and (b) that the OCs with R$_{RDP} < 3$  pc and 
R$_{core} < 0.6$ pc inside the Solar circle are older than 1 Gyr. 
As they lost their stellar content, they shrunk in size and mass with 
time. Nevertheless, they seem to survive against external shocks for a 
longer time, according to the simulations of \cite{sc73}. As one can see  
from  Fig.~18(b), there is no strong dependence of R$_{core}$ with 
R$_{GC}$.

\item As can be seen in Figs.~19(a) and (b), OCs with large dimensions are on 
the average more massive. There does not seem to be an age dependence 
for the relations in the panels~(a)$-$(b) of Fig~19. As can be seen from Fig.~20(a), 
massive and less massive OCs than m$_{overall} =2000\: 
M_{\odot}$ are located indistinctly in- or outwards of the Solar circle.  
Less massive OCs which are located outside the Solar circle appear to 
survive, because they are subject to less external dynamical processes: 
The OCs inside the Solar radius survive against the combined dynamical 
effects such as interactions with GMCs, tidal effects with the spiral 
arm and the Galactic disc, which are quite efficient in the Galactic 
center directions. As is seen in Fig.~20(b), less massive OCs older than 
1 Gyr are scarcer since they are dissolved into the field, i.e. the more 
massive and older OCs ($ > 1$ Gyr) survive.

\item The OCs with flat/steep positive overall MF slopes for $\chi_{core}<0$ in Fig.~21 show signs of 
a mild to large scale mass segregation, whereas the OCs with negative 
overall MF slopes for ~$\chi_{core} < 0$ ~indicate an advanced dynamical 
evolution. These MF slopes of ~$\chi_{core}<0$ ~in Fig.~21 can be explained 
by the external dynamical effects such as tidal stripping by tidal 
interactions (in the form of shocks) due to disc and bulge 
crossings, as well as encounters with GMCs.

\item As considered these MF slopes in Figs.~22(a) and (b), OCs are formed with 
flat core and Kroupa and Salpeter-like overall MFs, as stated by 
\cite{bon06a}. As is seen from Fig.~22, at cluster birth the core MF 
seems to be much flatter than the overall MF. Early core flattening may 
be partly linked to primordial processes associated to molecular-cloud 
fragmentation. 
Within the expected uncertainties the overall MF values are quite close to 
$\chi_\circ=1.30\pm0.30$ of \cite{kro01} and $\chi_\circ=1.35$ of  
\cite{sal55}. However, our core MF value are smaller than the ones of 
Kroupa and Salpeter. 
As is seen from  panels~(a)$-$(b), except for few MF slopes, the overall 
and core MF slopes tend to be negative values towards older ages,   
because of mild/large scale mass segregation, the presence of GMCs and 
tidal effects from disk and Bulge crossings as external processes. 

\item Most of the OCs of with positive overall slopes in Fig.~23(a) are  
mass-rich and present little or no signs of mass segregation. 
Apparently they retain their low-mass stars 
because they are strongly bounded to the clusters. The OCs with negative 
overall MF slopes in Fig.~23(a) seem to be in the phase of more advanced 
dynamical evolution. 
In panels~(a)$-$(b) of Fig.~23 there is no indication of age dependence 
that is seen between the MF slopes and  the core- or overall-masses.
In panels~(a)$-$(b) of Fig.~23, the OCs with steep overall and 
core MF slopes present signs of larger scale mass segregation in the core 
or halo region. As expected, there are no indications of age dependence 
among the positive/negative MF slopes.

\item From Fig.~27(a), one sees that for $\tau > 30$, the overall MF slopes of the 
OCs are negative, with one exception. For $\tau < 30$, the overall MF 
slopes of the remaining OCs fall in the range of $ +0.5< \chi_{overall} 
< +2.5$.  For $\tau>30$, as a result of the loss of low mass stars, $\chi_{overall}$ 
tends to negative values. As can be seen from panel~(b), the core MF slopes 
for the majority of OCs tend to be negative values after $\tau\approx 32$, 
with two exceptions. In panel~(b) there are two 
OCs with flat slopes for $\tau< 32$. 
For $\tau>32$, it is seen from panel~(b) that the OCs with dynamically 
evolved cores reveal a sign of strong mass segregation. 

From eleven OCs, \cite{bon05} detected the significant flattening in 
MF slopes for $\tau_{core} \leq 100$ and $\tau_{overall} \leq 7$, 
respectively. From their OCs, \cite{mn07} give for these values   
$\tau_{core} \le 1000$ and $\tau_{overall}\le 450$, respectively. 
Here we detect the flattening of MF slopes at $\tau_{core} \leq 32$ 
and $\tau_{overall} \leq 30$, respectively. However, these values 
are affected by the sample size with young and old OCs.
Note that our sample also contains the OCs with intermediate
and old ages.

The overall MF slopes of 31 OCs with $m > 0.5$,  M$_{\odot}$ could 
have been compared with the one given by \citet[$\chi=+1.3 \pm 0.3$]{kro01}.
As compared to the uncertainties of our MF slopes (Col.~5; Table 7) 
and the one ($\pm0.3$) of Kroupa, the overall MF slopes of 14 out of 
31 OCs are consistent with the one of  \citet[$\chi=+1.3\pm0.3$]{kro01}, 
which implies little or no dynamical evolution for these clusters. The 
remaining 17 OCs with MF slopes that depart from that of \cite{kro01} 
show mild to large scale mass segregation, due to the dynamical evolution.

\item We do not make an effort to determine the binary fractions of our sample OCs.
\cite{sol10} give the complete fractions of binaries as 35 $\%$ to 70 $\%$. 
They also give a minimum binary fraction, which is larger than  11 $\%$ within the core of OCs.
However, binaries of the OCs widen the main sequence of the OCs  by as much as 0.75 mag, so
theoretical isochrones are fitted to the mid-points of CMDs of the OCs, 
rather than the faint or blue sides, as emphasized by \cite{car01}. We have considered this issue
for isochrone fitting to CMDs (see Sect.4).
Binaries are indeed an effective way of storing energy in a cluster. Non-primordial
binary formation, especially the close ones, requires many encounters of at least 3
stars, 2 of which end up having orbits around each other and the 3rd one gets
"ejected". So, depending on the binary fraction, a cluster can get dynamically
swollen. As a consequence of the dynamical evolution in OCs, multiple systems
tend to concentrate in central regions \citep{tak00}. As indicated by \cite{bont05}, the main effect of 
a significant fraction of  binaries in central parts of OCs
is that the number of low-mass stars is underestimated with respect to the higher mass stars.

\end{enumerate}

\acknowledgements 

We thank the anonymous referee for her/his comments and suggestions on the manuscript.
We thank C. Chavarria for the correction of English in the text. 
This publication makes use of data products from the Two Micron All Sky Survey, 
which is a joint project of the University of Massachusetts and the 
Infrared Processing and Analysis Centre/California Institute of Technology, 
funded by the National Aeronautics and Space Administration and the National 
Science Foundation. This research has made use of the WEBDA database, operated 
at the Institute for Astronomy of the University of Vienna.

\clearpage

\section{Supplementary material}

Figures S5-S9: Observed decontaminated J, (J-H) CMDs of 40 OCs. The CMD filter shown 
with the shaded area is used to isolate cluster MS/evolved stars.

Figures S10-S13: Stellar RDPs (open circles) of 40 Ocs  built with CMD filtered photometry. 
Solid line shows the best-fit King profile. Horizontal red bar: stellar background level measured 
in the comparison field. Shaded region: $1\sigma$ King fit uncertainty.

Table S6: The number of stars, mass information, mass function slope, mass density, which correspond to cluster regions of available clusters 
for the cases of Evolved, Observed+Evolved, and Extrapolated+Evolved.

\clearpage

\begin{figure}
\renewcommand\thefigure{S5}
\centering
\includegraphics*[width = 14cm, height = 15cm]{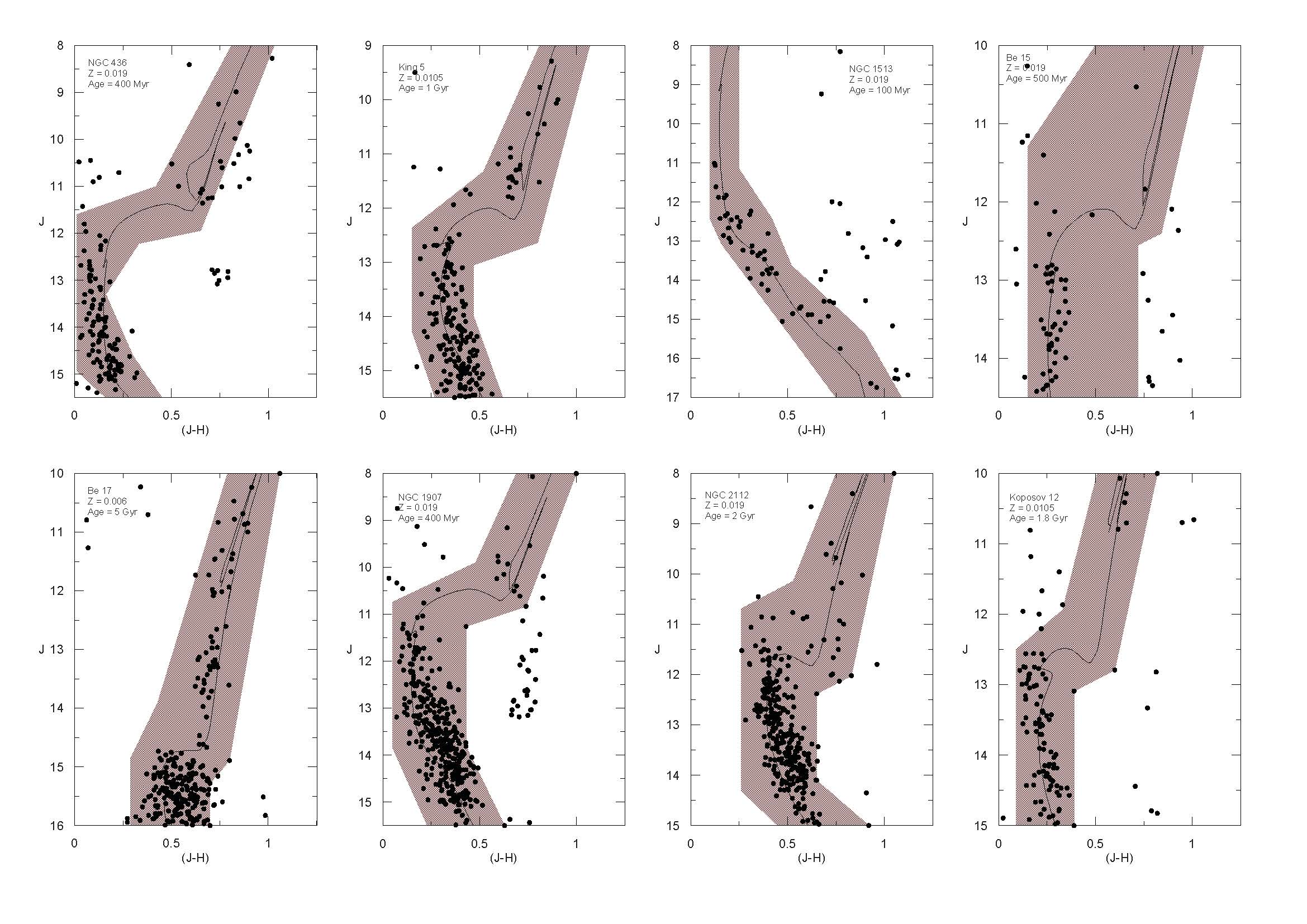}
\caption {Observed decontaminated J, (J-H) CMDs of 10 OCs. 
The CMD filter shown with the shaded area is used to isolate cluster MS/evolved stars. 
The OCs in the panels are  NGC~436, King~5, NGC~1513, Be~15, Be~17, NGC~1907, NGC~2112, and Koposov 12, respectively.}
\end{figure}

\clearpage
\begin{figure}
\renewcommand\thefigure{S6}
\centering
\includegraphics*[width = 14cm, height = 15cm]{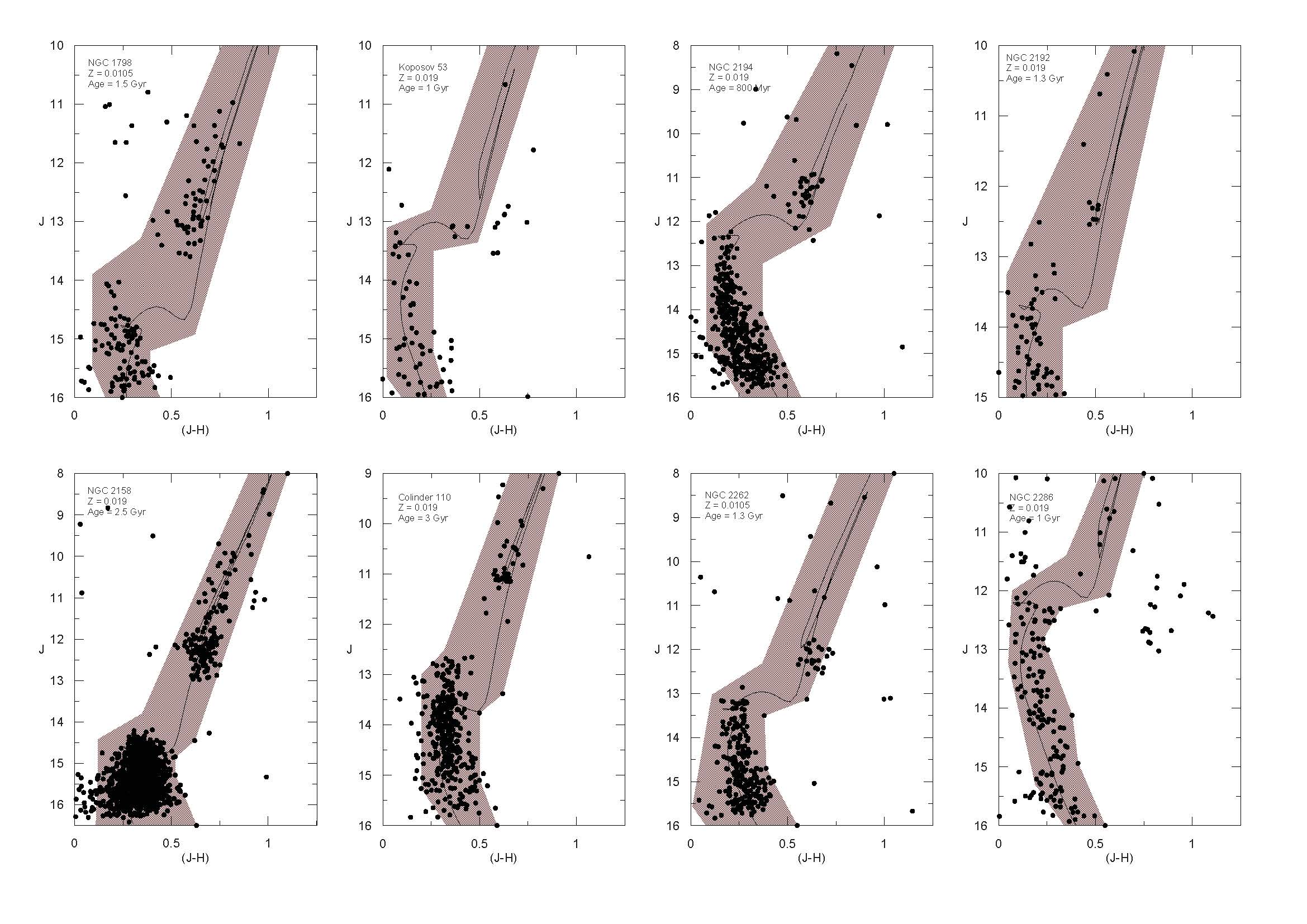}
\caption {Observed decontaminated J, (J-H) CMDs of NGC~1798, Koposov 53, NGC~2194, NGC~2192, NGC~2158, Col~110, NGC~2262, 
NGC~2286, respectively.
The symbols are the same as Fig.~S5.} 
\end{figure}

\clearpage
\begin{figure}
\renewcommand\thefigure{S7}
\centering
\includegraphics*[width = 14cm, height = 15cm]{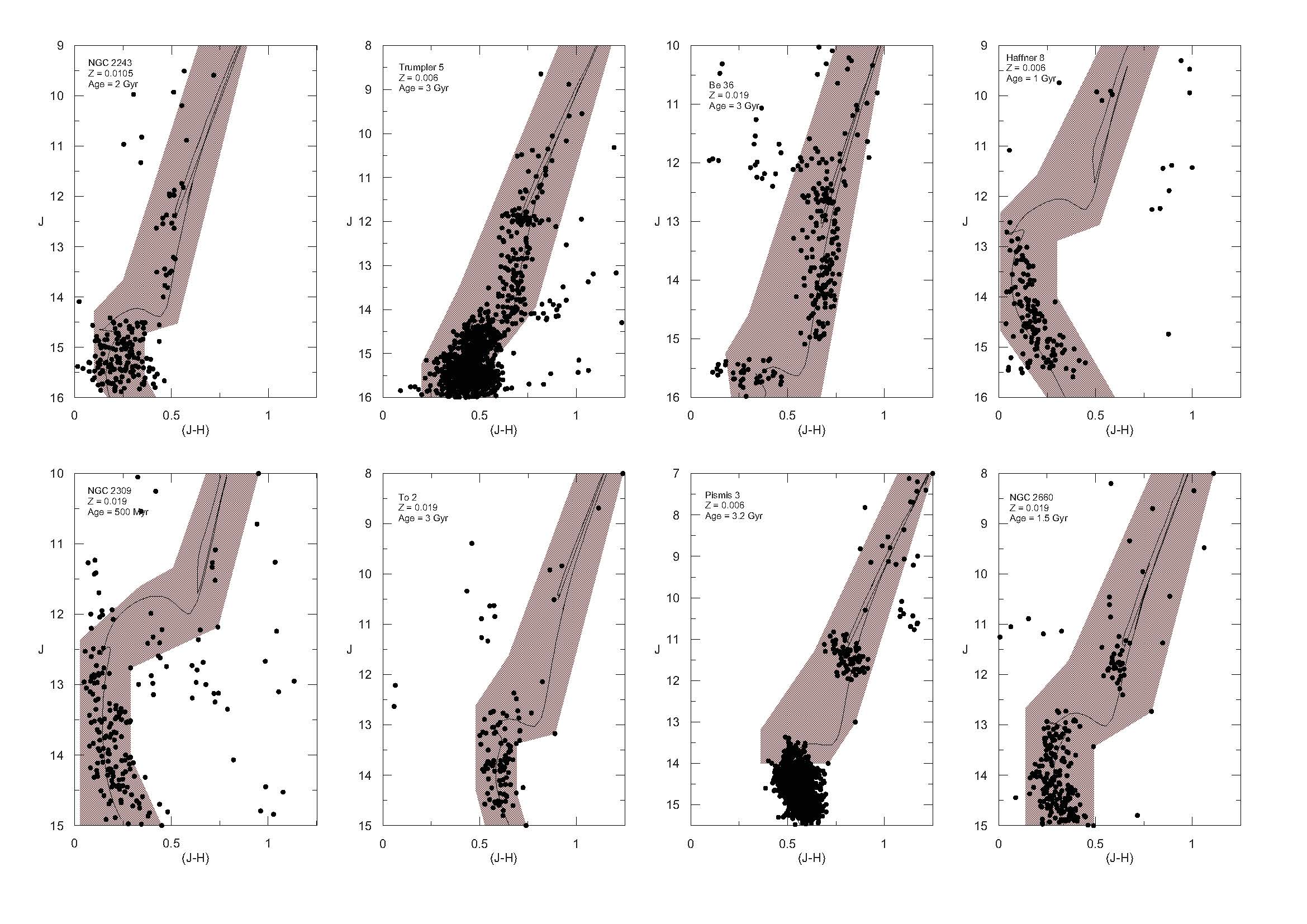}
\caption {Observed decontaminated J, (J-H) CMDs of NGC~2243, Trumpler 5, Be~36, Haffner 8, NGC 2309, To 2, Pismis 3, NGC~2660, respectively.
The symbols are the same as Fig.~S5.}
\end{figure}

\clearpage
\begin{figure}
\renewcommand\thefigure{S8}
\centering
\includegraphics*[width = 14cm, height = 15cm]{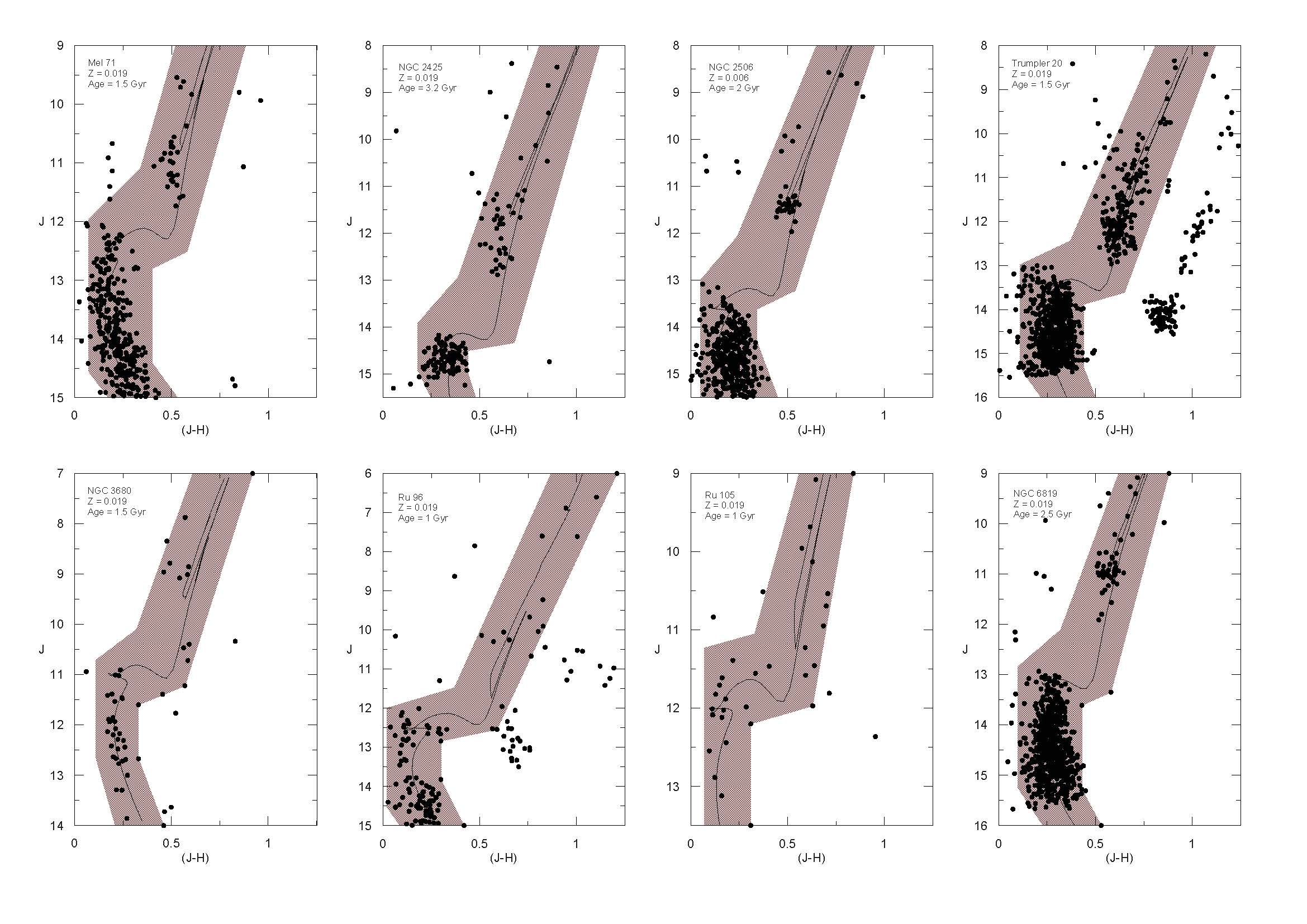}
\caption {Observed decontaminated J, (J-H) CMDs of  Mel 71, NGC 2425, NGC 2506, Trumpler 20, NGC 3680, Ru 96, Ru 105, NGC 6819, respectively.
The symbols are the same as Fig.~S5.} 
\end{figure}
\clearpage

\begin{figure}
\renewcommand\thefigure{S9}
\centering
\includegraphics*[width = 14cm, height = 15cm]{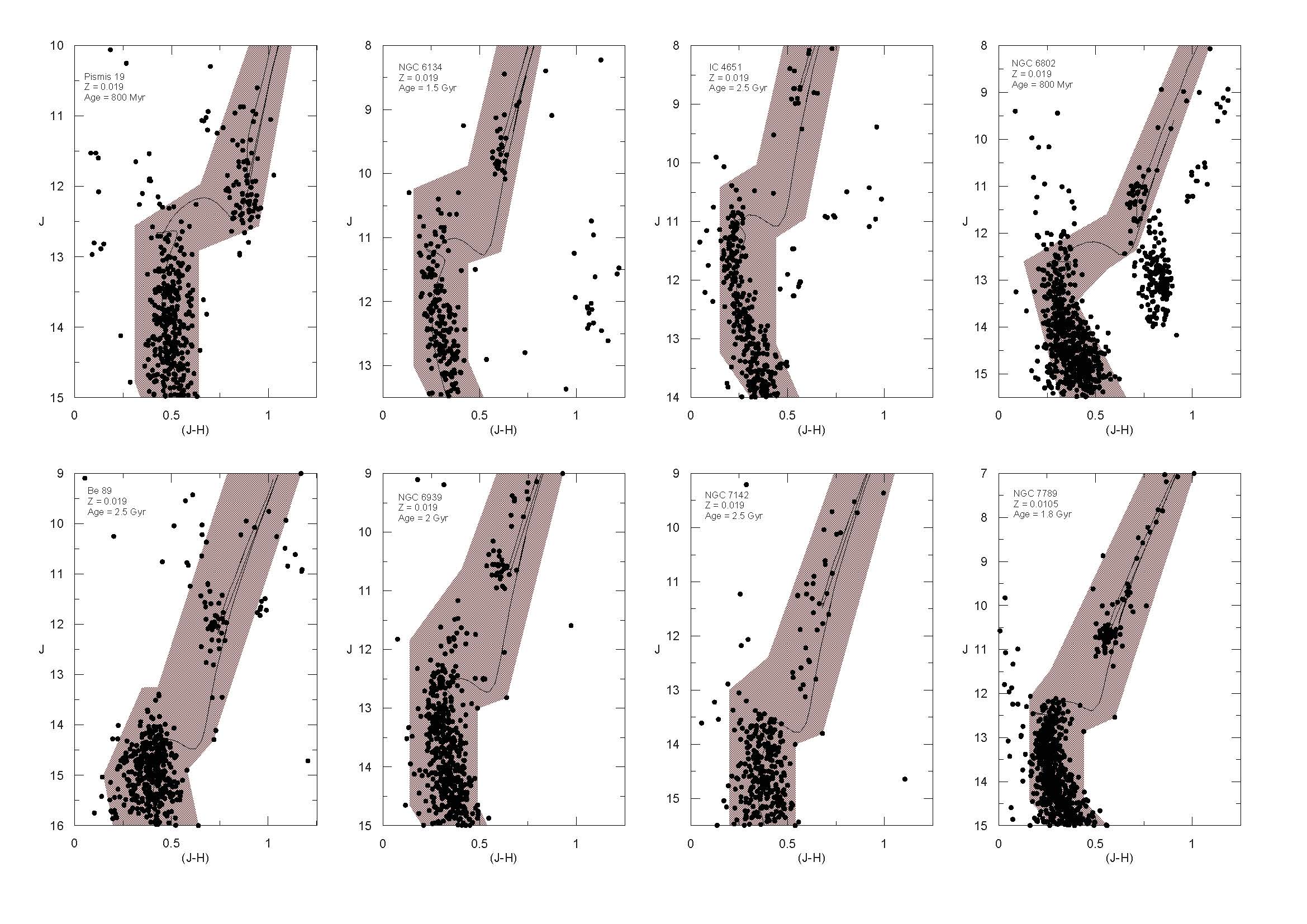}
\caption {Observed decontaminated J, (J-H) CMDs of Pismis 19, NGC 6134, IC 4651, NGC 6802, Be 89, NGC 6939, NGC 7142, NGC 7789, respectively.
The symbols are the same as Fig.~S5.} 
\end{figure}
\clearpage

\begin{figure}
\renewcommand\thefigure{S10}
\centering
\includegraphics*[width = 14cm, height = 15cm]{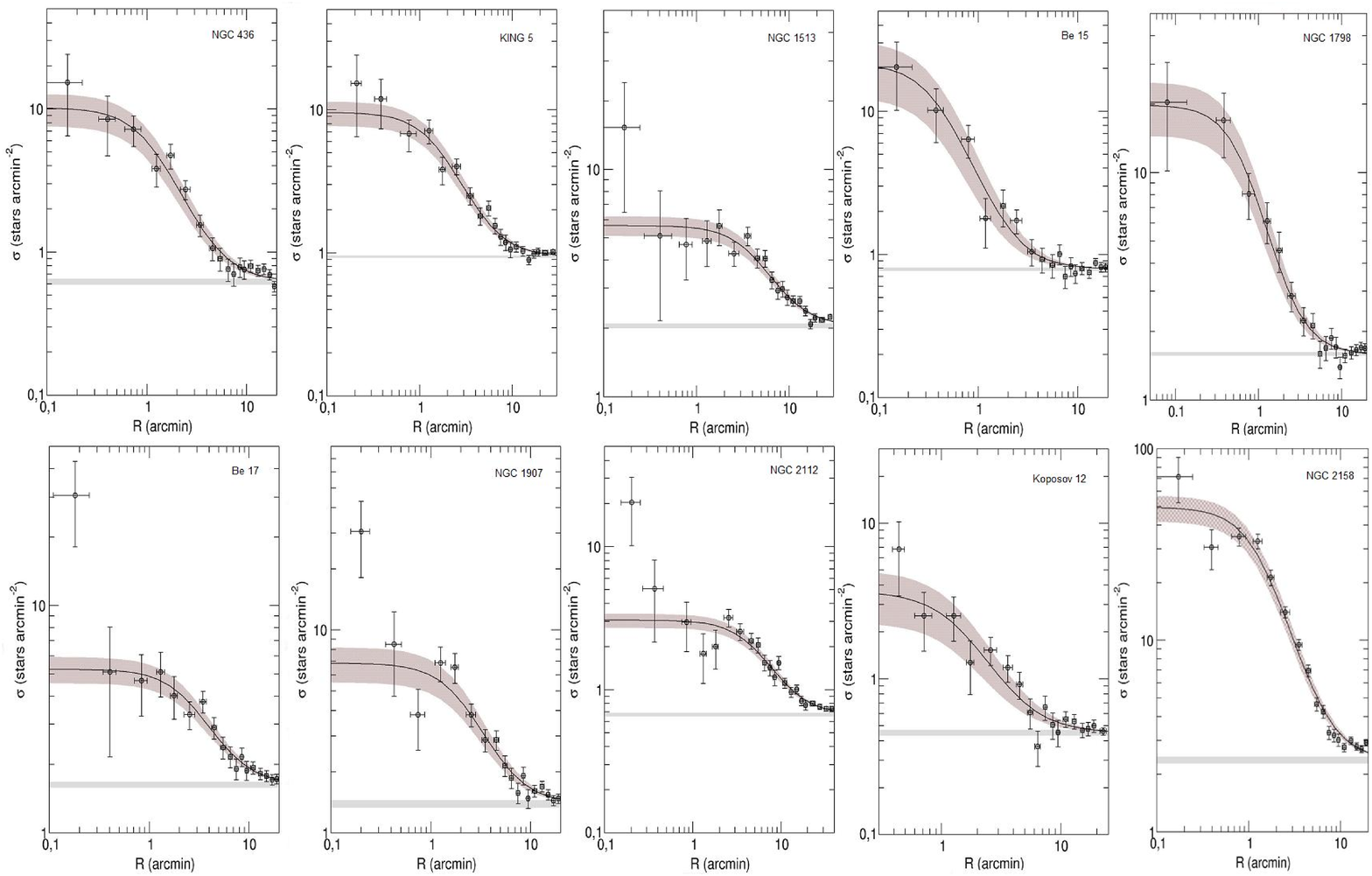}
\caption {Stellar RDPs (open circles) of 10 Ocs  built with CMD filtered photometry. 
Solid line shows the best-fit King profile. Horizontal red bar: stellar background level measured 
in the comparison field. Shaded region: $1\sigma$ King fit uncertainty.
The OCs in the panels are NGC~436, Ki~05, NGC~1513, Be~15, NGC~1798, NGC~2112, Koposov 12, and NGC~2158, respectively.}  
\end{figure}

\clearpage
\begin{figure}
\renewcommand\thefigure{S11}
\centering
\includegraphics*[width = 14cm, height = 15cm]{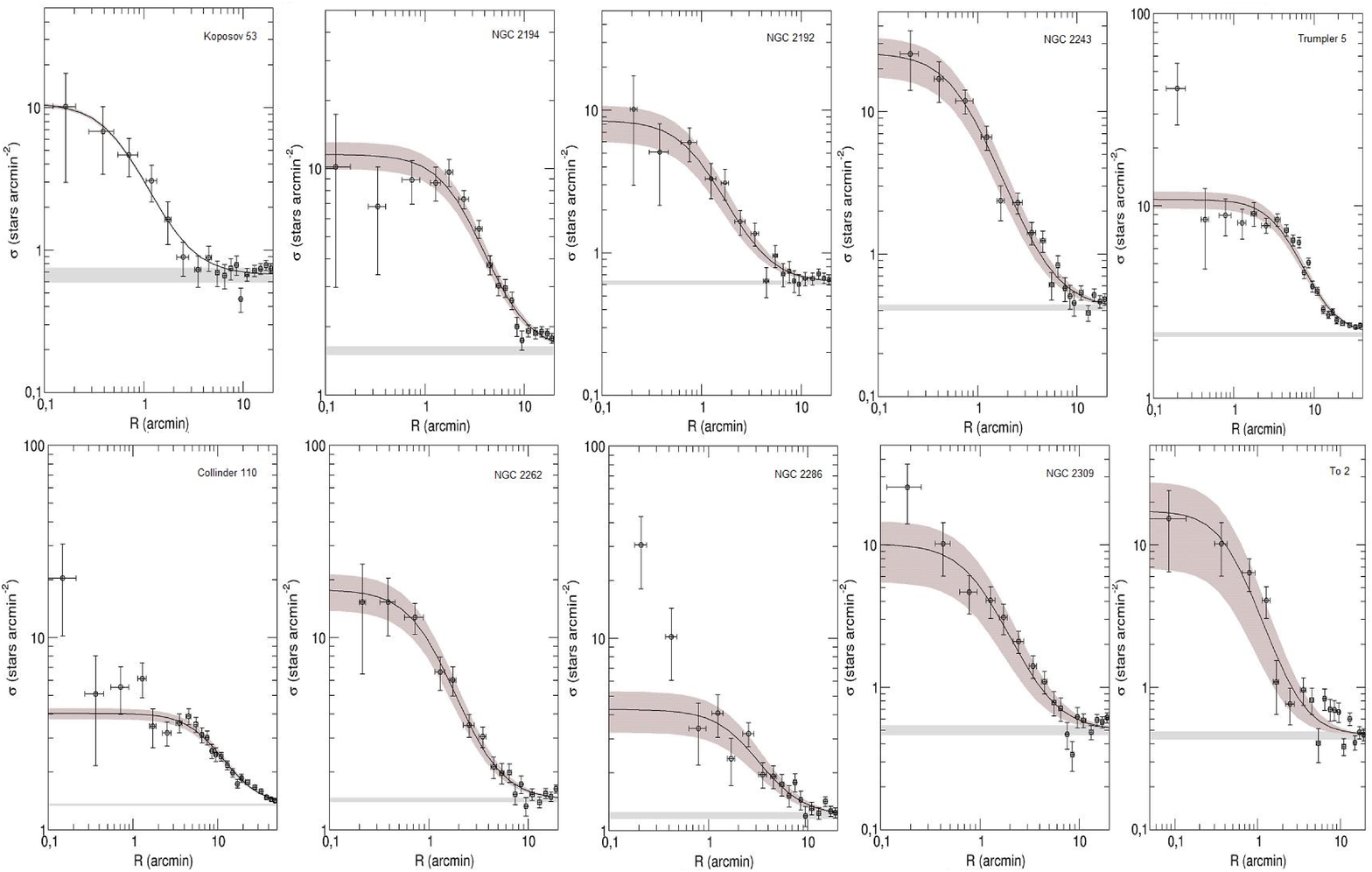}
\caption {Stellar RDPs (open circles) of Koposov 53, NGC~2194, NGC~2192, NGC~2243, Trumpler 5, Col~110, NGC~2262, NGC~2286, NGC~2309, To~2, respectively. 
The symbols are the same as Fig.~S10.} 
\end{figure}

\clearpage
\begin{figure}
\renewcommand\thefigure{S12}
\centering
\includegraphics*[width = 14cm, height = 15cm]{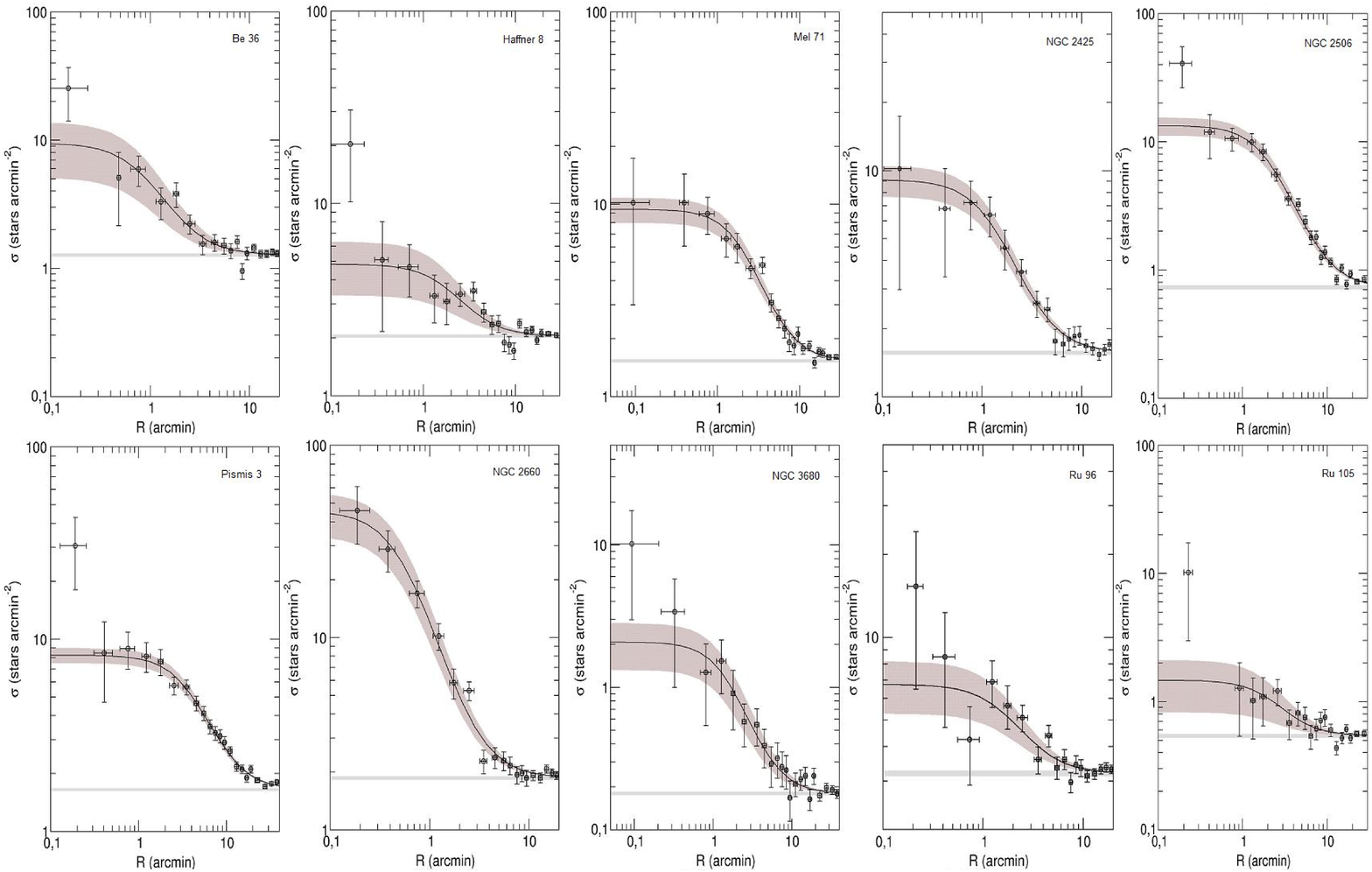}
\caption {Stellar RDPs (open circles) of Be~36, Haffner 8, Mel 71, NGC~2425, NGC~2506, Pismis 3, NGC~2660, NGC 3680, Ru 96, Ru 105, respectively.
The symbols are the same as Fig.~S10.} 
\end{figure}
\clearpage

\begin{figure}
\renewcommand\thefigure{S13}
\centering
\includegraphics*[width = 14cm, height = 15cm]{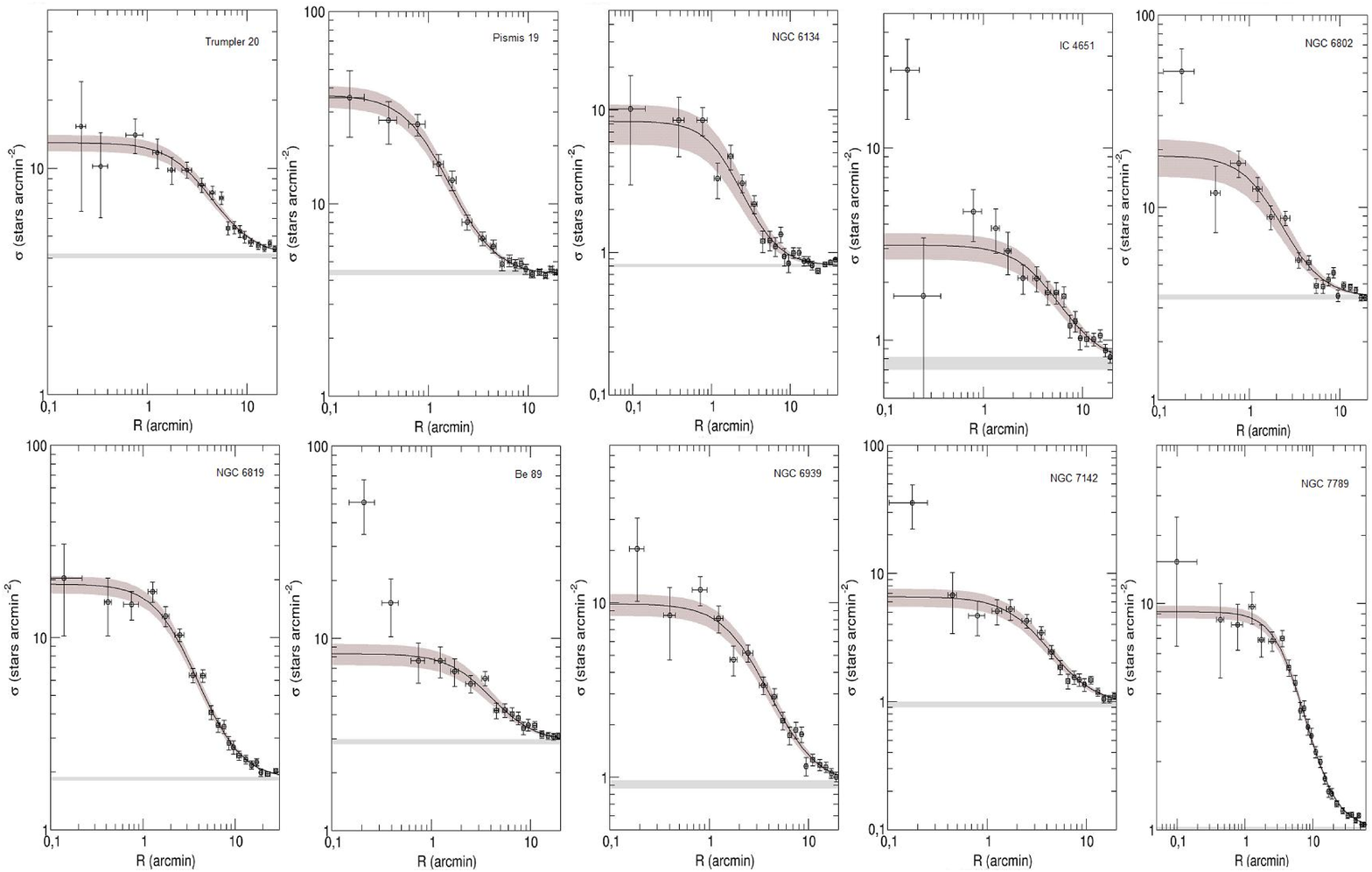}
\caption {Stellar RDPs (open circles) of Trumpler 20, Pismis 19, NGC 6134, IC 4651, NGC 6802, NGC 6819, Be 89, NGC 6939, NGC 7142, NGC 7789, respectively. 
The symbols are the same as Fig.~S10.} 
\end{figure}

\clearpage

\begin{table*}  
\renewcommand\thetable{S6}
  \centering
  \caption{The number of stars, mass information,  mass function slope, mass density, which correspond to cluster regions of available clusters 
  for the cases of Evolved, Observed+Evolved, and Extrapolated+Evolved.}

\tiny
\begin{tabular}{c|c|c|c|c|c|c|c|c|c|c}
\hline
\hline

\multicolumn{11}{c}{NGC 436}\\\cline {5-7} \\
      & \multicolumn{2}{c}{Evolved} & 
         \multicolumn{2}{c} {$\chi$} &  
          \multicolumn{3}{c}{Observed+Evolved} &    
           \multicolumn{3}{c}{Extrapolated+Evolved} \\\cline {2-11} \\
   
 {Region}  & {N*}    & {m$_{evol}$}        & {1.38-2.78} & {-} & {N*}             & {m$_{obs}$}         & {$\rho$} & {N*}& {m$_{tot}$  } & {$\rho$} \\
   (pc)     & (Stars) & ($10^1 M_{\odot}$)&             &     &  ($10^2 Stars$)   & ($10^2 M_{\odot}$)  & $M_{\odot} pc^{-3}$         & ($10^2 Stars$) & ($10^2 M_{\odot}$) & $M_{\odot} pc^{-3}$ \\
\hline
0.0-1.04 &  1$\pm$1   &  0.4$\pm$0.4 &  -1.46$\pm$0.47 &  {-} &  0.25$\pm$0.03 &  0.56$\pm$0.28 &  11.9$\pm$5.97 &  0.4$\pm$0.1 &  0.7$\pm$0.03 &  15.2$\pm$6.11 \\
1.04-6.97 &  12$\pm$6  &  3.5$\pm$1.8 &  1.74$\pm$0.36 &  {-} &  1.01$\pm$0.1 &  2$\pm$0.57 &  0.14$\pm$0.04 &  25.6$\pm$19.6 &  9.8$\pm$3.8 &  0.69$\pm$0.27 \\
0.0-6.97 &  14$\pm$6  &  3.9$\pm$1.9 &  0.86$\pm$0.29 &  {-} &  1.12$\pm$0.09 &  2.55$\pm$0.63 &  0.18$\pm$0.04 &  17.1$\pm$11.9 &  7.9$\pm$2.4 &  0.56$\pm$0.17 \\
\hline

\multicolumn{11}{c}{King 5}\\\cline {5-7} \\
      & \multicolumn{2}{c}{Evolved} & 
         \multicolumn{2}{c} {$\chi$} & 
          \multicolumn{3}{c}{Observed+Evolved} &  
           \multicolumn{3}{c}{Extrapolated+Evolved} \\\cline {2-11} \\  
 {Region}  & {N*}    & {m$_{evol}$}        & {1.28-1.78} & {-} & {N*}             & {m$_{obs}$}         & {$\rho$} & {N*}& {m$_{tot}$  } & {$\rho$} \\
   (pc)     & (Stars) & ($10^1 M_{\odot}$)&             &     &  ($10^2 Stars$)   & ($10^2 M_{\odot}$)  &  $M_{\odot} pc^{-3}$        & ($10^2 Stars$) & ($10^2 M_{\odot}$)& $M_{\odot} pc^{-3}$ \\
\hline
0.0-0.95 &  9$\pm$3   &  1.8$\pm$0.6 &  -3.06$\pm$0.96 &  {-} &  0.48$\pm$0.05 &  0.77$\pm$0.34 &  21.5$\pm$9.6 &  0.63$\pm$0.09 &  0.91$\pm$0.35 &  25.3$\pm$9.73 \\
0.95-5.62 &  27$\pm$8  &  5.5$\pm$1.7 &  2.65$\pm$0.41 &  {-} &  1.98$\pm$0.14 &  3.07$\pm$0.62 &  0.41$\pm$.0.08 &  80.2$\pm$61.7 &  27.3$\pm$11.7 &  3.69$\pm$1.58 \\
0.0-5.62 &  35$\pm$9  &  7.3$\pm$1.8 &  1.8$\pm$0.49 &  {-} &  2.44$\pm$0.16 &  3.84$\pm$0.89 &  0.52$\pm$0.12 &  86.9$\pm$67.2 &  30.0$\pm$12.8 &  4.04$\pm$1.73 \\
\hline

\multicolumn{11}{c}{NGC 1513}\\\cline {5-7} \\
      & \multicolumn{2}{c}{Evolved} & 
         \multicolumn{2}{c} {$\chi$} & 
          \multicolumn{3}{c}{Observed+Evolved} &  
           \multicolumn{3}{c}{Extrapolated+Evolved} \\\cline {2-11} \\  
 {Region}  & {N*}    & {m$_{evol}$}        & {0.68-0.98} & {0.98-3.33} & {N*}             & {m$_{obs}$}         & {$\rho$} & {N*}& {m$_{tot}$  } & {$\rho$} \\
   (pc)     & (Stars) & ($10^1 M_{\odot}$)&             &     &  ($10^2 Stars$)   & ($10^2 M_{\odot}$)  &  $M_{\odot} pc^{-3}$        & ($10^2 Stars$) & ($10^2 M_{\odot}$) & $M_{\odot} pc^{-3}$ \\
\hline
0.0-1.65 &  -     &  -     &  -3.37$\pm$0.52 &  1.12$\pm$0.24 &  1.77$\pm$0.11 &  2.31$\pm$0.27 &  12.3$\pm$1.41 &  10.8$\pm$7.65 &  4.54$\pm$1.43 &  24.1$\pm$7.6 \\
1.65-6.51 &  -     &  -     &  -     &  2.08$\pm$0.17 &  4.25$\pm$0.3 &  5$\pm$0.42 &  0.44$\pm$0.04 &  45.2$\pm$34.6 &  15.4$\pm$6.42 &  1.35$\pm$0.57 \\
0.0-6.51 &  -     &  -     &  -     &  1.9$\pm$0.12 &  5.9$\pm$0.32 &  7.13$\pm$0.46 &  0.62$\pm$0.04 &  58.6$\pm$44.4 &  20.4$\pm$8.23 &  1.77$\pm$0.71 \\
\hline

\multicolumn{11}{c}{Be 15}\\\cline {5-7} \\
      & \multicolumn{2}{c}{Evolved} & 
         \multicolumn{2}{c} {$\chi$} & 
          \multicolumn{3}{c}{Observed+Evolved} &  
           \multicolumn{3}{c}{Extrapolated+Evolved} \\\cline {2-11} \\  
 {Region}  & {N*}    & {m$_{evol}$}        & {2.03-2.68} & {-} & {N*}             & {m$_{obs}$}         & {$\rho$} & {N*}& {m$_{tot}$  } & {$\rho$} \\
   (pc)     & (Stars) & ($10^1 M_{\odot}$)&             &     &  ($10^2 Stars$)   & ($10^2 M_{\odot}$)  & $M_{\odot} pc^{-3}$         & ($10^2 Stars$) & ($10^2 M_{\odot}$) & $M_{\odot} pc^{-3}$ \\
\hline
0.0-1.16 &  -     &  -     &  -     &  {-} &  -     &  -     &  -     &  -     &  -     &  - \\
1.16-5.73 &  -     &  -     &  -3.07$\pm$1.07 &  {-} &  0.35$\pm$0.09 &  0.83$\pm$1.11 &  0.15$\pm$0.21 &  0.61$\pm$0.28 &  1.23$\pm$1.19 &  0.23$\pm$0.22 \\
0.0-5.73 &  -     &  -     &  -1.54$\pm$1.15 &  {-} &  0.42$\pm$0.09 &  1$\pm$1.4 &  0.19$\pm$0.26 &  1.2$\pm$0.83 &  1.96$\pm$1.73 &  0.36$\pm$0.32 \\
\hline

\multicolumn{11}{c}{NGC 1907}\\\cline {5-7} \\
      & \multicolumn{2}{c}{Evolved} & 
         \multicolumn{2}{c} {$\chi$} & 
          \multicolumn{3}{c}{Observed+Evolved} &  
           \multicolumn{3}{c}{Extrapolated+Evolved} \\\cline {2-11} \\  
 {Region}  & {N*}    & {m$_{evol}$}        & {1.18-2.83} & {-} & {N*}             & {m$_{obs}$}         & {$\rho$} & {N*}& {m$_{tot}$  } & {$\rho$} \\
   (pc)     & (Stars) & ($10^1 M_{\odot}$)&             &     &  ($10^2 Stars$)   & ($10^2 M_{\odot}$)  &  $M_{\odot} pc^{-3}$        & ($10^2 Stars$) & ($10^2 M_{\odot}$) & $M_{\odot} pc^{-3}$ \\
\hline
0.0-1.28 &  9$\pm$3   &  2.5$\pm$0.9 &  -0.76$\pm$0.4 &  {-} &  0.78$\pm$0.07 &  1.68$\pm$0.61 &  19.2$\pm$6.99 &  1.62$\pm$2.02 &  2.24$\pm$0.84 &  25.5$\pm$9.55 \\
1.28-4.26 &  3$\pm$3   &  0.9$\pm$1 &  -0.73$\pm$0.43 &  {-} &  1.6$\pm$0.11 &  2.81$\pm$0.93 &  0.89$\pm$0.3 &  3.69$\pm$8.12 &  4$\pm$2.42 &  1.27$\pm$0.77 \\
0.0-4.26 &  12$\pm$5  &  3.5$\pm$1.3 &  0.0$\pm$0.23 &  {-} &  2.36$\pm$0.14 &  4.56$\pm$0.91 &  1.41$\pm$0.28 &  -     &  -     &  - \\
\hline

\multicolumn{11}{c}{NGC 2112}\\\cline {5-7} \\
      & \multicolumn{2}{c}{Evolved} & 
         \multicolumn{2}{c} {$\chi$} & 
          \multicolumn{3}{c}{Observed+Evolved} &  
           \multicolumn{3}{c}{Extrapolated+Evolved} \\\cline {2-11} \\  
 {Region}  & {N*}    & {m$_{evol}$}        & {0.93-1.63} & {-} & {N*}             & {m$_{obs}$}         & {$\rho$} & {N*}& {m$_{tot}$  } & {$\rho$} \\
   (pc)     & (Stars) & ($10^1 M_{\odot}$)&             &     &  ($10^2 Stars$)   & ($10^2 M_{\odot}$)  &  $M_{\odot} pc^{-3}$        & ($10^2 Stars$) & ($10^2 M_{\odot}$) & $M_{\odot} pc^{-3}$ \\
\hline
0.0-1.64 &  19$\pm$5  &  3$\pm$0.8 &  -1.28$\pm$0.51 &  -     &  1.56$\pm$0.09 &  2.06$\pm$0.34 &  11.1$\pm$1.86 &  2.8$\pm$3.7 &  2.73$\pm$1.17 &  14.8$\pm$6.36 \\
1.64-5.92 &  10$\pm$8  &  1.6$\pm$1.4 &  0.26$\pm$0.39 &  -     &  3.59$\pm$0.23 &  4.33$\pm$0.5 &  0.51$\pm$0.06 &  -     &  -     &  - \\
0.0-5.92 &  28$\pm$10 &  4.6$\pm$1.6 &  0.5$\pm$0.42 &  -     &  5.11$\pm$0.24 &  6.38$\pm$0.78 &  0.73$\pm$0.09 &  44.9$\pm$31.5 &  18.1$\pm$5.96 &  2.08$\pm$0.69 \\
\hline

\multicolumn{11}{c}{NGC 2158}\\\cline {5-7} \\
      & \multicolumn{2}{c}{Evolved} & 
         \multicolumn{2}{c} {$\chi$} & 
          \multicolumn{3}{c}{Observed+Evolved} &  
           \multicolumn{3}{c}{Extrapolated+Evolved} \\\cline {2-11} \\  
 {Region}  & {N*}    & {m$_{evol}$}        & {1.28-1.53} & {-} & {N*}             & {m$_{obs}$}         & {$\rho$} & {N*}& {m$_{tot}$  } & {$\rho$} \\
   (pc)     & (Stars) & ($10^1 M_{\odot}$)&             &     &  ($10^2 Stars$)   & ($10^2 M_{\odot}$)  & $M_{\odot} pc^{-3}$         & ($10^2 Stars$) & ($10^2 M_{\odot}$) & $M_{\odot} pc^{-3}$ \\
\hline
0.0-1.74 &  57$\pm$8  &  8.7$\pm$1.2 &  -4.24$\pm$1 &  {-} &  1.9$\pm$0.11 &  2.74$\pm$0.91 &  12.4$\pm$4.12 &  4.21$\pm$0.82 &  5.21$\pm$1.27 &  23.6$\pm$5.74 \\
1.74-14.03 &  203$\pm$22 &  30.8$\pm$0.3.3 &  -0.08$\pm$0.6 &  {-} &  8.39$\pm$0.31 &  12$\pm$2.56 &  0.1$\pm$.0.02 &  -     &  -     &  - \\
0.0-14.03 &  260$\pm$23 &  39.6$\pm$3.5 &  -1.55$\pm$0.71 &  {-} &  10.3$\pm$0.33 &  14.7$\pm$3.69 &  1.27$\pm$0.04 &  -     &  33.4$\pm$26.6 &  0.29$\pm$0.23 \\
\hline

\multicolumn{11}{c}{Koposov 53}\\\cline {5-7} \\
      & \multicolumn{2}{c}{Evolved} & 
         \multicolumn{2}{c} {$\chi$} & 
          \multicolumn{3}{c}{Observed+Evolved} &  
           \multicolumn{3}{c}{Extrapolated+Evolved} \\\cline {2-11} \\  
 {Region}  & {N*}    & {m$_{evol}$}        & {1.28-2.08} & {-} & {N*}             & {m$_{obs}$}         & {$\rho$} & {N*}& {m$_{tot}$  } & {$\rho$} \\
   (pc)     & (Stars) & ($10^1 M_{\odot}$)&             &     &  ($10^2 Stars$)   & ($10^2 M_{\odot}$)  & $M_{\odot} pc^{-3}$         & ($10^2 Stars$) & ($10^2 M_{\odot}$) & $M_{\odot} pc^{-3}$ \\
\hline
0.0-0.66 &  2$\pm$1   &  0.4.$\pm$0.3 &  -3.96 $\pm$3.4 &  {-} &  -     &  -     &  -     &  -     &  -     &  - \\
0.66-4.18 &  1$\pm$3   &  0.3$\pm$0.6 &  1.36$\pm$0.92 &  {-} &  0.26$\pm$0.05 &  0.43$\pm$0.26 &  0.14$\pm$0.08 &  7.33$\pm$6.22 &  2.62$\pm$1.3 &  0.86$\pm$0.43 \\
0.0-4.18 &  3$\pm$3   &  0.5$\pm$0.7 &  0.93$\pm$0.81 &  {-} &  0.33$\pm$0.05 &  0.54$\pm$0.28 &  0.18$\pm$0.09 &  5.94$\pm$5.15 &  2.37$\pm$1.1 &  0.78$\pm$0.36 \\
\hline

\multicolumn{11}{c}{NGC 2194}\\\cline {5-7} \\
      & \multicolumn{2}{c}{Evolved} & 
         \multicolumn{2}{c} {$\chi$} & 
          \multicolumn{3}{c}{Observed+Evolved} &  
           \multicolumn{3}{c}{Extrapolated+Evolved} \\\cline {2-11} \\  
 {Region}  & {N*}    & {m$_{evol}$}        & {1.23-1.98} & {-} & {N*}             & {m$_{obs}$}         & {$\rho$} & {N*}& {m$_{tot}$  } & {$\rho$} \\
   (pc)     & (Stars) & ($10^1 M_{\odot}$)&             &     &  ($10^2 Stars$)   & ($10^2 M_{\odot}$)  &  $M_{\odot} pc^{-3}$        & ($10^2 Stars$) & ($10^2 M_{\odot}$) & $M_{\odot} pc^{-3}$ \\
\hline
0.0-1.66 &  18$\pm$5  &  4.1$\pm$1.0 &  0.38$\pm$0.42 &  {-} &  1.39$\pm$0.09 &  2.27$\pm$0.5 &  11.9$\pm$2.6 &  14.5$\pm$9.81 &  6.84$\pm$1.96 &  35.7$\pm$10.2 \\
1.66-6.55 &  19$\pm$7  &  4.3$\pm$1.6 &  3.65$\pm$0.48 &  {-} &  3.8$\pm$0.17 &  5.69$\pm$1.38 &  0.49$\pm$.0.12 &  159$\pm$123 &  53.2$\pm$23.3 &  4.6$\pm$2.01 \\
0.0-6.55 &  38$\pm$9  &  8.4$\pm$1.9 &  2.52$\pm$0.37 &  {-} &  5.18$\pm$0.2 &  8.03$\pm$1.56 &  0.68$\pm$0.13 &  170$\pm$129 &  58.4$\pm$24.3 &  4.96$\pm$2.07 \\
\hline

\end{tabular}
\end{table*}

\begin{table*}
\renewcommand\thetable{S5}
  \centering
\tiny
\begin{tabular}{c|c|c|c|c|c|c|c|c|c|c}

\multicolumn{11}{c}{NGC 2192}\\\cline {5-7} \\
      & \multicolumn{2}{c}{Evolved} & 
         \multicolumn{2}{c} {$\chi$} & 
          \multicolumn{3}{c}{Observed+Evolved} &  
           \multicolumn{3}{c}{Extrapolated+Evolved} \\\cline {2-11} \\  
 {Region}  & {N*}    & {m$_{evol}$}        & {1.43-1.68} & {-} & {N*}             & {m$_{obs}$}         & {$\rho$} & {N*}& {m$_{tot}$  } & {$\rho$} \\
   (pc)     & (Stars) & ($10^1 M_{\odot}$)&             &     &  ($10^2 Stars$)   & ($10^2 M_{\odot}$)  &  $M_{\odot} pc^{-3}$        & ($10^2 Stars$) & ($10^2 M_{\odot}$) & $M_{\odot} pc^{-3}$ \\
\hline
0.0-1.11 &  2$\pm$1   &  0.3$\pm$0.3 &  -2.78$\pm$0.96 &  {-} &  0.17$\pm$0.03 &  0.27$\pm$0.16 &  4.71$\pm$2.77 &  0.27$\pm$0.06 &  0.37$\pm$0.17 &  6.49$\pm$2.89 \\
1.11-5.47 &  17$\pm$5  &  3.3$\pm$1 &  -3.07$\pm$0.76 &  {-} &  0.62$\pm$0.07 &  1.03$\pm$0.36 &  0.15$\pm$0.05 &  1.32$\pm$0.26 &  1.78$\pm$0.45 &  0.26$\pm$0.07 \\
0.0-5.47 &  19$\pm$6  &  3.5$\pm$1 &  -3.12$\pm$0.43 &  {-} &  0.79$\pm$0.08 &  1.28$\pm$0.28 &  0.19$\pm$0.04 &  1.7$\pm$0.21 &  2.27$\pm$0.35 &  0.33$\pm$0.05 \\
\hline

\multicolumn{11}{c}{NGC 2243}\\\cline {5-7} \\
      & \multicolumn{2}{c}{Evolved} & 
         \multicolumn{2}{c} {$\chi$} & 
          \multicolumn{3}{c}{Observed+Evolved} &  
           \multicolumn{3}{c}{Extrapolated+Evolved} \\\cline {2-11} \\  
 {Region}  & {N*}    & {m$_{evol}$}        & {1.43-1.63} & {-} & {N*}             & {m$_{obs}$}         & {$\rho$} & {N*}& {m$_{tot}$  } & {$\rho$} \\
   (pc)     & (Stars) & ($10^1 M_{\odot}$)&             &     &  ($10^2 Stars$)   & ($10^2 M_{\odot}$)  &  $M_{\odot} pc^{-3}$        & ($10^2 Stars$) & ($10^2 M_{\odot}$) & $M_{\odot} pc^{-3}$ \\
\hline
0.0-0.89 &  6$\pm$3   &  1.$\pm$0.4 &  -     &  {-} &  -     &  -     &  -     &  -     &  -     &  - \\
0.89-12.94 &  29$\pm$10 &  4.8$\pm$1.6 &  4.81$\pm$1.16 &  {-} &  1.33$\pm$0.08 &  1.99$\pm$1.31 &  0.03$\pm$0.01 &  198$\pm$182 &  65.1$\pm$39.5 &  0.72$\pm$0.43 \\
0.0-12.94 &  36$\pm$10 &  5.8.1$\pm$1.7 &  2.09$\pm$1.01 &  {-} &  1.49$\pm$0.08 &  2.26$\pm$1.35 &  0.03$\pm$0.01 &  154$\pm$136 &  51.7$\pm$29.0 &  0.57$\pm$0.32 \\
\hline

\multicolumn{11}{c}{Trumpler 5}\\\cline {5-7} \\
      & \multicolumn{2}{c}{Evolved} & 
         \multicolumn{2}{c} {$\chi$} & 
          \multicolumn{3}{c}{Observed+Evolved} &  
           \multicolumn{3}{c}{Extrapolated+Evolved} \\\cline {2-11} \\  
 {Region}  & {N*}    & {m$_{evol}$}        & {1.18-1.43} & {-} & {N*}             & {m$_{obs}$}         & {$\rho$} & {N*}& {m$_{tot}$  } & {$\rho$} \\
   (pc)     & (Stars) & ($10^1 M_{\odot}$)&             &     &  ($10^2 Stars$)   & ($10^2 M_{\odot}$)  & $M_{\odot} pc^{-3}$         & ($10^2 Stars$) & ($10^2 M_{\odot}$) & $M_{\odot} pc^{-3}$ \\
\hline
0.0-3.86 &  93$\pm$12 &  13.2$\pm$1.7 &  0.42$\pm$0.75 &  {-} &  4.35$\pm$0.18 &  5.75$\pm$1.23 &  2.39$\pm$0.51 &  879$\pm$693 &  340$\pm$138 &  14.1$\pm$5.72 \\
3.86-15.18 &  169$\pm$34 &  24$\pm$4.9 &  1.64$\pm$1.43 &  {-} &  11.2$\pm$0.5 &  14.7$\pm$6.37 &  0.1$\pm$0.04 &  530$\pm$444 &  169$\pm$88.8 &  1.19$\pm$0.62 \\
0.0-15.18 &  262$\pm$37 &  37.2$\pm$5.2 &  1.32$\pm$1.18 &  {-} &  15.6$\pm$0.54 &  20.4$\pm$7.15 &  0.14$\pm$0.05 &  699$\pm$568 &  223$\pm$111 &  1.55$\pm$0.76 \\
\hline

\multicolumn{11}{c}{Col 110}\\\cline {5-7} \\
      & \multicolumn{2}{c}{Evolved} & 
         \multicolumn{2}{c} {$\chi$} & 
          \multicolumn{3}{c}{Observed+Evolved} &  
           \multicolumn{3}{c}{Extrapolated+Evolved} \\\cline {2-11} \\  
 {Region}  & {N*}    & {m$_{evol}$}        & {1.08-1.48} & {-} & {N*}             & {m$_{obs}$}         & {$\rho$} & {N*}& {m$_{tot}$  } & {$\rho$} \\
   (pc)     & (Stars) & ($10^1 M_{\odot}$)&             &     &  ($10^2 Stars$)   & ($10^2 M_{\odot}$)  &  $M_{\odot} pc^{-3}$        & ($10^2 Stars$) & ($10^2 M_{\odot}$) & $M_{\odot} pc^{-3}$ \\
\hline
0.0-6.25 &  93$\pm$12 &  13.1$\pm$1.7 &  -2.58$\pm$0.21 &  {-} &  4.57$\pm$0.19 &  5.91$\pm$0.41 &  0.58$\pm$0.04 &  7.97$\pm$1 &  8.54$\pm$0.56 &  0.83$\pm$0.05 \\
6.25-12.12 &  70$\pm$16 &  10$\pm$2.3 &  -1.84$\pm$0.51 &  {-} &  4.5$\pm$0.26 &  5.78$\pm$0.86 &  0.09$\pm$0.01 &  10$\pm$6.74 &  9.67$\pm$2.45 &  0.15$\pm$0.04 \\
0.0-12.12 &  163$\pm$21 &  23.1$\pm$3 &  -2.84$\pm$0.57 &  {-} &  9.06$\pm$0.34 &  11.7$\pm$1.84 &  0.16$\pm$0.02 &  15.1$\pm$3.95 &  16.5$\pm$2.44 &  0.22$\pm$0.03 \\
\hline

\multicolumn{11}{c}{NGC 2262}\\\cline {5-7} \\
      & \multicolumn{2}{c}{Evolved} & 
         \multicolumn{2}{c} {$\chi$} & 
          \multicolumn{3}{c}{Observed+Evolved} &  
           \multicolumn{3}{c}{Extrapolated+Evolved} \\\cline {2-11} \\  
 {Region}  & {N*}    & {m$_{evol}$}        & {1.33-1.83} & {-} & {N*}             & {m$_{obs}$}         & {$\rho$} & {N*}& {m$_{tot}$  } & {$\rho$} \\
   (pc)     & (Stars) & ($10^1 M_{\odot}$)&             &     &  ($10^2 Stars$)   & ($10^2 M_{\odot}$)  &  $M_{\odot} pc^{-3}$        & ($10^2 Stars$) & ($10^2 M_{\odot}$) & $M_{\odot} pc^{-3}$ \\
\hline
0.0-0.85 &  9$\pm$3   &  0. 16$\pm$0.06 &  -1.49$\pm$0.80 &  -     &  0.26$\pm$0.04 &  0.43$\pm$0.16 &  21.8$\pm$8.3 &  0.5$\pm$0.2 &  0.6$\pm$0.2 &  31.8$\pm$9.59 \\
0.85-6.37 &  22$\pm$8  &  0.42$\pm$0.15 &  1.32$\pm$0.61 &  -     &  2.12$\pm$0.14 &  3.35$\pm$1.12 &  0.31$\pm$0.1 &  79.5$\pm$63 &  27.6$\pm$12.4 &  2.6$\pm$1.1 \\
0.0-6.37 &  31$\pm$8  &  0.59$\pm$0.16 &  1.01$\pm$0.44 &  -     &  2.4$\pm$0.14 &  3.8$\pm$0.9 &  0.35$\pm$0.08 &  64.4$\pm$49 &  24.1$\pm$9.4 &  2.2$\pm$0.87 \\
\hline

\multicolumn{11}{c}{NGC 2286}\\\cline {5-7} \\
      & \multicolumn{2}{c}{Evolved} & 
         \multicolumn{2}{c} {$\chi$} & 
          \multicolumn{3}{c}{Observed+Evolved} &  
           \multicolumn{3}{c}{Extrapolated+Evolved} \\\cline {2-11} \\  
 {Region}  & {N*}    & {m$_{evol}$}        & {0.98-2.03} & {-} & {N*}             & {m$_{obs}$}         & {$\rho$} & {N*}& {m$_{tot}$  } & {$\rho$} \\
   (pc)     & (Stars) & ($10^1 M_{\odot}$)&             &     &  ($10^2 Stars$)   & ($10^2 M_{\odot}$)  & $M_{\odot} pc^{-3}$         & ($10^2 Stars$) & ($10^2 M_{\odot}$) & $M_{\odot} pc^{-3}$ \\
\hline
0.0-1.59 &  2$\pm$2   &  0.4$\pm$0.4 &  1.3$\pm$0.5 &  {-} &  0.46$\pm$0.05 &  0.62$\pm$0.17 &  3.93$\pm$1.02 &  6$\pm$4.6 &  2.3$\pm$0.9 &  13.7$\pm$5.31 \\
1.59-6.39 &  3$\pm$5   &  0.6$\pm$1 &  1.76$\pm$0.33 &  {-} &  1.55$\pm$0.14 &  2.08$\pm$0.34 &  0.19$\pm$0.03 &  27$\pm$20.4 &  9.3$\pm$3.8 &  0.86$\pm$0.35 \\
0.0-6.39 &  5$\pm$5   &  1$\pm$1   &  1.45$\pm$0.3 &  {-} &  1.98$\pm$0.15 &  2.7$\pm$0.39 &  0.25$\pm$0.04 &  31.5$\pm$23.6 &  11.1$\pm$4.4 &  1.01$\pm$0.4 \\
\hline

\multicolumn{11}{c}{NGC 2309}\\\cline {5-7} \\
      & \multicolumn{2}{c}{Evolved} & 
         \multicolumn{2}{c} {$\chi$} & 
          \multicolumn{3}{c}{Observed+Evolved} &  
           \multicolumn{3}{c}{Extrapolated+Evolved} \\\cline {2-11} \\  
 {Region}  & {N*}    & {m$_{evol}$}        & {1.68-2.53} & {-} & {N*}             & {m$_{obs}$}         & {$\rho$} & {N*}& {m$_{tot}$  } & {$\rho$} \\
   (pc)     & (Stars) & ($10^1 M_{\odot}$)&             &     &  ($10^2 Stars$)   & ($10^2 M_{\odot}$)  &  $M_{\odot} pc^{-3}$        & ($10^2 Stars$) & ($10^2 M_{\odot}$) & $M_{\odot} pc^{-3}$ \\
\hline
0.0-0.84 &  -     &  -     &  -1.52$\pm$1.03 &  {-} &  0.19$\pm$0.03 &  0.41$\pm$0.46 &  -     &  0.36$\pm$0.18 &  0.58$\pm$0.48 &  23.4$\pm$19.3 \\
0.84-7.5 &  -     &  -     &  -0.84$\pm$0.93 &  {-} &  0.98$\pm$0.08 &  2.05$\pm$2 &  0.12$\pm$0.11 &  -     &  3.87$\pm$3.46 &  0.22$\pm$0.2 \\
0.0-7.5 &  -     &  -     &  -0.89$\pm$0.6 &  {-} &  1.16$\pm$0.09 &  2.43$\pm$1.54 &  0.14$\pm$0.09 &  -     &  4.5$\pm$2.51 &  0.25$\pm$0.14 \\
\hline

\multicolumn{11}{c}{Be 36}\\\cline {5-7} \\
      & \multicolumn{2}{c}{Evolved} & 
         \multicolumn{2}{c} {$\chi$} & 
          \multicolumn{3}{c}{Observed+Evolved} &  
           \multicolumn{3}{c}{Extrapolated+Evolved} \\\cline {2-11} \\  
 {Region}  & {N*}    & {m$_{evol}$}        & {-} & {-} & {N*}             & {m$_{obs}$}         & {$\rho$} & {N*}& {m$_{tot}$  } & {$\rho$} \\
   (pc)     & (Stars) & ($10^1 M_{\odot}$)&             &     &  ($10^2 Stars$)   & ($10^2 M_{\odot}$)  & $M_{\odot} pc^{-3}$         & ($10^2 Stars$) & ($10^2 M_{\odot}$) & $M_{\odot} pc^{-3}$ \\
\hline
0.0-1.32 &  15$\pm$4  &  2.1$\pm$0.6 &  -     &  {-} &  -     &  -     &  -     &  -     &  -     &  - \\
1.32-13.39 &  83$\pm$19 &  11.8$\pm$2.7 &  -     &  {-} &  -     &  -     &  -     &  -     &  -     &  - \\
0.0-13.39 &  98$\pm$20 &  13.8$\pm$2.8 &  -     &  {-} &  -     &  -     &  -     &  -     &  -     &  - \\
\hline

\multicolumn{11}{c}{Haffner 8}\\\cline {5-7} \\
      & \multicolumn{2}{c}{Evolved} & 
         \multicolumn{2}{c} {$\chi$} & 
          \multicolumn{3}{c}{Observed+Evolved} &  
           \multicolumn{3}{c}{Extrapolated+Evolved} \\\cline {2-11} \\  
 {Region}  & {N*}    & {m$_{evol}$}        & {1.08-1.88} & {-} & {N*}             & {m$_{obs}$}         & {$\rho$} & {N*}& {m$_{tot}$  } & {$\rho$} \\
   (pc)     & (Stars) & ($10^1 M_{\odot}$)&             &     &  ($10^2 Stars$)   & ($10^2 M_{\odot}$)  & $M_{\odot} pc^{-3}$         & ($10^2 Stars$) & ($10^2 M_{\odot}$) & $M_{\odot} pc^{-3}$ \\
\hline
0.0-1.47 &  -     &  -     &  1.28$\pm$0.77 &  {-} &  0.29$\pm$0.05 &  0.4$\pm$0.16 &  2.27$\pm$0.9 &  5.98$\pm$5.23 &  2.07$\pm$1.05 &  11.6$\pm$5.88 \\
1.47-6.93 &  -     &  -     &  1.99$\pm$0.71 &  {-} &  1.02$\pm$0.15 &  1.39$\pm$0.48 &  0.1$\pm$0.03 &  25$\pm$19.6 &  8.43$\pm$3.75 &  0.61$\pm$0.27 \\
0.0-6.93 &  -     &  -     &  1.82$\pm$0.59 &  {-} &  1.23$\pm$0.16 &  1.69$\pm$0.5 &  0.12$\pm$0.04 &  29.1$\pm$22.5 &  9.85$\pm$4.27 &  0.71$\pm$0.31 \\
\hline

\end{tabular}
\end{table*}

\begin{table*}
\renewcommand\thetable{S5}
  \centering
\tiny
\begin{tabular}{c|c|c|c|c|c|c|c|c|c|c}

\multicolumn{11}{c}{Mel 71}\\\cline {5-7} \\
      & \multicolumn{2}{c}{Evolved} & 
         \multicolumn{2}{c} {$\chi$} & 
          \multicolumn{3}{c}{Observed+Evolved} &  
           \multicolumn{3}{c}{Extrapolated+Evolved} \\\cline {2-11} \\  
 {Region}  & {N*}    & {m$_{evol}$}        & {1.08-1.78} & {-} & {N*}             & {m$_{obs}$}         & {$\rho$} & {N*}& {m$_{tot}$  } & {$\rho$} \\
   (pc)     & (Stars) & ($10^1 M_{\odot}$)&             &     &  ($10^2 Stars$)   & ($10^2 M_{\odot}$)  &  $M_{\odot} pc^{-3}$        & ($10^2 Stars$) & ($10^2 M_{\odot}$) & $M_{\odot} pc^{-3}$ \\
\hline
0.0-1.27 &  13$\pm$4  &  2.4$\pm$0.7 &  0.3$\pm$1.04 &  -     &  0.77$\pm$0.06 &  1.13$\pm$0.44 &  13.1$\pm$5.11 &  -     &  -     &  - \\
1.27-5.0 &  18$\pm$7  &  3.2$\pm$1.3 &  1.82$\pm$0.42 &  -     &  2.4$\pm$0.15 &  3.39$\pm$0.63 &  0.66$\pm$0.12 &  62.9$\pm$47.8 &  21.4$\pm$8.95 &  4.16$\pm$1.74 \\
0.0-5.0 &  31$\pm$8  &  5.5$\pm$1.4 &  1.29$\pm$0.4 &  -     &  3.17$\pm$0.16 &  4.44$\pm$0.72 &  0.85$\pm$0.14 &  64.1$\pm$49.0 &  21.8$\pm$9.21 &  4.26$\pm$1.76 \\
\hline

\multicolumn{11}{c}{NGC 2506}\\\cline {5-7} \\
      & \multicolumn{2}{c}{Evolved} & 
         \multicolumn{2}{c} {$\chi$} & 
          \multicolumn{3}{c}{Observed+Evolved} &  
           \multicolumn{3}{c}{Extrapolated+Evolved} \\\cline {2-11} \\  
 {Region}  & {N*}    & {m$_{evol}$}        & {1.23-1.53} & {-} & {N*}             & {m$_{obs}$}         & {$\rho$} & {N*}& {m$_{tot}$  } & {$\rho$} \\
   (pc)     & (Stars) & ($10^1 M_{\odot}$)&             &     &  ($10^2 Stars$)   & ($10^2 M_{\odot}$)  & $M_{\odot} pc^{-3}$         & ($10^2 Stars$) & ($10^2 M_{\odot}$) & $M_{\odot} pc^{-3}$ \\
\hline
0.0-1.65 &  17$\pm$4  &  2.8$\pm$0.7 &  4.11$\pm$1.63 &  {-} &  1.14$\pm$0.07 &  1.69$\pm$1.21 &  8.96$\pm$6.42 &  75.8$\pm$74.4 &  25.1$\pm$16.5 &  133$\pm$87.9 \\
1.65-10.76 &  45$\pm$13 &  7.4$\pm$2.2 &  1.48$\pm$0.6 &  {-} &  4.43$\pm$0.21 &  6.16$\pm$1.45 &  0.12$\pm$.0.03 &  207$\pm$161 &  68.2$\pm$30.6 &  1.31$\pm$0.59 \\
0.0-10.76 &  63$\pm$14 &  10.2$\pm$2.3 &  0.97$\pm$0.63 &  {-} &  5.58$\pm$0.22 &  7.77$\pm$1.9 &  0.15$\pm$0.04 &  188$\pm$151 &  65.8$\pm$29.4 &  1.26$\pm$0.56 \\
\hline

\multicolumn{11}{c}{Pismis 3}\\\cline {5-7} \\
      & \multicolumn{2}{c}{Evolved} & 
         \multicolumn{2}{c} {$\chi$} & 
          \multicolumn{3}{c}{Observed+Evolved} &  
           \multicolumn{3}{c}{Extrapolated+Evolved} \\\cline {2-11} \\  
 {Region}  & {N*}    & {m$_{evol}$}        & {1.13-1.38} & {-} & {N*}             & {m$_{obs}$}         & {$\rho$} & {N*}& {m$_{tot}$  } & {$\rho$} \\
   (pc)     & (Stars) & ($10^1 M_{\odot}$)&             &     &  ($10^2 Stars$)   & ($10^2 M_{\odot}$)  & $M_{\odot} pc^{-3}$         & ($10^2 Stars$) & ($10^2 M_{\odot}$) & $M_{\odot} pc^{-3}$ \\
\hline
0.0-2.1 &  63$\pm$9  &  8.6$\pm$1.3 &  -1.6$\pm$0.84 &  {-} &  2.77$\pm$0.14 &  3.54$\pm$0.74 &  6.77$\pm$1.41 &  8.34$\pm$16.6 &  7.45$\pm$5.43 &  14.2$\pm$10.4 \\
2.1-8.58 &  29$\pm$21 &  4$\pm$0.2.8 &  2.94$\pm$0.56 &  {-} &  6.37$\pm$0.34 &  7.9$\pm$1.32 &  0.31$\pm$.0.05 &  345$\pm$267 &  109$\pm$49.8 &  4.19$\pm$1.92 \\
0.0-8.58 &  92$\pm$23 &  12.5$\pm$3.1 &  1.71$\pm$0.49 &  {-} &  9.14$\pm$0.38 &  11.4$\pm$1.59 &  0.43$\pm$0.06 &  418$\pm$321 &  133$\pm$59.8 &  5.03$\pm$2.26 \\
\hline

\multicolumn{11}{c}{NGC 3680}\\\cline {5-7} \\
      & \multicolumn{2}{c}{Evolved} & 
         \multicolumn{2}{c} {$\chi$} & 
          \multicolumn{3}{c}{Observed+Evolved} &  
           \multicolumn{3}{c}{Extrapolated+Evolved} \\\cline {2-11} \\  
 {Region}  & {N*}    & {m$_{evol}$}        & {1.08-1.48} & {-} & {N*}             & {m$_{obs}$}         & {$\rho$} & {N*}& {m$_{tot}$  } & {$\rho$} \\
   (pc)     & (Stars) & ($10^1 M_{\odot}$)&             &     &  ($10^2 Stars$)   & ($10^2 M_{\odot}$)  & $M_{\odot} pc^{-3}$         & ($10^2 Stars$) & ($10^2 M_{\odot}$) & $M_{\odot} pc^{-3}$ \\
\hline
0.0-0.47 &  4$\pm$2   &  0.7$\pm$0.4 &  -     &  {-} &  -     &  -     &  -     &  -     &  -     &  - \\
0.47-2.98 &  11$\pm$4  &  1.9$\pm$0.8 &  0.24$\pm$1.04 &  {-} &  0.36$\pm$0.06 &  0.44$\pm$0.16 &  0.57$\pm$0.16 &  -     &  -     &  - \\
0.0-1.72 &  14$\pm$5  &  2.6$\pm$0.8 &  -1.2$\pm$2.03 &  {-} &  0.55$\pm$0.08 &  0.77$\pm$0.35 &  0.7$\pm$0.32 &  -     &  -     &  - \\
\hline

\multicolumn{11}{c}{Ru 96}\\\cline {5-7} \\
      & \multicolumn{2}{c}{Evolved} & 
         \multicolumn{2}{c} {$\chi$} & 
          \multicolumn{3}{c}{Observed+Evolved} &  
           \multicolumn{3}{c}{Extrapolated+Evolved} \\\cline {2-11} \\  
 {Region}  & {N*}    & {m$_{evol}$}        & {1.33-1.93} & {-} & {N*}             & {m$_{obs}$}         & {$\rho$} & {N*}& {m$_{tot}$  } & {$\rho$} \\
   (pc)     & (Stars) & ($10^1 M_{\odot}$)&             &     &  ($10^2 Stars$)   & ($10^2 M_{\odot}$)  &  $M_{\odot} pc^{-3}$        & ($10^2 Stars$) & ($10^2 M_{\odot}$) & $M_{\odot} pc^{-3}$ \\
\hline
0.0-1.43 &  7$\pm$3   &  1.4$\pm$0.7 &  4.58$\pm$0.98 &  {-} &  0.36$\pm$0.05 &  0.71$\pm$0.37 &  5.75$\pm$2.99 &  23.9$\pm$21 &  8.15$\pm$4.46 &  66.5$\pm$36.4 \\
1.43-2.6 &  9$\pm$4   &  1.8$\pm$0.9 &  6.18$\pm$0.72 &  {-} &  0.52$\pm$0.07 &  0.82$\pm$0.29 &  1.34$\pm$0.48 &  32.5$\pm$26.6 &  10.9$\pm$5.3 &  17.7$\pm$8.63 \\
0.0-2.6 &  15$\pm$6  &  3.1$\pm$1.1 &  4.55$\pm$0.65 &  {-} &  0.92$\pm$0.09 &  1.49$\pm$0.48 &  2.02$\pm$0.66 &  46.5$\pm$37.3 &  15.8$\pm$7.36 &  21.4$\pm$10 \\
\hline

\multicolumn{11}{c}{Ru 105}\\\cline {5-7} \\
      & \multicolumn{2}{c}{Evolved} & 
         \multicolumn{2}{c} {$\chi$} & 
          \multicolumn{3}{c}{Observed+Evolved} &  
           \multicolumn{3}{c}{Extrapolated+Evolved} \\\cline {2-11} \\  
 {Region}  & {N*}    & {m$_{evol}$}        & {-} & {-} & {N*}             & {m$_{obs}$}         & {$\rho$} & {N*}& {m$_{tot}$  } & {$\rho$} \\
   (pc)     & (Stars) & ($10^1 M_{\odot}$)&             &     &  ($10^2 Stars$)   & ($10^2 M_{\odot}$)  &  $M_{\odot} pc^{-3}$        & ($10^2 Stars$) & ($10^2 M_{\odot}$) & $M_{\odot} pc^{-3}$\\
\hline
0.0-1.79 &  10$\pm$4  &  2.2$\pm$0.8 &  -     &  {-} &  -     &  -     &  -     &  -     &  -     &  - \\
1.79-3.83 &  9$\pm$5   &  1.8$\pm$1.1 &  -     &  {-} &  -     &  -     &  -     &  -     &  -     &  - \\
0.0-3.83 &  19$\pm$7  &  4$\pm$1.4 &  -     &  {-} &  -     &  -     &  -     &  -     &  -     &  - \\
\hline

\multicolumn{11}{c}{Trumpler 20}\\\cline {5-7} \\
      & \multicolumn{2}{c}{Evolved} & 
         \multicolumn{2}{c} {$\chi$} & 
          \multicolumn{3}{c}{Observed+Evolved} &  
           \multicolumn{3}{c}{Extrapolated+Evolved} \\\cline {2-11} \\  
 {Region}  & {N*}    & {m$_{evol}$}        & {1.38-1.78} & {-} & {N*}             & {m$_{obs}$}         & {$\rho$} & {N*}& {m$_{tot}$  } & {$\rho$} \\
   (pc)     & (Stars) & ($10^1 M_{\odot}$)&             &     &  ($10^2 Stars$)   & ($10^2 M_{\odot}$)  &  $M_{\odot} pc^{-3}$        & ($10^2 Stars$) & ($10^2 M_{\odot}$) & $M_{\odot} pc^{-3}$\\
\hline
0.0-3.12 &  69$\pm$10 &  12.3$\pm$1.8 &  -1.07$\pm$0.5 &  {-} &  2.18$\pm$1.35 &  3.58$\pm$0.79 &  2.81$\pm$0.62 &  6.7$\pm$10.8 &  6.94$\pm$3.32 &  5.45$\pm$2.61 \\
3.12-13.98 &  188$\pm$32 &  33.6$\pm$5.7 &  2.85$\pm$0.95 &  {-} &  7.9$\pm$0.45 &  12.6$\pm$5.46 &  0.11$\pm$0.05 &  379$\pm$326 &  130$\pm$68.2 &  1.15$\pm$0.6 \\
0.0-13.98 &  256$\pm$34 &  45.9$\pm$6 &  2.06$\pm$0.68 &  {-} &  10.1$\pm$0.47 &  16.2$\pm$4.96 &  0.14$\pm$0.04 &  434$\pm$350 &  150$\pm$69.9 &  1.31$\pm$0.61 \\
\hline

\multicolumn{11}{c}{Pismis 19}\\\cline {5-7} \\
      & \multicolumn{2}{c}{Evolved} & 
         \multicolumn{2}{c} {$\chi$} & 
          \multicolumn{3}{c}{Observed+Evolved} &  
           \multicolumn{3}{c}{Extrapolated+Evolved} \\\cline {2-11} \\  
 {Region}  & {N*}    & {m$_{evol}$}        & {1.38-2.18} & {-} & {N*}             & {m$_{obs}$}         & {$\rho$} & {N*}& {m$_{tot}$  } & {$\rho$} \\
   (pc)     & (Stars) & ($10^1 M_{\odot}$)&             &     &  ($10^2 Stars$)   & ($10^2 M_{\odot}$)  & $M_{\odot} pc^{-3}$         & ($10^2 Stars$) & ($10^2 M_{\odot}$) & $M_{\odot} pc^{-3}$\\
\hline
0.0-0.54 &  14$\pm$4  &  3.1$\pm$0.9 &  -2.42$\pm$1.07 &  {-} &  0.63$\pm$0.06 &  1.16$\pm$0.71 &  175$\pm$108 &  0.95$\pm$0.23 &  1.47$\pm$0.74 &  222$\pm$112 \\
0.54-6.16 &  26$\pm$12 &  5.9$\pm$2.7 &  1.57$\pm$0.4 &  {-} &  5.16$\pm$0.3 &  8.91$\pm$2.55 &  0.91$\pm$0.26 &  173$\pm$133 &  61.5$\pm$25.8 &  6.35$\pm$2.64 \\
0.0-6.16 &  40$\pm$13 &  9$\pm$2.9 &  1.18$\pm$0.38 &  {-} &  5.38$\pm$0.28 &  9.19$\pm$2.69 &  1.03$\pm$0.28 &  158$\pm$119 &  59.1$\pm$22.9 &  6.04$\pm$2.34 \\
\hline

\multicolumn{11}{c}{NGC 6134}\\\cline {5-7} \\
      & \multicolumn{2}{c}{Evolved} & 
         \multicolumn{2}{c} {$\chi$} & 
          \multicolumn{3}{c}{Observed+Evolved} &  
           \multicolumn{3}{c}{Extrapolated+Evolved} \\\cline {2-11} \\  
 {Region}  & {N*}    & {m$_{evol}$}        & {1.28-1.78} & {-} & {N*}             & {m$_{obs}$}         & {$\rho$} & {N*}& {m$_{tot}$  } & {$\rho$} \\
   (pc)     & (Stars) & ($10^1 M_{\odot}$)&             &     &  ($10^2 Stars$)   & ($10^2 M_{\odot}$)  &  $M_{\odot} pc^{-3}$        & ($10^2 Stars$) & ($10^2 M_{\odot}$) & $M_{\odot} pc^{-3}$\\
\hline
0.0-0.45 &  9$\pm$3   &  1.5$\pm$0.5 &  -0.95$\pm$0.9 &  {-} &  0.34$\pm$0.04 &  0.52$\pm$0.20 &  138$\pm$53.6 &  -     &  0.99$\pm$0.64 &  259$\pm$167 \\
0.45-3.08 &  42$\pm$21 &  7.6$\pm$1.5 &  -1.3$\pm$1.33 &  {-} &  1.54$\pm$0.14 &  2.47$\pm$1.39 &  2.03$\pm$.1.14 &  -     &  3.97$\pm$2.41 &  3.26$\pm$1.97 \\
0.0-3.08 &  51$\pm$9  &  9.1$\pm$1.6 &  -0.83$\pm$1.11 &  {-} &  1.87$\pm$0.15 &  2.98$\pm$1.39 &  2.44$\pm$1.13 &  -     &  -     &  - \\
\hline

\end{tabular}
\end{table*}

\begin{table*}
\renewcommand\thetable{S5}
  \centering
\tiny
\begin{tabular}{c|c|c|c|c|c|c|c|c|c|c}

\multicolumn{11}{c}{IC 4651}\\\cline {5-7} \\
      & \multicolumn{2}{c}{Evolved} & 
         \multicolumn{2}{c} {$\chi$} & 
          \multicolumn{3}{c}{Observed+Evolved} &  
           \multicolumn{3}{c}{Extrapolated+Evolved} \\\cline {2-11} \\  
 {Region}  & {N*}    & {m$_{evol}$}        & {0.93-1.53} & {-} & {N*}             & {m$_{obs}$}         & {$\rho$} & {N*}& {m$_{tot}$  } & {$\rho$} \\
   (pc)     & (Stars) & ($10^1 M_{\odot}$)&             &     &  ($10^2 Stars$)   & ($10^2 M_{\odot}$)  &  $M_{\odot} pc^{-3}$        & ($10^2 Stars$) & ($10^2 M_{\odot}$) & $M_{\odot} pc^{-3}$\\
\hline
0.0-1.02 &  16$\pm$4  &  2.4$\pm$0.7 &  -2.78$\pm$0.75 &  -     &  0.97$\pm$0.07 &  1.27$\pm$0.28 &  28.5$\pm$6.27 &  1.24$\pm$0.21 &  1.45$\pm$0.29 &  32.6$\pm$6.53 \\
1.02-2.35 &  19$\pm$6  &  2.9$\pm$0.9 &  0.26$\pm$0.7 &  -     &  1.59$\pm$0.12 &  1.93$\pm$0.3 &  3.87$\pm$0.6 &  -     &  -     &  - \\
0.0-2.35 &  35$\pm$7  &  5.4$\pm$1.1 &  -0.6$\pm$0.41 &  -     &  2.55$\pm$0.14 &  3.21$\pm$0.35 &  5.91$\pm$0.65 &  -     &  -     &  - \\
\hline

\multicolumn{11}{c}{NGC 6802}\\\cline {5-7} \\
      & \multicolumn{2}{c}{Evolved} & 
         \multicolumn{2}{c} {$\chi$} & 
          \multicolumn{3}{c}{Observed+Evolved} &  
           \multicolumn{3}{c}{Extrapolated+Evolved} \\\cline {2-11} \\  
 {Region}  & {N*}    & {m$_{evol}$}        & {1.28-2.08} & {-} & {N*}             & {m$_{obs}$}         & {$\rho$} & {N*}& {m$_{tot}$  } & {$\rho$} \\
   (pc)     & (Stars) & ($10^1 M_{\odot}$)&             &     &  ($10^2 Stars$)   & ($10^2 M_{\odot}$)  &  $M_{\odot} pc^{-3}$        & ($10^2 Stars$) & ($10^2 M_{\odot}$) & $M_{\odot} pc^{-3}$\\
\hline
0.0-1.03 &  18$\pm$4  &  3.9$\pm$0.9 &  -0.46$\pm$0.84 &  {-} &  1.02$\pm$0.07 &  1.8$\pm$0.88 &  39.4$\pm$19.2 &  -     &  -     &  - \\
1.03-4.24 &  25$\pm$7  &  5.3$\pm$1.4 &  2.4$\pm$0.24 &  {-} &  3.59$\pm$0.17 &  5.77$\pm$0.85 &  1.83$\pm$0.27 &  121$\pm$91.2 &  42.1$\pm$17 &  13.4$\pm$5.42 \\
0.0-4.24 &  43$\pm$8  &  9.2$\pm$1.7 &  1.66$\pm$0.24 &  {-} &  4.59$\pm$0.18 &  7.53$\pm$1.09 &  2.36$\pm$0.34 &  130$\pm$97.7 &  46.4$\pm$18.3 &  14.5$\pm$5.72 \\
\hline

\multicolumn{11}{c}{NGC 6819}\\\cline {5-7} \\
      & \multicolumn{2}{c}{Evolved} & 
         \multicolumn{2}{c} {$\chi$} & 
          \multicolumn{3}{c}{Observed+Evolved} &  
           \multicolumn{3}{c}{Extrapolated+Evolved} \\\cline {2-11} \\  
 {Region}  & {N*}    & {m$_{evol}$}        & {1.03-1.53} & {-} & {N*}             & {m$_{obs}$}         & {$\rho$} & {N*}& {m$_{tot}$  } & {$\rho$} \\
   (pc)     & (Stars) & ($10^1 M_{\odot}$)&             &     &  ($10^2 Stars$)   & ($10^2 M_{\odot}$)  & $M_{\odot} pc^{-3}$         & ($10^2 Stars$) & ($10^2 M_{\odot}$) & $M_{\odot} pc^{-3}$\\
\hline
0.0-1.5 &  32$\pm$6  &  4.9$\pm$0.9 &  -1.07$\pm$0.55 &  {-} &  1.94$\pm$0.1 &  2.6$\pm$0.45 &  18.4$\pm$3.2 &  -     &  4.57$\pm$4.19 &  32.3$\pm$29.6 \\
1.5-12.92 &  68$\pm$23 &  10.3$\pm$3.5 &  0.96$\pm$0.42 &  {-} &  8.71$\pm$0.41 &  11$\pm$1.4 &  0.12$\pm$0.01 &  152$\pm$114 &  53.6$\pm$21.4 &  0.59$\pm$0.24 \\
0.0-12.92 &  100$\pm$24 &  15.1$\pm$3.7 &  0.47$\pm$0.4 &  {-} &  10.6$\pm$0.42 &  13.6$\pm$1.63 &  0.15$\pm$0.02 &  124$\pm$87.5 &  49.1$\pm$16.5 &  0.54$\pm$0.18 \\
\hline

 \multicolumn{11}{c}{Be 89}\\\cline {5-7} \\
      & \multicolumn{2}{c}{Evolved} & 
         \multicolumn{2}{c} {$\chi$} & 
          \multicolumn{3}{c}{Observed+Evolved} &  
           \multicolumn{3}{c}{Extrapolated+Evolved} \\\cline {2-11} \\

{Region}  & {N*}    & {m$_{evol}$}  & {1.13-1.63} & {-} & {N*}             & {m$_{obs}$}         & {$\rho$} & {N*}& {m$_{tot}$  } & {$\rho$} \\
(pc)     & (Stars) & ($10^1 M_{\odot}$)&             &     &  ($10^2 Stars$)   & ($10^2 M_{\odot}$)  & $M_{\odot} pc^{-3}$         & ($10^2 Stars$) & ($10^2 M_{\odot}$) & $M_{\odot} pc^{-3}$\\
\hline
 0.0-2.75 &  11$\pm$5  &  1.9$\pm$0.8 &  0.18$\pm$0.66 &  {-} &  1.15$\pm$0.09 &  1.56$\pm$0.37 &  1.79$\pm$0.43 &  -     &  -     &  - \\
2.75-7.48 &  12$\pm$10 &  2$\pm$1.7 &  1.27$\pm$1.05 &  {-} &  2.12$\pm$0.21 &  2.84$\pm$1.13 &  0.17$\pm$0.07 &  62.5$\pm$60.5 &  20.9$\pm$12.3 &  1.25$\pm$0.74 \\
 0.0-7.48 &  23$\pm$12 &  3.8$\pm$1.9 &  1.64$\pm$0.93 &  {-} &  3.26$\pm$0.23 &  4.42$\pm$1.57 &  0.25$\pm$0.09 &  96$\pm$75.8 &  32$\pm$14.6 &  1.83$\pm$0.83 \\  
\hline

\multicolumn{11}{c}{NGC 6939}\\\cline {5-7} \\
      & \multicolumn{2}{c}{Evolved} & 
         \multicolumn{2}{c} {$\chi$} & 
          \multicolumn{3}{c}{Observed+Evolved} &  
           \multicolumn{3}{c}{Extrapolated+Evolved} \\\cline {2-11} \\  
 {Region}  & {N*}    & {m$_{evol}$}        & {1.03-1.63} & {-} & {N*}             & {m$_{obs}$}         & {$\rho$} & {N*}& {m$_{tot}$  } & {$\rho$} \\
  (pc)     & (Stars) & ($10^1 M_{\odot}$)&             &     &  ($10^2 Stars$)   & ($10^2 M_{\odot}$)  &  $M_{\odot} pc^{-3}$        & ($10^2 Stars$) & ($10^2 M_{\odot}$) & $M_{\odot} pc^{-3}$\\
\hline
0.0-1.16 &  25$\pm$5  &  4.$\pm$0.8 &  -2.17 $\pm$0.66 &  {-} &  1.01$\pm$0.07 &  1.44$\pm$0.31 &  22$\pm$4.7 &  1.45$\pm$0.44 &  1.75$\pm$0.35 &  26.7$\pm$5.53 \\
1.16-4.92 &  43$\pm$9  &  7$\pm$1.5 &  -0.51$\pm$0.46 &  {-} &  3.3$\pm$0.16 &  4.47$\pm$0.71 &  1.41$\pm$0.22 &  -     &  -     &  - \\
0.0-4.92 &  68$\pm$10 &  11.1$\pm$1.7 &  -0.84$\pm$0.46 &  {-} &  4.3$\pm$0.17 &  5.89$\pm$0.91 &  1.82$\pm$0.28 &  -     &  -     &  - \\
\hline

\multicolumn{11}{c}{NGC 7142}\\\cline {5-7} \\
      & \multicolumn{2}{c}{Evolved} & 
         \multicolumn{2}{c} {$\chi$} & 
          \multicolumn{3}{c}{Observed+Evolved} &  
           \multicolumn{3}{c}{Extrapolated+Evolved} \\\cline {2-11} \\  
 {Region}  & {N*}    & {m$_{evol}$}        & {1.18-1.53} & {-} & {N*}             & {m$_{obs}$}         & {$\rho$} & {N*}& {m$_{tot}$  } & {$\rho$} \\
  (pc)     & (Stars) & ($10^1 M_{\odot}$)&             &     &  ($10^2 Stars$)   & ($10^2 M_{\odot}$)  &  $M_{\odot} pc^{-3}$        & ($10^2 Stars$) & ($10^2 M_{\odot}$) & $M_{\odot} pc^{-3}$\\
\hline
0.0-1.98 &  22$\pm$5  &  3.4$\pm$0.8 &  -1.97 $\pm$1.44 &  {-} &  0.96$\pm$0.07 &  1.36$\pm$0.68 &  4.19$\pm$2.09 &  -     &  2.48$\pm$1.19 &  7.61$\pm$3.67 \\
1.98-11.19 &  85$\pm$18 &  12.9$\pm$2.7 &  -0.58$\pm$0.68 &  {-} &  4.24$\pm$0.25 &  5.86$\pm$1.35 &  0.1$\pm$0.02 &  -     &  -     &  - \\
0.0-11.19 &  107$\pm$19 &  16.3$\pm$2.9 &  -1.06$\pm$0.56 &  {-} &  5.21$\pm$0.25 &  7.22$\pm$1.38 &  0.12$\pm$0.02 &  -     &  -     &  - \\
\hline

\multicolumn{11}{c}{NGC 7789}\\\cline {5-7} \\
      & \multicolumn{2}{c}{Evolved} & 
         \multicolumn{2}{c} {$\chi$} & 
          \multicolumn{3}{c}{Observed+Evolved} &  
           \multicolumn{3}{c}{Extrapolated+Evolved} \\\cline {2-11} \\  
 {Region}  & {N*}    & {m$_{evol}$}        & {1.08-1.38} & {-} & {N*}             & {m$_{obs}$}         & {$\rho$} & {N*}& {m$_{tot}$  } & {$\rho$} \\
 (pc)     & (Stars) & ($10^1 M_{\odot}$)&             &     &  ($10^2 Stars$)   & ($10^2 M_{\odot}$)  &  $M_{\odot} pc^{-3}$        & ($10^2 Stars$) & ($10^2 M_{\odot}$) & $M_{\odot} pc^{-3}$\\
\hline
0.0-2.32 &  74$\pm$9  &  12$\pm$1.5 &  -0.42$\pm$0.45 &  {-} &  3.87$\pm$0.14 &  5.4$\pm$0.82 &  10.3$\pm$1.56 &  -     &  -     &  - \\
2.32-26.88 &  147$\pm$60 &  23.6$\pm$9.7 &  1.03$\pm$0.66 &  {-} &  19.9$\pm$1.02 &  24.7$\pm$4.25 &  0.03$\pm$0.005 &  574$\pm$471 &  192$\pm$90.5 &  0.24$\pm$0.11 \\
0.0-26.88 &  221$\pm$61 &  35.6$\pm$9.8 &  0.79$\pm$0.65 &  {-} &  23.8$\pm$1.03 &  29.7$\pm$4.87 &  0.04$\pm$0.006 &  556$\pm$444 &  194$\pm$85.5 &  0.24$\pm$0.1 \\
\hline
\hline

\end{tabular}
\\
Col.~1: the distance from the core. Cols.~2,6,9 : cluster stars for the regions in Col.~1.
Col.~4 gives the MF slopes ($\chi$),  derived for the low-mass and high-mass ranges. 
The masses of $m_{evol}$, $m_{obs}$, and $m_{tot}$ are listed in Cols.~3, 7 and 10, respectively.
The mass densities are given in Cols.~8 and 11.
\\
\end{table*}

\end{document}